\documentclass{article}
\usepackage{jheppub}

\usepackage{xcolor}

\definecolor{darkmagenta}{rgb}{0.55, 0.0, 0.55}
\definecolor{darkblue}{rgb}{0.0, 0.0, 0.55}
\definecolor{darkred}{rgb}{0.7, 0.0, 0.3}

\usepackage{bbm}
\usepackage[T1]{fontenc}
\usepackage{graphicx,amssymb,amstext}
\usepackage{subcaption}
\usepackage{mathrsfs}
\usepackage{appendix}
\usepackage{float}
\usepackage{empheq}
\usepackage{mathtools}

\hypersetup{
  colorlinks=true,
  linkcolor=darkblue,
  citecolor=darkmagenta,
  urlcolor=darkred,
  linktoc=all
}

\newcommand{\Hajicek}{Háj{\'{i}}\u{c}ek }
\def\scri{\mathscr{I}}
\let\oldsqrt\sqrt
\def\sqrt{\mathpalette\DHLhksqrt}
\def\DHLhksqrt#1#2{%
  \setbox0=\hbox{$#1\oldsqrt{#2\,}$}\dimen0=\ht0
  \advance\dimen0-0.2\ht0
  \setbox2=\hbox{\vrule height\ht0 depth -\dimen0}%
  {\box0\lower0.9pt\box2}}

\def\d{{\rm d}}

\def\ed{\end{document}}

\def\bea{\begin{eqnarray}}
\def\eea{\end{eqnarray}}
\def\ba{\begin{array}}
\def\ea{\end{array}}

\def\N{{\cal N}}

\makeatletter
\allowdisplaybreaks

\newcommand{\be}{\begin{equation}}
\newcommand{\ee}{\end{equation}}

\newcommand{\vast}{\bBigg@{4}}
\newcommand{\Vast}{\bBigg@{5}}

\preprint{}

\title{\boldmath Gravitational Memory Beyond Null Infinity through Finite-Distance Carrollian Screens}

\author[a]{Felipe Diaz}
\author[b,c]{Sercan H\"usn\"ugil}
\author[a]{Oriana Labrin}
\author[d]{Leonardo Sanhueza}

\emailAdd{felipe.diaz@uss.cl}
\emailAdd{shusnugil@perimeterinstitute.ca}
\emailAdd{olabrinz@correo.uss.cl}
\emailAdd{lsanhueza@udec.cl}

\affiliation[a]{Facultad de Ingenier\'ia, Universidad San Sebasti\'an, Santiago 8420524, Chile}
\affiliation[b]{Department of Physics and Astronomy, University of Waterloo, 200 University Avenue West, Waterloo, Ontario, N2L 3G1, Canada}
\affiliation[c]{Perimeter Institute for Theoretical Physics, 31 Caroline Street North, Waterloo, Ontario N2L 2Y5, Canada}
\affiliation[d]{Departamento de F\'isica, Universidad de Concepci\'on, Casilla 160-C, Concepci\'on, Chile}

\abstract{
We investigate gravitational memory beyond null infinity by studying finite-distance null hypersurfaces endowed with Carrollian geometry. We show that the intrinsic degenerate geometry, optical data, and null Brown--York tensor of a finite null screen define a quasilocal Carrollian dissipative system, providing a natural framework to characterize the residual geometric response after the passage of radiation. To make this construction explicit and compare it with the standard asymptotic description, we use Robinson--Trautman spacetimes as an exactly solvable radiative setting.
For asymptotically flat Robinson--Trautman geometries, we transform the solution to Bondi gauge and extract the asymptotic data directly in terms of the Robinson--Trautman field. In the linearized sector, the Bondi shear is purely electric and produces the standard displacement-memory effect associated with the relaxation toward the final Schwarzschild geometry. We show that the leading large-radius tracefree component of the finite-screen memory reduces to the Bondi displacement memory, while finite-distance corrections retain additional focusing, embedding dependence, angular drift, Coulombic data, and near-zone information. Thus, Bondi memory emerges as the universal asymptotic projection of a broader quasilocal Carrollian response. Late-time screens approaching the final Schwarzschild horizon exhibit exponentially decaying non-isotropic Carrollian data, leaving only the isotropic null Brown--York stress.
As a by-product, we construct the Robinson--Trautman solution with nonzero cosmological constant to second order in the radiative amplitude and use the holographic dictionary to study the associated energy fluxes and find that the resulting charge does not obey a universal monotonicity property.
}

\begin{document}

\maketitle

%\tableofcontents

\flushbottom

\section{Introduction}

Asymptotically flat gravity has acquired a new interpretation in recent years \cite{Strominger:2017zoo}. The radiative data at future null infinity are not only the far-zone limit of gravitational waves, but also the dynamical data of a Carrollian boundary geometry and the building blocks of celestial observables. In Bondi--Sachs language \cite{Bondi:1960jsa,Sachs:1961zz,Bondi:1962px, Sachs:1962wk}, radiation is encoded in the Bondi shear $C_{AB}$ and in its retarded-time derivative, the news tensor $N_{AB}$. In the infrared description, the same data enter the flux-balance laws for BMS charges, the Ward identities associated with soft graviton theorems~\cite{He:2014laa,Donnay:2021wrk}, and the light-ray or celestial operators built from null-infinity fields~\cite{Pasterski:2021raf,Raclariu:2021zjz,Gonzo:2020xza}.
Gravitational radiation can be viewed simultaneously as a bulk wave-zone phenomenon, as Carrollian dynamics on $\scri^+$, and as part of the celestial organization of asymptotically flat quantum gravity.

Gravitational memory is one of the most concrete observables in this structure. Classically, the displacement memory effect is the permanent relative displacement of freely falling detectors after a burst of radiation has passed \cite{Zeldovich:1974gvh,Braginsky:1985vlg,Braginsky:1987kwo,Christodoulou:1991cr,Thorne:1992sdb}. In Bondi language, it is the finite change of the asymptotic shear between two cuts of future null infinity, or equivalently the retarded-time integral of the news. In the infrared formulation, the same observable describes a transition between asymptotic vacua related by a supertranslation and is tied to the leading soft graviton theorem through the BMS Ward identity \cite{Strominger:2014pwa,Strominger:2017zoo,Speziale:2025zjp}. Memory therefore provides a useful meeting point between gravitational-wave physics, asymptotic symmetries, and the quantum-infrared structure of gravity.

The standard memory construction is intrinsically asymptotic. It isolates universal radiative data by pushing the observation surface to $\scri^+$ and by choosing a Bondi conformal frame. This is precisely what makes Bondi memory universal, but it also removes finite-distance information. A detector at finite radius does not see a purely asymptotic observable; it sees radiation mixed with Coulombic data, focusing, near-zone fields, and the choice of frame used to compare early and late configurations. This motivates a quasilocal question: what is the finite-distance geometric response whose large-radius radiative projection gives the usual Bondi memory?

A natural arena for this question is a finite null hypersurface. Finite-distance hypersurfaces have proved useful in several areas of gravitational physics. In the fluid/gravity correspondence, for example, timelike cutoff surfaces provide effective descriptions interpolating between horizon and asymptotic dynamics \cite{Bredberg:2010ky}. This suggests that finite null hypersurfaces likewise provide a natural setting for quasilocal observables associated with gravitational radiation and memory.

Since the intrinsic geometry of null hypersurfaces is Carrollian, we shall refer to such finite-distance null hypersurfaces as \emph{Carrollian screens}. They may be viewed as a finite-distance counterpart of future null infinity: both carry a degenerate metric on their spatial cuts, a preferred null evolution vector, and optical data. After a choice of rigging, Carrollian screens also admit a null Brown--York tensor with the structure of a Carrollian fluid stress tensor \cite{Chandrasekaran:2018aop,Chandrasekaran:2021hxc,Chandrasekaran:2021vyu,Chandrasekaran:2020wwn} (see \autoref{App:CarrollManifolds} for a brief review and further references).\footnote{We use the terminology ``Carrollian fluid'' to emphasize the analogy between the null Brown--York tensor and a fluid stress tensor: the null constraint equations organize the screen data into a set of Carrollian variables with fluid-like conservation equations. This terminology is not meant to imply a standard thermodynamic fluid description with a temperature or equation of state. The relation between the null Brown--York data and Carrollian hydrodynamic variables is discussed in \autoref{App:CarrollManifolds}.} This geometric interpretation is also supported by symmetry. The BMS group at null infinity can be understood as a conformal Carrollian symmetry \cite{Duval:2014uva},\footnote{The Carroll group was originally introduced by L\'evy-Leblond as the ultra-relativistic ($c\to0$) contraction of the Poincar\'e group \cite{LevyLeblond:1965,SenGupta:1966}. Its conformal extension was later shown to be naturally related to the BMS group acting at null infinity \cite{Duval:2014uva}.} while more general null hypersurfaces admit boundary-preserving diffeomorphisms with supertranslation-like sectors and localized charges and fluxes  \cite{Chandrasekaran:2018aop,Chandrasekaran:2021hxc,Odak:2023pga}. This viewpoint suggests that the Bondi news and displacement memory arise as the universal wave-zone limit of a more general geometric response of finite Carrollian screens. At finite-distance, however, this response is not expected to be universal. It depends on the chosen screen, on the normalization of its generator, and on the prescription used to compare cuts. This screen dependence is not a defect, but rather the finite-distance information discarded by the strict asymptotic limit. 
The central proposal of this work is that gravitational memory admits a natural quasilocal formulation in terms of finite Carrollian screens, with the Bondi displacement memory emerging as its universal large-radius limit. This quasilocal perspective also connects naturally with several active directions in asymptotically flat gravity. The Carrollian formulation of null infinity recasts the asymptotic Einstein equations as Carrollian Ward identities \cite{Donnay:2022aba,Donnay:2022wvx}, while celestial holography organizes the same radiative data into celestial correlators and light-ray operators \cite{Gonzo:2020xza,Hu:2023geb}. 
Since the finite-screen memory introduced here reduces to the Bondi memory in the large-radius limit, it furnishes a concrete classical setting in which these structures can be followed continuously from finite null hypersurfaces to $\scri^+$.

{ The usefulness of the Carrollian description goes beyond the observation that null surfaces carry a degenerate geometry. The Carrollian generator provides a geometrically preferred flow along which the transverse geometry of the screen evolves and neighboring cuts can be compared. The corresponding optical data organize the evolution into scalar, tensorial, and momentum sectors, governed by the null gravitational balance laws. Gravitational memory can thus be understood as the accumulated geometric response of this Carrollian system during the passage of radiation.

}

We develop this proposal in Robinson--Trautman spacetimes \cite{Robinson:1962zz}, which provide an ideal laboratory for studying gravitational memory beyond the asymptotic regime. Their analytic tractability allows us to determine both the asymptotic Bondi data and the finite-screen Carrollian geometry explicitly. 
They are exact vacuum radiative solutions admitting a preferred geodesic, shear-free, twist-free, expanding null congruence. Their entire radiative dynamics is encoded in a single scalar function obeying the nonlinear Robinson--Trautman equation. Under suitable regularity assumptions, they relax at late retarded time to the Schwarzschild solution \cite{Chrusciel:1991vxx,Chrusciel:1992rv}, providing an analytically controlled interpolation between genuinely radiative and stationary configurations.

Besides providing an ideal laboratory for gravitational memory, Robinson--Trautman spacetimes have also proved valuable in holography, where they provide analytically tractable models of fluid/gravity duality, nonequilibrium relaxation, and Carrollian or flat limits of holographic fluids; see, for example, \cite{Bakas:2014kfa,BernardideFreitas:2014eoi, Gath:2015nxa, Bakas:2015hdc, Ciambelli:2017wou, Skenderis:2017dnh, Adami:2024mtu, Arenas-Henriquez:2025rpt, Castro:2025itb}.

Our first goal is to connect the Robinson--Trautman radiative field to standard asymptotic memory observables. The natural Robinson--Trautman coordinates are adapted to the preferred null congruence, not to the Bondi luminosity-radius condition. We therefore construct the asymptotic coordinate transformation to Bondi gauge and extract the Bondi shear, news tensor, mass aspect, and angular-momentum aspect directly in terms of the Robinson--Trautman function. This gives an explicit map between the Robinson--Trautman radiative field and the Bondi radiative data at $\scri^+$. 
A related asymptotic analysis of Robinson--Trautman memory was recently given in \cite{Barnich:2026wpw} within the Newman--Penrose/Newman--Unti framework. Here, instead, we work directly in Bondi--Sachs gauge { to establish the connection between the finite-screen geometric response and the standard asymptotic displacement memory, which emerges from its leading tracefree large-radius component.}

Having established the Bondi description, we then specialize to linearized Robinson--Trautman modes around Schwarzschild. In this sector the physical content becomes transparent. A single radiative mode produces a purely electric Bondi shear and hence an ordinary displacement-memory effect. The magnetic-parity potential vanishes, in agreement with the Newman--Penrose analysis of \cite{Barnich:2026wpw}, where the asymptotic Robinson--Trautman shear is purely real in a locally asymptotically flat frame. The memory records the relaxation from an initially radiative cut to the final Schwarzschild cut, with higher multipoles suppressed by their faster Robinson--Trautman decay rates. The same exponential profiles also give closed-form classical configurations for retarded-time moments of the news, and therefore provide a simple classical testbed for the relation between memory, Carrollian dynamics, and celestial light-ray observables.

We then move away from null infinity and turn to Carroll screens ${\cal N}$ in Robinson--Trautman spacetime. For screens described by $r=\rho(u,x^A)$, the nullity condition gives an evolution equation for the embedding function $\rho$, sourced by the Robinson--Trautman field and supplemented by suitable initial or boundary data. The induced Carrollian geometry comprises the metric on the spatial cuts, the null generator, and the associated optical and extrinsic data, including the expansion, shear, inaffinity, and \Hajicek momentum one-form. These quantities combine into the null Brown--York tensor, whose projected conservation equations take the form of Carrollian fluid equations intrinsic to the screen. For the Robinson--Trautman screens considered here, the nontrivial dynamical content of these conservation laws reduces precisely to the Robinson--Trautman equation. Thus, the same parabolic relaxation that governs the radiative bulk geometry also controls the quasilocal Carrollian response of finite null screens.

The finite-screen memory defined here is the residual change of the intrinsic Carrollian data of a prescribed null screen between two cuts. The change of the cut metric gives a tensorial memory, its trace gives an area-density or focusing memory, and the change of the \Hajicek one-form gives a momentum component. At finite-distance these quantities contain more than the Bondi displacement memory: they retain focusing, embedding dependence, angular drift, Coulombic data, and near-zone information. The relation to the standard asymptotic observable is obtained by pushing the screen to large radius in Bondi gauge. We show that the leading tracefree part of the finite-screen metric deformation, divided by the screen radius, reproduces the usual Bondi memory. In this sense, Bondi memory is the universal wave-zone projection of a broader quasilocal Carrollian response.

We also study screens that approach the final Schwarzschild horizon at late retarded time. These should not be confused with the exact Robinson--Trautman event horizon, which is a global and teleological object, nor with the apparent horizon, which is determined quasilocally by the Penrose--Tod equation on each cut. Instead, we consider prescribed null screens satisfying the local nullity condition and settling to $r=2m$ as $u\to+\infty$. Their non-perfect Carrollian data decay exponentially: the expansion, tracefree shear, and \Hajicek momentum vanish, while the isotropic Schwarzschild null Brown--York stress remains. This late-time relaxation is in line with the Carrollian-fluid interpretation of dynamical black-hole horizons, where horizon equilibration, teleological boundary conditions, and nonlinear relaxation can be described in Carrollian hydrodynamic language \cite{Redondo-Yuste:2022czg, Husnugil:2025edm}. The resulting memory is therefore a transition from an initially distorted null screen to the final round Schwarzschild screen, not permanent nonspherical horizon hair.

Finally, we discuss Robinson--Trautman spacetimes with nonzero cosmological constant. When $\Lambda\neq0$, the conformal boundary is no longer null and therefore, the standard displacement-memory construction is not directly available. Finite null screens in the bulk, however, remain well defined and still carry Carrollian data. In the corresponding null Brown--York tensor, $\Lambda$ appears as an isotropic background curvature pressure in the scalar sector, modifying the screen embedding and scalar response without introducing a new dissipative channel. Separately, using the holographic dictionary \cite{Poole:2018koa,Compere:2019bua,compere2020lambda}, we construct the Robinson--Trautman solution to second order in the radiative amplitude and analyze the evolution of the associated holographic charge. The resulting finite-$\Lambda$ charge depends on the choice of boundary generator and frame, and the charge considered here does not obey a universal monotonicity property.

The paper is organized as follows. In \autoref{Sec:RTintro} we review Robinson--Trautman radiation and its horizon structure. In \autoref{Sec:BondiMemoryRT} we transform the asymptotically flat family to Bondi gauge and extract the Bondi radiative data. In \autoref{Sec:LinearRT} we specialize to linearized modes and compute the displacement memory. In \autoref{Sec:CarrFluidRT} we introduce finite null screens, their Carrollian geometry, and their null Brown--York fluid data. In \autoref{Sec:FiniteScreenMemory} we define finite-screen memory and show that its large-radius limit reproduces Bondi memory. In \autoref{Sec:RTLambda} we discuss nonzero cosmological constant and the associated holographic flux diagnostic. We conclude with open directions. Technical details are collected in the appendices.

%%%%%%%
%%%%%%%
%%%%%%%
%%%%%%%
%%%%%%%
\section{Robinson--Trautman Waves}\label{Sec:RTintro}

Robinson--Trautman spacetimes provide one of the simplest exact laboratories for studying gravitational radiation beyond perturbation theory \cite{Robinson:1962zz,Podolsky:2016sff}. They describe vacuum radiative geometries which, under suitable regularity assumptions, relax at late retarded times to a Schwarzschild black hole \cite{Chrusciel:1991vxx,Chrusciel:1992rv}. Their importance stems from the fact that they retain enough structure to be analytically tractable while still capturing genuinely nonlinear aspects of gravitational-wave emission.

Geometrically, the Robinson--Trautman class is singled out by the existence of a preferred null congruence which is geodesic, shear-free, twist-free, and expanding \cite{Podolsky:2016sff}. In vacuum, the preferred congruence is a repeated principal null direction, and the generic radiative solution is algebraically special of Petrov type II. These properties allow the Einstein equations to reduce to a single nonlinear fourth-order parabolic equation on the transverse two-sphere. The corresponding scalar field controls the time-dependent conformal geometry of the angular sections and therefore encodes the radiative relaxation of the spacetime.

These features make Robinson--Trautman waves particularly suitable for our purposes. Their natural coordinates are adapted to the preferred null congruence, while an asymptotic transformation to Bondi gauge exposes the standard radiative observables at $\scri^+$. The same solution also admits finite null screens on which the Robinson--Trautman dynamics governs a quasilocal Carrollian response. The family therefore permits a direct comparison between asymptotic Bondi memory and its finite-distance extension.

Robinson--Trautman spacetimes are most naturally written in Newman--Unti coordinates, namely coordinates adapted to the preferred outgoing null congruence. In these coordinates $u_R$ labels null hypersurfaces, $r_R$ is an affine parameter along their null generators, and the angular coordinates $x^A_R$ label the transverse two-dimensional sections. Equivalently, the metric satisfies
\begin{align}
g_{r_R r_R}=0~,\qquad g_{r_R A}=0~,\qquad g_{u_R r_R}=-1~,
\end{align}
so that the preferred null vector is simply
\begin{align}
\ell^\mu\partial_\mu=\partial_{r_R}~,
\end{align}
with $\ell^\nu\nabla_\nu\ell^\mu=0$. This should be distinguished from Bondi--Sachs gauge, where the radial coordinate is instead fixed by the luminosity-radius condition on the determinant of the angular metric. Thus, while the Robinson--Trautman form is adapted to the algebraically special null congruence, a further asymptotic coordinate transformation is required in order to express the solution in standard Bondi gauge and extract the Bondi shear, news tensor, mass aspect, and memory observables.

Using complex stereographic coordinates $x_R^A = (z_R,\bar{z}_R)$ on the transverse sections, the four-dimensional vacuum Robinson--Trautman line element reads\footnote{Throughout the paper, Greek indices $\mu,\nu,\ldots$ denote bulk spacetime directions, lower-case Latin indices $a,b,\ldots$ denote directions tangent to a fixed-$r$ hypersurface, and capital Latin indices $A,B,\ldots$ denote angular directions on the spatial sections of the null hypersurface.}
\begin{align}\label{RTgenLineElement}
   \d s^2 = -F(r_R,x_R^a)\d u^2_R - 2\d u_R \d r_R + 2r_R^2e^{\Phi(x^a_R)}\d z_R\d \bar{z}_R~,
\end{align}
where 
\begin{align}
    F = r_R \partial_{u_R}\Phi - \Delta_R \Phi -\frac{2m}{r_R}~,\qquad \Delta_R = e^{-\Phi}\partial_{z_R}\partial_{\bar{z}_R}~.
\end{align}
The constant $m$ is an integration constant identified with the Robinson--Trautman mass parameter. More generally, one may allow $m=m(u_R)$, but the reparametrization freedom of the line element can be used to set it to a constant (see for instance \cite{Bakas:2014kfa}).

The remaining Einstein equations constrain $\Phi$ to satisfy a parabolic fourth-order nonlinear differential
equation on the transverse sphere
\begin{align}
\Delta_R\Delta_R\Phi + 3m\partial_{u_R}\Phi = 0~,
\end{align}
which is referred to as the Robinson--Trautman equation. 
This equation also admits a geometric interpretation as the Calabi flow of the transverse angular metric. For $g_{AB}=r^2_R h_{AB}$, i.e.\footnote{Here, the notation $2dz_Rd\bar z_R$ means $dz_R\otimes d\bar z_R+d\bar z_R\otimes dz_R$, so that $g_{z_R\bar z_R}=r_R^2e^\Phi$.}
\begin{align}
    h_{AB}\d x_R^A \d x_R^B = 2e^\Phi \d z_R\d\bar z_R~,
\end{align}
the scalar curvature is $\mathcal R_R=-2\Delta_R\Phi$. The Robinson--Trautman equation can be written as
\begin{align}
\partial_{u_R}h_{AB}=\frac{1}{6m}\left(\Delta_R\mathcal R_R\right)h_{AB}~.
\end{align}
Thus the retarded-time evolution of the conformal factor is driven by the Laplacian of the scalar curvature of the angular metric. Up to the conventional normalization and orientation of the flow parameter, this is precisely the Calabi flow on the transverse two-sphere \cite{Tod1989}.
%%%%%%

The radiative character of the solution is also manifest in the Newman--Penrose description
(see \autoref{App:BondiSachs} for our conventions and for the physical interpretation of the Weyl scalars). 
Using a principal null tetrad adapted to the Robinson--Trautman congruence,\footnote{In the Carrollian language used throughout this work, $\ell^\mu$ is the null normal and generator of the hypersurfaces $u_R=\text{constant}$, while $n^\mu$ plays the role of the auxiliary transverse null rigging vector. See \autoref{App:CarrollManifolds} for the corresponding embedded-null-surface geometry and projector conventions.}
\begin{align}
    \ell^\mu\partial_\mu &= \partial_{r_R}~,\\
    n^\mu\partial_\mu &= \partial_{u_R}-\frac{1}{2}F\,\partial_{r_R}~,\\
    m^\mu\partial_\mu &= \frac{e^{-\Phi/2}}{r_R}\partial_{z_R}~,\\
    \bar m^\mu\partial_\mu &= \frac{e^{-\Phi/2}}{r_R}\partial_{\bar z_R}~,
\end{align}
with $\ell\cdot n=-1$ and $m\cdot\bar m=1$ and all other contractions vanishing so that the spacetime metric decomposes as
\begin{align}
g_{\mu\nu}=-2\ell_{(\mu}n_{\nu)}+2m_{(\mu}\bar m_{\nu)}~.
\end{align}

With this tetrad, the Weyl scalars are
\begin{align}
    \Psi_2 &= -\frac{m}{r_R^3}~,\\
    \Psi_3 &= \frac{e^{-\Phi/2}}{2r_R^2}\,
    \partial_{z_R}\!\left(\Delta_R\Phi\right)~,\\
    \Psi_4 &=
    -\frac{1}{2r_R}\,
    \partial_{\bar{z}_R}\!\left(e^{-\Phi}\partial_{\bar{z}_R}\partial_{u_R}\Phi\right)
    -
    \frac{1}{2r_R^2}\,
    \partial_{\bar{z}_R}\!\left(e^{-\Phi}\partial_{\bar{z}_R}\Delta_R\Phi\right)~,
\end{align}
while
\begin{align}
    \Psi_0=0=\Psi_1~.
\end{align}
The vanishing of $\Psi_0$ and $\Psi_1$ reflects the fact that $\ell^\mu$ is a repeated principal null direction, so the generic radiative Robinson--Trautman spacetime is algebraically special of Petrov type II. The Coulombic part is carried by $\Psi_2$, while the leading $O(r_R^{-1})$ term in $\Psi_4$ encodes the outgoing radiative degree of freedom. In particular, when $\partial_{u_R}\Phi=0$ and the angular metric is the round-sphere representative, the radiative scalars vanish and the solution reduces to the Schwarzschild member of the family.

%%%%%
%%%%%
%%%%%
It is often convenient to trade the conformal factor $\Phi$ for the Robinson--Trautman function $P(u_R,x_R^A)$, defined by
\begin{align}
\Phi=-2\log P~.
\end{align}
In terms of $P$, the transverse metric is written as
\begin{align}
2r_R^2e^\Phi \d z_R\d\bar z_R=\frac{2r_R^2}{P^2}\d z_R\d\bar z_R~,
\end{align}
and the deformed sphere Laplacian is
\begin{align}
\Delta_R=P^2\partial_{z_R}\partial_{\bar z_R}~.
\end{align}
The Robinson--Trautman equation then becomes
\begin{align}\label{RTeqWithP}
\Delta_R\Delta_R \log P+3m\partial_{u_R}\log P=0~.
\end{align}
The round unit two-sphere is recovered for the time-independent representative
\begin{align}
P_\circ := 1+\frac{1}{2}z_R\bar z_R~,
\qquad
e^{\Phi_\circ}=P_\circ^{-2}~,
\qquad
\Phi_\circ=-2\log P_\circ~.
\end{align}
With this choice,
\begin{align}
2e^{\Phi_\circ}\d z_R\d\bar z_R=\frac{2\d z_R\d\bar z_R}{\left(1+\frac{1}{2}z_R\bar z_R\right)^2}~,
\end{align}
which is the metric on the unit round sphere in stereographic coordinates, and the solution becomes the Schwarzschild black hole in outgoing Eddington--Finkelstein coordinates.
It is useful to recall the corresponding horizon structure. For a generic radiative Robinson--Trautman spacetime, the future event horizon is not located by the local condition $F=0$, nor is it represented within the Robinson--Trautman exterior patch by a generic finite-$u_R$ graph $r_R=r_{\cal H}(u_R,x_R^A)$. Rather, the Robinson--Trautman geometry relaxes toward Schwarzschild as $u_R\to+\infty$, and the future null boundary of the Robinson--Trautman coordinate region becomes the future event horizon after a suitable extension of the spacetime. Its global structure and regularity have been studied extensively in the literature~\cite{lukacs_perjes_sebestyen_porter_1983,Chrusciel:1992rv,Chrusciel:1992tj,Podolsky:2009an}.

The local nullity equation nevertheless defines a distinguished family of finite null hypersurfaces. Writing such a screen as $r_R=\rho(u_R,x_R^A)$, one finds
\begin{align}
    F(\rho,u_R,x_R^A)
    +2\partial_{u_R}\rho
    +\frac{2e^{-\Phi}}{\rho^2}
    \partial_{z_R}\rho\partial_{\bar z_R}\rho
    =0~.
    \label{RTHorizonNullity}
\end{align}
These hypersurfaces describe ingoing null screens in the Robinson--Trautman exterior. For solutions relaxing to Schwarzschild, an appropriate choice of screen can approach $r_R=2m$ as $u_R\to+\infty$ and hence asymptote to the future event horizon.

There is also a distinct quasilocal horizon notion associated with marginally trapped surfaces on the null slices $u_R=\text{constant}$. Writing the past apparent horizon as $r_R=r_{\rm AH}(u_R,x_R^A)$, the vanishing of the relevant null expansion gives the Penrose--Tod equation~\cite{Tod1989,chow1995apparenthorizonsvacuumrobinsontrautman}
\begin{align}
    \Delta_R\Phi
    +\Delta_R\log r_{\rm AH}
    +\frac{2m}{r_{\rm AH}}
    =0~.
    \label{RTPenroseTodEquation}
\end{align}
For the stationary round representative this reduces to $r_{\rm AH}=2m$, whereas in the radiative case it is an elliptic equation on each constant-$u_R$ cut. It is therefore conceptually distinct both from the global future event horizon and from the null evolution equation defining our finite screens. In the analysis below, we do not attempt to construct either the exact event horizon or the marginally trapped tube. Instead, we study finite ingoing null screens satisfying the local nullity condition and later specialize to screens that asymptote to the final Schwarzschild horizon.

%%%%%%%
%%%%%%%
%%%%%%%
%%%%%%%
%%%%%%%
\section{Bondi Data and Asymptotic Memory}
\label{Sec:BondiMemoryRT}
The Robinson--Trautman coordinates used in the previous section are adapted to the preferred algebraically special null congruence. In particular, $u_R$ labels the null hypersurfaces, $r_R$ is an affine parameter along the generators, and the angular metric on the cuts is conformal to the unit sphere, with conformal factor determined by the Robinson--Trautman field. This is not yet the standard Bondi frame. In Bondi gauge the radial coordinate is instead fixed by the luminosity-radius condition, and the leading angular metric at future null infinity is chosen to be the unit round-sphere metric. Therefore, to identify the standard asymptotic observables, one must perform an asymptotic change of coordinates from the Robinson--Trautman frame to Bondi gauge. 
A related asymptotic analysis of Robinson--Trautman memory was recently given in \cite{Barnich:2026wpw}, using the Newman--Penrose/Newman--Unti formalism and a combined frame rotation and coordinate transformation to a locally asymptotically flat frame. Our treatment below is complementary: we perform a direct Bondi--Sachs metric gauge fixing, keep the resulting Bondi data explicitly in terms of the Robinson--Trautman function, and use the electric/magnetic parity decomposition of the shear as the asymptotic anchor for the finite-screen construction developed later. In the linearized sector, the two descriptions agree on the purely electric character of the Robinson--Trautman shear, as discussed in \autoref{Sec:LinearRT}.
{The Bondi transformation also provides the asymptotic reference required to compare the gravitational memory of finite Carrollian screens with he standard displacement memory at null infinity. Moreover, the resulting Bondi data allow us to investigate the parity decomposition of the shear and celestial observables associated with moments of the news.}

We refer the reader to \autoref{App:BondiSachs} for a brief review of the Bondi--Sachs formalism, our conventions for the asymptotic expansion, and the definition of the Bondi shear, news tensor, mass aspect, and memory observables.

\subsection{Asymptotic Map to Bondi Gauge}
We denote Robinson--Trautman coordinates by $x_R^\mu=(u_R,r_R,x_R^A)$ and Bondi coordinates by $x^\mu=(u,r,x^A)$.
We look for an asymptotic diffeomorphism of the form
\begin{equation}
\begin{aligned}
u_R&=\sum_{n\geq 0} U_{(n)}(u,x^A)r^{-n}~,\\ r_R&=rR_{\rm L}(u,x^A)+\sum_{n\geq 0} R_{(n)}(u,x^A)r^{-n}~, \\ x_R^A&=x^A+\sum_{n\geq 1} Z^A_{(n)}(u,x^A)r^{-n}~.
\end{aligned}
\end{equation}
Up to the residual boundary freedom encoded in the leading time map $U_{(0)}$, the coefficients in this expansion are fixed recursively by the Bondi gauge conditions
\begin{align}
g_{rr}=0~,\qquad g_{rA}=0~,\qquad\partial_r\left(\frac{\det g_{AB}}{r^4}\right)=0~,
\end{align}
together with the choice of round-sphere representative at null infinity,
\begin{align}
q_{AB}dx^Adx^B=\frac{2 \d z \d\bar z}{P_\circ^2}~,
\end{align}
where $P_\circ = 1+\frac12 z\bar z$ is the round-sphere representative defined in the previous section.

We now introduce the leading pullback of the Robinson--Trautman function to Bondi variables,
\begin{align}
\widehat P(u,x^A):=P\big(U_{(0)}(u,x^A),x^A\big)~.
\end{align}
The determinant condition at leading order fixes the relation between the Robinson--Trautman affine radius and the Bondi luminosity radius. One finds
\begin{align}
R_{\rm L}=\frac{\widehat P}{P_\circ}~.
\label{RLleading}
\end{align}
Thus the leading conformal factor in the Robinson--Trautman angular metric is absorbed into the radial rescaling, so that the leading angular metric in Bondi gauge is the round-sphere metric.

The normalization of the retarded time is fixed by the condition $g_{ur}=-1+{\cal O}(r^{-2})$. At leading order this gives\footnote{A similar leading relation between Robinson--Trautman time and asymptotically flat retarded time appears in \cite{Barnich:2026wpw}. Their choice of fixed integration limits fixes the reference Bondi cut, equivalently setting the angular integration function to zero. We keep this function explicit, as it encodes the residual supertranslation-type freedom of the leading time map once the Bondi conformal representative is chosen.}

\begin{align}
\dot U_{(0)}=\frac{P_\circ}{\widehat P}~.
\label{U0condition}
\end{align}
Here $\dot U_{(0)}:=\partial_uU_{(0)}$. Therefore $U_{(0)}$ is determined by the Robinson--Trautman field up to one integration function of the angles, which parametrizes the residual choice of cuts in the leading boundary map. This residual freedom reduces to an ordinary BMS supertranslation after a Bondi conformal frame has been chosen, in particular in the round-sphere stationary limit where the leading time map becomes additive.

Since $P$ is naturally a function of $(u_R,x_R^A)$, while the Bondi data are written in terms of $(u,x^A)$, angular derivatives must distinguish whether $u$ or $u_R$ is held fixed. At leading order near null infinity, the Robinson--Trautman time derivative pulls back to the Bondi variables as
\begin{align}
\widehat\partial_u:=(\partial_{u_R})_{(0)}=\frac{1}{\partial_u U_{(0)}}\partial_u=\frac{\widehat P}{P_\circ}\partial_u~,
\end{align}

where the subscript emphasizes that this is the derivative induced by the leading boundary map $u_R=U_{(0)}(u,x^A)$, $x_R^A=x^A$. Similarly, the angular Robinson--Trautman derivatives at fixed $u_R$ pull back at leading order to
\begin{align}
\widehat\partial_A:=(\partial_{x_R^A})_{(0)}=\partial_A-\frac{\partial_A U_{(0)}}{\partial_u U_{(0)}}\partial_u~.
\end{align}
These operators are adapted to the leading Robinson--Trautman coordinates in the sense that
\begin{align}
\widehat\partial_u U_{(0)}=1~,\qquad\widehat\partial_A U_{(0)}=0~.
\end{align}
They are simply the Robinson--Trautman coordinate derivatives rewritten in Bondi variables. In particular, for any pulled-back Robinson--Trautman scalar $\widehat F$, they form a holonomic basis,
\begin{align}
[\widehat\partial_a,\widehat\partial_b]\widehat F=0~.
\end{align}
They should nevertheless be treated as differential operators: repeated hatted derivatives also act on the $U_{(0)}$-dependent coefficients appearing in them. 

In what follows we use the shorthand
\begin{align}
\partial:=\partial_z~,\qquad\bar\partial:=\partial_{\bar z}~.
\end{align}
Correspondingly, for the pulled-back Robinson--Trautman derivatives we write
\begin{align}
\widehat\partial:=\widehat\partial_z~,\qquad \widehat{\bar\partial}:=\widehat\partial_{\bar z}~.
\end{align}

At the next order, the conditions $g_{rA}=0$ and $g_{rr}=0$ determine the first angular and time corrections. Then, we find 
\begin{align}
U_{(1)}=-P_\circ \widehat P\,
\partial U_{(0)}\,\bar\partial U_{(0)}~,\qquad Z^A_{(1)}=-\frac{\widehat{P}}{P_\circ}D^AU_{(0)}~,
\label{FirstBondiMapCoefficients}
\end{align}
where $D^A = \gamma^{AB}D_B$ and $D_A$ is the covariant derivative compatible with the unit round-sphere metric. The determinant condition also fixes the first radial coefficient,
\begin{align}
R_{(0)}=\widehat P^2\partial \bar\partial U_{(0)}~.
\label{R0BondiMap}
\end{align}

The next order gives
\begin{align}
    Z^z_{(2)}
    &=
    P_\circ^2\widehat P\,\bar\partial U_{(0)}
    \left[
        \widehat P\,\partial\bar\partial U_{(0)}
        +
        \bar\partial U_{(0)}\,\partial\widehat P
    \right]~,
    \\
    Z^{\bar z}_{(2)}
    &=
    P_\circ^2\widehat P\,\partial U_{(0)}
    \left[
        \widehat P\,\partial\bar\partial U_{(0)}
        +
        \partial U_{(0)}\,\bar\partial\widehat P
    \right]~,
    \\
    U_{(2)}
    &=
    P_\circ^2 \widehat P\,
    \partial U_{(0)}\bar\partial U_{(0)}
    \left[
        \widehat P\,\partial\bar\partial U_{(0)}
        +
        \frac12
        \partial U_{(0)}\bar\partial U_{(0)}
        \widehat\partial_u\widehat P
    \right]~,
\end{align}
and
\begin{align}
R_{(1)}
={}&
\frac12 P_\circ \widehat P^{\,2}
\Bigg\{
\bar\partial^2 U
\left[
2\,\widehat\partial\widehat P\,\partial U
+\widehat P\,\partial^2 U
+(\partial U)^2\,\widehat\partial_u\widehat P
\right]+\bar\partial U
\Big[
\partial^2 U
\left(
2\,\widehat{\bar\partial}\widehat P
+\bar\partial U\,\widehat\partial_u\widehat P
\right)
\nonumber\\
&\hspace{3.1cm}
+2\,\partial U
\left(
2\,\widehat\partial\widehat{\bar\partial}\widehat P
+\bar\partial U\,\widehat\partial_u\widehat\partial\widehat P
\right) +(\partial U)^2
\left(
2\,\widehat\partial_u\widehat{\bar\partial}\widehat P
+\bar\partial U\,\widehat\partial_u^2\widehat P
\right)
\Big]
\Bigg\}~.
\label{R1BondiMap}
\end{align}

The coefficients displayed above are the only ones needed to extract the Bondi shear and mass aspect. Higher-order coefficients are fixed recursively by continuing the Bondi gauge conditions order by order. Although their explicit expressions become lengthy and will not be displayed, we have carried out the recursion beyond the orders needed for the mass aspect as a consistency check. With the resulting asymptotic map, the transformed metric satisfies
\begin{align}
g_{rr} &={\cal O}\left(r^{-3}\right)~,\\
g_{rA} &={\cal O}\left(r^{-3}\right)~,\\
g_{ur} &=-1+{\cal O}\left(r^{-2}\right)~,\\
g_{uu} &=-1+{\cal O}\left(r^{-1}\right)~,\\
g_{uA} &=\frac12 D^B C_{AB}+{\cal O}\left(r^{-1}\right)~,
\end{align}
while the angular metric obeys the luminosity-radius condition
\begin{align}
\frac{\det g_{AB}}{r^4}
=
\det \gamma_{AB}~.
\end{align}
Equivalently, the angular sector takes the standard Bondi form
\begin{align}
g_{AB}=r^2\gamma_{AB}+rC_{AB}+{\cal O}\left(r^0\right)~,
\qquad
\gamma^{AB}C_{AB}=0~.
\end{align}
This confirms that, to the required order, the asymptotic diffeomorphism brings the Robinson--Trautman metric to Bondi gauge, with $C_{AB}$ identified as the Bondi shear.

The leading effect of the map is to trade the time-dependent conformal geometry of the Robinson--Trautman cuts for a Bondi shear on a fixed round-sphere representative. In this sense the Bondi shear is not an independent degree of freedom; it is the asymptotic image of the Robinson--Trautman radiative field after the luminosity-radius and boundary-frame conditions have been imposed.

\subsection{Radiative Data and Flux Balance}

We now use the asymptotic map described above to read off the Bondi data of the Robinson--Trautman solution. For the Robinson--Trautman family, the only nonvanishing components of the Bondi shear are
\begin{align}
C_{zz}&=-\frac{4}{P_\circ}\left[\widehat\partial\widehat P\,\partial U_{(0)}+\frac12\widehat P\,\partial^2 U_{(0)}+\frac12\left(\partial U_{(0)}\right)^2\widehat\partial_u\widehat P\right]~,\\C_{\bar z\bar z}&=-\frac{4}{P_\circ}\left[\widehat{\bar\partial}\widehat P\,\bar\partial U_{(0)}+\frac12\widehat P\,\bar\partial^2 U_{(0)}+\frac12\left(\bar\partial U_{(0)}\right)^2\widehat\partial_u\widehat P\right]~.
\label{RTShear}
\end{align}
The tracefree condition is therefore automatic. 

The shear depends on the residual choice of cuts encoded in the angular integration function in $U_{(0)}$, as expected from the supertranslation ambiguity of the Bondi shear. The news removes the purely $u$-independent shift of the shear: in a fixed Bondi frame, adding a standard supertranslation changes $C_{AB}$ by a time-independent electric-parity tensor and therefore leaves the news tensor unchanged. In the present Robinson--Trautman-to-Bondi map, however, the leading time map obeys \eqref{U0condition}, so changing the integration function in $U_{(0)}$ also changes the slicing of the Robinson--Trautman solution. The compact expression
\begin{align}
N_{zz}=\frac{2}{\widehat P}\widehat\partial^{2}\widehat P~,
\qquad
N_{\bar z\bar z}=\frac{2}{\widehat P}\widehat{\bar\partial}^{2}\widehat P~
\end{align}
should therefore be understood as the Bondi news evaluated on the chosen Bondi cut. The associated flux,
\begin{align}
\frac18 N_{AB}N^{AB}=\frac{P_\circ^4}{\widehat P^{2}}
\widehat\partial^{2}\widehat P\,
\widehat{\bar\partial}^{2}\widehat P~,
\end{align}
is independent of the additive supertranslation shift of the shear, but it is still a function on the chosen cut through the pulled-back field $\widehat P$.

For instance, let us consider a time-independent Robinson--Trautman representative,
\begin{align}
\widehat\partial_u \widehat P=0~.
\end{align}
Then the leading Bondi condition \eqref{U0condition} can be integrated exactly as
\begin{align}
U_{(0)}(u,z,\bar z)
=
\frac{P_\circ}{\widehat P}u
+\beta(z,\bar z)~,
\end{align}
where $\beta(z,\bar z)$ is the residual supertranslation function. Substituting this expression into \eqref{RTShear}, one finds that the shear depends explicitly on $\beta$,
\begin{align}
C_{zz}={} 2u\frac{\partial^2 P}{P} - \frac{2}{P_\circ}\left(P\partial^2 \beta + 2\partial P\partial \beta\right)~,
\qquad 
C_{\bar z\bar z}={}
2u\frac{\bar\partial^2 P}{P} - \frac{2}{P_\circ}\left(P\bar\partial^2 \beta + 2\bar\partial P\bar\partial \beta\right)~.
\end{align}
Thus the residual choice of cuts shifts the Bondi shear, as expected for a supertranslation-type ambiguity. The news tensor, however, is insensitive to the $u$-independent function $\beta$. Taking one $u$-derivative gives
\begin{align}
N_{zz}=2\frac{\partial^2 P}{P}~,
\qquad 
N_{\bar z\bar z}=2\frac{\bar\partial^2 P}{P}~. 
\label{TimeIndependentPNews}
\end{align}
Equivalently, the residual supertranslation changes the shear by a $u$-independent term and therefore leaves the news invariant. This provides a simple check that the construction has the standard Bondi behavior: the shear carries the cut ambiguity, while the radiative data encoded in $N_{AB}$ do not. Writing $P=e^{-\Phi/2}$, the news tensor \eqref{TimeIndependentPNews} becomes
\begin{align}
N_{zz}=-\partial^2\Phi
+\frac12(\partial\Phi)^2~,
\qquad 
N_{\bar z\bar z}&=-\bar\partial^2\Phi
+\frac12(\bar\partial\Phi)^2~.
\end{align}
Therefore, up to the conventional overall normalization, the non-vanishing components of the news are given by the chiral components of the classical Liouville stress tensor associated with the conformal factor of the celestial metric. This parallels the structure found in \cite{Compere:2018ylh}, where the superboost field contributes to the leading news through the tracefree stress tensor of a Euclidean Liouville theory. Here the corresponding field is the pulled-back Robinson--Trautman conformal factor $\widehat\Phi=-2\log\widehat P$. Of course, a generic time-independent representative need not solve the regular Robinson--Trautman equation on the sphere; the discussion here is meant as a check of the cut dependence of the Bondi map. For a regular stationary Robinson--Trautman solution the representative is equivalent to the round sphere, and the news vanishes in the corresponding Bondi frame.

The Bondi mass aspect can be written as
\begin{align}\label{BondiMassAspectCompact}
m_{\rm B}={}& m\frac{P_\circ^3}{\widehat P^3} + m_{\rm B}^{\rm der}~,
\end{align}
where
\begin{align}
m_{\rm B}^{\rm der}\equiv\frac{P_\circ^3}{2}\left[2\partial U_{(0)}\widehat\partial\widehat{\bar\partial}^{2}\widehat P+\bar\partial^{2}U_{(0)}\widehat\partial^{2}\widehat P+\widehat{\bar\partial}^{2}\widehat P\partial^{2}U_{(0)}+2\,\bar\partial U_{(0)}\widehat\partial^{2}\widehat{\bar\partial}\widehat P+(\partial U_{(0)})^2\widehat\partial_u\widehat{\bar\partial}^{2}\widehat P+(\bar\partial U_{(0)})^2\widehat\partial_u\widehat\partial^{2}\widehat P\right]~,
\end{align}
%%%%%
is generated by the subleading terms in the coordinate transformation to Bondi gauge. It is built out of angular derivatives of the leading map and of the pulled-back Robinson--Trautman conformal factor, and it is independent of the Robinson--Trautman mass parameter $m$. Although it is derivative in origin, it is not useful to assign an independent invariant meaning to $m_{\rm B}^{\rm der}$ by itself. The geometrically meaningful quantity is the complete Bondi mass aspect $m_{\rm B}$ obtained from the transformed metric. 
Substituting the Bondi shear, news, and mass aspect into the Bondi mass-loss equation yields
\begin{align}
\partial_u m_{\rm B}=\frac14 D_A D_B N^{AB}-T_{uu}~,\qquad T_{uu}:=\frac18 N_{AB}N^{AB}~,
\label{BondiMassLossRT}
\end{align}
provided that
\begin{align}
\widehat\Delta_R\widehat\Delta_R\log\widehat P+3m\widehat\partial_u\log\widehat P=0~.
\end{align}
Here $\widehat\Delta_R=\widehat P^2\widehat\partial\widehat{\bar\partial}$ denotes the Robinson--Trautman Laplacian expressed in Bondi variables. Thus, the Bondi mass-loss equation is satisfied precisely when the Robinson--Trautman equation holds, or equivalently, its residual is proportional to the pullback of the Robinson--Trautman equation. The asymptotic diffeomorphism therefore maps the Robinson--Trautman dynamics directly into the Bondi evolution equation.

Integrating over a smooth cut of future null infinity, the last term in \eqref{BondiMassLossRT} drops out. Therefore the total Bondi mass,
\begin{align}
M_{\rm B}(u)=\frac{1}{4\pi G}\int_{S^2}\d\Omega_\circ m_{\rm B}~,
\end{align}
where $\d\Omega_\circ$ is the volume element of the round two-sphere, satisfies the standard global mass-loss formula
\begin{align}
\frac{\d M_{\rm B}}{\d u}=-\frac{1}{32\pi G}\int_{S^2}\d\Omega_\circ N_{AB}N^{AB}\leq 0~.
\label{GlobalBondiMassLossRT}
\end{align}
The single Robinson--Trautman equation is precisely the condition ensuring that the Bondi mass decreases by the flux carried by the news tensor obeying the standard asymptotic Einstein constraint.

The angular momentum aspect is also fixed by the same asymptotic expansion. In the
general nonlinear Robinson--Trautman-to-Bondi map we find a nontrivial local expression
for $N_A$. Its explicit form is considerably longer than the mass aspect and is not
illuminating, so we do not display it here. Instead, we use it as a consistency check of
the construction. In particular, we have verified that the extracted $N_A$ satisfies the
Bondi angular-momentum flux equation
\begin{align}
\partial_u N_A={}&-\frac14 D_B\left(D^B D_C C_A{}^C-D_A D_C C^{BC}\right)+u\partial_A\left(T_{uu}-\frac14 D_BD_C N^{BC}\right)-T_{uA}~,
\end{align}
with
\begin{align}
T_{uA}=\frac14D_B\left(C^{BC}N_{CA}\right)-\frac14D_A\left(C_{BC}N^{BC}\right)-\frac12 C_{AB}D_C N^{BC}~,
\end{align}
modulo the pullback of the Robinson--Trautman equation and its angular derivatives. This is a nontrivial check of the Bondi expansion, since the angular-momentum equation involves higher angular derivatives of the Bondi data and therefore requires differential consequences of the Robinson--Trautman equation.

We should nevertheless be careful in interpreting this nonzero local aspect.
Robinson--Trautman spacetimes are not expected to carry intrinsic angular momentum in
the usual stationary sense. The expression obtained for $N_A$ is a Bondi-frame quantity
and may contain contributions induced by the asymptotic diffeomorphism used to bring the
metric to standard Bondi gauge. This caveat is closely related to the well-known
subtleties in defining angular momentum at null infinity 
\cite{Ashtekar:1981bq,Wald:1999wa,Barnich:2011mi}. Unlike the Bondi mass, angular momentum is sensitive to the choice of Bondi frame and to the choice of cuts of
$\mathscr I^+$, leading to the standard supertranslation ambiguity \cite{Madler:2016xju,Madler:2017umy,Chen:2021kug,Riva:2023xxm, Javadinezhad:2023mtp, Mao:2024urq}. In the present construction this issue is sharpened by the fact that the Robinson--Trautman-to-Bondi map combines a residual supertranslation with a Weyl rescaling needed to bring the leading angular metric to the unit-sphere representative. This is naturally described in the Weyl-BMS framework \cite{Freidel:2021fxf,Flanagan:2023jio}. As a result, the local aspect $N_A$ can acquire frame-dependent contributions even when the underlying Robinson--Trautman geometry is not expected to carry intrinsic angular momentum. For this reason, we distinguish the
local Bondi aspect from the associated angular momentum charge, whose interpretation requires fixing the asymptotic frame and the relevant BMS generator. In the linearized Robinson--Trautman sector studied below, which is the main regime used to extract the memory observable, these subtleties simplify: the magnetic-parity part of the shear vanishes and the angular momentum charge associated with the extracted aspect is zero. We therefore keep $N_A$ as a useful nonlinear consistency check of the Bondi expansion, while postponing its physical interpretation to a fixed choice of asymptotic frame.
We now turn to the electric/magnetic decomposition of the Bondi shear and to scalar diagnostics that distinguish the associated memory channels.

\subsection{Electric and Magnetic Parity of the Bondi Shear}
We now decompose the Bondi shear into its electric- and magnetic-parity components. This is the natural language for separating the ordinary displacement-memory channel from possible parity-odd contributions to the radiative data. On the unit two-sphere, the Bondi shear can be decomposed into electric- and magnetic-parity tensor harmonics as
\begin{align}\label{ShearEBdecomp}
C_{AB}=-2D_A D_B C+\gamma_{AB}D^2 C+\epsilon_{C(A}D_{B)}D^C\widetilde C~,
\end{align}
where $C$ is the electric-parity scalar potential and $\widetilde C$ is the magnetic-parity pseudoscalar potential which are defined modulo the $\ell=0,1$ harmonics, which lie in the kernel of the spin-two map and therefore do not contribute to $C_{AB}$.

The two parity sectors can be isolated by taking the double divergence and the curl-divergence of the shear,
\begin{align}
{\cal E}&:=D_A D_B C^{AB}= -D^2(D^2+2)C~,
\\ {\cal B}&:=\epsilon^{AC}D_A D^B C_{BC}=\frac12 D^2(D^2+2)\widetilde C~.
\label{ElectricMagneticScalars}
\end{align}
The kernel of $D^2(D^2+2)$ consists of the $\ell=0,1$ harmonics, which do not contribute to a symmetric tracefree shear tensor. Thus ${\cal E}$ and ${\cal B}$ provide scalar diagnostics of the electric and magnetic parts of the Bondi shear. In particular, ${\cal B}=0$ is equivalent to the absence of a magnetic-parity shear component.

This decomposition is particularly useful in the present Robinson--Trautman construction. In the original Robinson--Trautman frame the radiative field appears as a time-dependent conformal factor on the angular metric. After transforming to Bondi gauge, the same field is encoded in the shear \eqref{RTShear}. The electric/magnetic decomposition therefore tells us which parity channels are generated by the asymptotic diffeomorphism and, in particular, whether the nontrivial local angular-momentum aspect found above is accompanied by a genuine magnetic-parity shear factor. Substituting \eqref{RTShear} into \eqref{ShearEBdecomp}, we obtain
\begin{align}
2\partial^2 C
+P_\circ^2\partial^2\widetilde C
+4\left(
\partial C+\frac12P_\circ^2\partial\widetilde C
\right)\partial\log P_\circ
&=
\frac{4}{P_\circ}\left(\widehat\partial\widehat P\,\partial U_{(0)}
+\frac12\widehat P\,\partial^2 U_{(0)}
+\frac12\left(\partial U_{(0)}\right)^2
\widehat\partial_u\widehat P\right)~, \nonumber
\\
2\bar\partial^2 C
-P_\circ^2\bar\partial^2\widetilde C
+4\left(
\bar\partial C-\frac12P_\circ^2\bar\partial\widetilde C
\right)\bar\partial\log P_\circ
&=
\frac{4}{P_\circ}\left(\widehat{\bar\partial}\widehat P\,\bar\partial U_{(0)}
+\frac12\widehat P\,\bar\partial^2 U_{(0)}
+\frac12\left(\bar\partial U_{(0)}\right)^2
\widehat\partial_u\widehat P\right)~,
\label{RTShearElectricMagneticSystem}
\end{align}
where we used $\epsilon_{z\bar z }=1$.
In full generality, this nonlinear system does not appear to admit a simple closed-form solution for $C$ and $\widetilde C$. We therefore use the decomposition mainly as a diagnostic of the parity content of the Bondi shear rather than as an explicit reconstruction of the scalar potentials. This distinction is important: the shear itself is known from the asymptotic map, while solving for the potentials requires inverting the spin-two map on the sphere.

The result shows that the Robinson--Trautman-to-Bondi map can produce a shear with nontrivial electric/magnetic structure. Thus, although the angular metric in the original Robinson--Trautman frame is conformal to the sphere, the corresponding Bondi shear need not be purely electric in the fully nonlinear Bondi frame. This is consistent with the frame-dependent features already encountered in the local angular-momentum aspect.

Rather than solving the coupled system \eqref{RTShearElectricMagneticSystem} directly for the potentials $C$ and $\widetilde C$, it is useful to compute the scalar diagnostics ${\cal E}$ and ${\cal B}$ defined in \eqref{ElectricMagneticScalars}. These quantities are obtained directly from the shear and therefore determine the electric and magnetic parity content without inverting the spin-two map on the sphere. We get
\begin{align}
{\cal E}=-4P_\circ^3\left[\bar\partial^2\left(\widehat\partial\widehat P\,\partial U_{(0)}+\frac12\widehat P\,\partial^2 U_{(0)}+\frac12\left(\partial U_{(0)}\right)^2\widehat\partial_u\widehat P\right) +\partial^2
\left(\widehat{\bar\partial}\widehat P\,\bar\partial U_{(0)}+\frac12\widehat P\,\bar\partial^2 U_{(0)}+\frac12\left(\bar\partial U_{(0)}\right)^2\widehat\partial_u\widehat P\right)\right]~,
\label{ElectricScalar}
\end{align}
for the electric scalar, and 
\begin{align}
{\cal B}=-4P_\circ^3\left[\bar\partial^2\left(\widehat\partial\widehat P\,\partial U_{(0)}+\frac12\widehat P\,\partial^2 U_{(0)}+\frac12\left(\partial U_{(0)}\right)^2\widehat\partial_u\widehat P\right) -\partial^2
\left(\widehat{\bar\partial}\widehat P\,\bar\partial U_{(0)}+\frac12\widehat P\,\bar\partial^2 U_{(0)}+\frac12\left(\bar\partial U_{(0)}\right)^2\widehat\partial_u\widehat P\right)\right]~,
\label{MagneticScalar}
\end{align}
for the magnetic pseudoscalar. In the fully nonlinear map this condition is not automatic, reflecting the fact that the asymptotic diffeomorphism can mix the conformal Robinson--Trautman data into both parity sectors of the Bondi shear. The magnetic diagnostic is a property of the Bondi-frame shear after the asymptotic diffeomorphism has been performed, and its relation to an intrinsic angular-momentum charge depends on the choice of asymptotic frame and cuts. In the linearized Robinson--Trautman sector studied below, however, the situation simplifies: the magnetic diagnostic vanishes, and the memory is entirely carried by the electric-parity displacement-memory channel.

\section{Linearized Robinson--Trautman Memory}\label{Sec:LinearRT}

The expressions obtained in the previous section give the Bondi data of the full Robinson--Trautman family in terms of the pulled-back Robinson--Trautman function. Although this form is exact, the general nonlinear expressions are not very transparent. In order to isolate the physical content of the memory effect, it is useful to specialize to the linearized Robinson--Trautman modes around the Schwarzschild solution. This linearized solution was originally obtained by Newman and Foster \cite{FosterNewman}, and provides a simple setting in which the relaxation of the spacetime, the Bondi news, and the associated displacement memory can be written explicitly.

We therefore consider a small deformation of the round-sphere Robinson--Trautman representative,
\begin{align}
P(u_R,x^A_R)=P_\circ(x^A_R)
\left[1-\frac12 K(u_R,x^A_R)\right]+{\cal O}(K^2)~,
\end{align}
or equivalently
\begin{align}
\Phi(u_R,x^A_R)=\Phi_\circ(x^A_R)+ K(u_R,x^A_R)
+{\cal O}(K^2)~.
\end{align}
At zeroth order the metric is Schwarzschild in outgoing Eddington--Finkelstein coordinates, while the function $K$ encodes the radiative Robinson--Trautman perturbation.

Linearizing the Robinson--Trautman equation gives
\begin{align}
\partial_{u_R} K
+\frac{1}{12m}
\Delta_\circ
\left(
\Delta_\circ+2
\right)K = 0~,
\end{align}
where $\Delta_\circ$ is the Laplacian on the unit round two-sphere. Expanding in spherical harmonics,
\begin{align}\label{eq:KSphHar}
K(u_R,x^A_R)=\sum_{\ell=0}^\infty \sum_{n=-\ell}^\ell K_{\ell n}(u_R)Y_{\ell n}(x^A_R)~,
\end{align}
one obtains
\begin{align}
K_{\ell n}(u_R)=K_{\ell n}^{(0)}\exp\left[-\frac{(\ell-1)\ell(\ell+1)(\ell+2)}{12m}u_R\right]~.
\end{align}
The $\ell=0$ mode changes the mass parameter, while the $\ell=1$ modes belong to the dipole/translation sector. The genuinely radiative Robinson--Trautman perturbations therefore start at $\ell\geq2$ and decay exponentially in retarded time. This realizes explicitly the late-time relaxation of the spacetime to Schwarzschild.

For later use, and also to make contact with the original Newman--Foster form of the solution, it is useful to display the corresponding line element explicitly. We therefore pass from stereographic coordinates $(z_R,\bar z_R)$ to spherical coordinates $(\theta_R,\phi_R)$ on the Robinson--Trautman angular sections, so that the round metric is written as
\begin{align}
\d\Omega_R^2=\d\theta_R^2+\sin^2\theta_R\,\d\phi_R^2~.
\end{align}
Then, we specialize to the axisymmetric sector, keeping only the $n=0$ harmonics. Since, in spherical coordinates, $Y_{\ell0}(\theta_R,\phi_R)$ is proportional to $P_\ell(\cos\theta_R)$, the normalization of the spherical harmonic can be absorbed into the mode amplitude such that the axisymmetric linearized perturbation can be written as
\begin{align}\label{FNmode}
K(u_R,\theta_R)=\varepsilon_\ell e^{-\omega_\ell u_R}P_\ell(\cos\theta_R)~,
\qquad \omega_\ell = \frac{(\ell-1)\ell(\ell+1)(\ell+2)}{12m}~.
\end{align}
Notice that $K$ is not the fractional perturbation of $P$, but rather the fractional perturbation of the angular metric.
Therefore, to first order in the radiative amplitude, the Robinson--Trautman line element in these coordinates reads
\begin{align}\label{FNmetric}
\d s^2=-\left[-\left(\frac12\Delta_\circ+1-r_R\partial_{u_R}\right)K+1-\frac{2m}{r_R}\right]\d u_R^2-2\d u_R\d r_R+r_R^2(1+K)\d\Omega_R^2~,
\end{align}
where $\Delta_\circ$ is normalized as
\begin{align}
\Delta_\circ P_\ell(\cos\theta_R)=-\ell(\ell+1)P_\ell(\cos\theta_R)~.
\end{align}

For further details on the local and global structure of Robinson--Trautman spacetimes, as well as their geometric and physical properties, see for instance \cite{Singleton2020,Chrusciel:1991vxx,Chrusciel:1992rv,Stephani:2003tm,Griffiths:2009dfa,Bakas:2014kfa}.

\subsection{Bondi Data and Displacement Memory}

We now insert the Newman--Foster modes into the asymptotic Robinson--Trautman-to-Bondi map constructed in the previous section. This gives the linearized Bondi shear and the corresponding displacement memory in a form where the late-time relaxation to Schwarzschild is explicit.

In the conventions of \autoref{Sec:LinearRT}, the angular perturbation is defined by
\begin{align}
e^\Phi=e^{\Phi_\circ}(1+K)~,
\qquad P=\frac{P_\circ}{\sqrt{1+K}}~.
\label{NFConventionP}
\end{align}
After pullback to Bondi variables this gives
\begin{align}
\widehat P=\frac{P_\circ}{\sqrt{1+\widehat K}}~,\qquad\widehat K(u,x^A):=K(U_{(0)}(u,x^A),x^A)~.
\end{align}
The leading time map is fixed by \eqref{U0condition}. Therefore
\begin{align}
\partial_u U_{(0)}=\frac{P_\circ}{\widehat P}=\sqrt{1+\widehat K}~.
\label{ExactU0EquationNF}
\end{align}
For a single axisymmetric Newman--Foster mode \eqref{FNmode}, the equation for $U_{(0)}$ is exactly solvable at each angle. We find
\begin{align}
U_{(0)}(u,\theta)=u+\beta(\theta)+\frac{2}{\omega_\ell}\log\left[1-\frac{\varepsilon_\ell}{4}e^{-\omega_\ell(u+\beta(\theta))}P_\ell(\cos\theta)\right]~,
\label{ExactU0SolutionMode}
\end{align}
where $\beta(\theta)$ is the residual integration function. This branch has been chosen so that
\begin{align}
U_{(0)}(u,\theta)\to u+\beta(\theta)\qquad\text{as}\qquad u\to+\infty~,
\end{align}
namely after the Robinson--Trautman mode has decayed and the spacetime has relaxed to Schwarzschild.

Expanding \eqref{ExactU0SolutionMode} at small amplitude gives
\begin{align}
U_{(0)}(u,\theta)=u+\beta(\theta)-\frac{\varepsilon_\ell}{2\omega_\ell}
e^{-\omega_\ell(u+\beta(\theta))}
P_\ell(\cos\theta)+{\cal O}(\varepsilon_\ell^2)~.
\label{LinearizedU0Solution}
\end{align}
Thus, in the cut frame $\beta=0$,
\begin{align}
U_{(0)}(u,\theta)=u+C^{(1)}+{\cal O}(\varepsilon_\ell^2)~,
\label{LinearizedU0SolutionBetaZero}
\end{align}
where
\begin{align}
    C^{(1)}:=-\frac{\varepsilon_\ell}{2\omega_\ell}
e^{-\omega_\ell u}P_\ell(\cos\theta)~.
\end{align}
This exact solution of the leading time-map equation will be used below to evaluate the linearized Bondi shear and memory. It should be understood as exact for the leading asymptotic map associated with a single Newman--Foster mode; the Robinson--Trautman spacetime itself is still being treated at linear order in the radiative amplitude.

At linear order, the exact shear \eqref{RTShear} reduces to the standard electric-parity form generated by the leading angular-dependent time shift. Then, the shear reads
\begin{align}\label{LinearShearO1}
C_{AB}^{(1)}=-2\left(D_AD_B-\frac12\gamma_{AB}D^2\right)C^{(1)}~.
\end{align}
For the Newman--Foster mode, using \eqref{LinearizedU0SolutionBetaZero}, this gives
\begin{align}\label{LinearShear02}
C_{AB}^{(1)}=\frac{\varepsilon_\ell}{\omega_\ell}e^{-\omega_\ell u}
\left(D_AD_B-\frac12\gamma_{AB}D^2
\right)P_\ell(\cos\theta)~,
\end{align}
which shows that, at linear order, the nontrivial map between the Robinson--Trautman retarded time and the Bondi retarded time generates the electric-parity part of the Bondi shear. In this frame the magnetic-parity potential vanishes,
\begin{align}
\widetilde C^{(1)}=0
\qquad\Longrightarrow\qquad
{\cal B}^{(1)}=0~,
\end{align}
for the linearized Robinson--Trautman solution. This agrees with the Newman--Penrose analysis of \cite{Barnich:2026wpw}, where the asymptotic Robinson--Trautman shear is purely real in a locally asymptotically flat frame. In the Bondi Hodge decomposition used here, this is precisely the statement that the magnetic-parity potential vanishes, so the linearized Robinson--Trautman memory lies entirely in the ordinary electric displacement-memory channel. Beyond the linearized sector, a direct comparison of the two descriptions requires matching the full asymptotic frame, including the leading cut map, the luminosity-radius rescaling, and the tetrad rotation; our direct Bondi gauge fixing should therefore be viewed as a complementary way of extracting the Bondi data adapted to the finite-screen analysis.

The only nonzero components of the linearized shear are
\begin{align}
C_{\theta\theta}^{(1)}=\frac{\varepsilon_\ell}{2\omega_\ell}
e^{-\omega_\ell u}
\left(\partial_\theta^2-\cot\theta\,\partial_\theta
\right)P_\ell(\cos\theta)~,\qquad  C_{\phi\phi}^{(1)}=-\sin^2\theta\,C_{\theta\theta}^{(1)}~.
\end{align}
Hence
\begin{align}
\gamma^{AB}C_{AB}^{(1)}=C_{\theta\theta}^{(1)}+\frac{1}{\sin^2\theta}C_{\phi\phi}^{(1)}=0~,
\end{align}
as required by the Bondi expansion.

The displacement memory is obtained by comparing the Bondi shear on two cuts $u_i$ and $u_f$ of future null infinity,
\begin{align}
\Delta C_{AB}^{(1)}
:=
C_{AB}^{(1)}(u_f)-C_{AB}^{(1)}(u_i)~.
\end{align}
Using the linearized Robinson--Trautman shear found above, this gives
\begin{align}
\Delta C_{AB}^{(1)}
=
\frac{\varepsilon_\ell}{\omega_\ell}
\left(
e^{-\omega_\ell u_f}
-
e^{-\omega_\ell u_i}
\right)
\left(
D_AD_B-\frac12\gamma_{AB}D^2
\right)
P_\ell(\cos\theta)~.
\label{LinearRTMemoryTensor}
\end{align}
Equivalently, in terms of the electric shear potential,
\begin{align}
\Delta C^{(1)}
:=
C^{(1)}(u_f)-C^{(1)}(u_i)
=
-\frac{\varepsilon_\ell}{2\omega_\ell}
\left(
e^{-\omega_\ell u_f}
-
e^{-\omega_\ell u_i}
\right)
P_\ell(\cos\theta)~,
\end{align}
so that
\begin{align}
\Delta C_{AB}^{(1)}
=
-2
\left(
D_AD_B-\frac12\gamma_{AB}D^2
\right)
\Delta C^{(1)}~.
\end{align}
The relative displacement of two nearby freely falling detectors separated by $s^A$ at large radius $r$ is therefore
\begin{align}
\Delta s^A
=
\frac{1}{2r}\gamma^{AC}\Delta C_{BC}^{(1)}s^B
+
{\cal O}(r^{-2})~.
\label{LinearRTDisplacementMemory}
\end{align}

In particular, if the final cut is taken after the Robinson--Trautman mode has decayed, $u_f\to+\infty$, then
\begin{align}
\Delta C_{AB}^{(1)}
=
-\frac{\varepsilon_\ell}{\omega_\ell}
e^{-\omega_\ell u_i}
\left(
D_AD_B-\frac12\gamma_{AB}D^2
\right)
P_\ell(\cos\theta)~.
\label{LinearRTMemoryLateTime}
\end{align}
Thus the memory measures the finite relaxation from an initially radiative Robinson--Trautman cut to the final Schwarzschild cut.

The angular factor can also be written as
\begin{align}
\left(
\partial_\theta^2-\cot\theta\,\partial_\theta
\right)
P_\ell(\cos\theta) = P_\ell^2(\cos\theta)~,
\end{align}
where $P_\ell^2(\cos\theta)$ denotes the associated Legendre polynomial of order two. Thus the action of the angular operator on the axisymmetric scalar harmonic $P_\ell(\cos\theta)$ extracts the corresponding spin-two angular profile.
Therefore,
\begin{align}
\Delta C_{\theta\theta}^{(1)}
={}&
\frac{\ell+1}{2\omega_\ell\sin^2\theta}\left[\left((\ell+2)\cos^2\theta-\ell\right)P_\ell(\cos\theta)-2\cos\theta\,P_{\ell+1}(\cos\theta)\right]\varepsilon_\ell\left(
e^{-\omega_\ell u_f}
-
e^{-\omega_\ell u_i}
\right)~, \\ \Delta C_{\phi\phi}^{(1)}
={}&
-\sin^2\theta\,\Delta C_{\theta\theta}^{(1)}~,
\end{align}
and the corresponding relative displacement of two nearby detectors separated by $s^A$ at large radius $r$ is
\begin{align}
\Delta s^\theta&=\frac{1}{2r}\Delta C_{\theta\theta}^{(1)}s^\theta~,
\\
\Delta s^\phi&=
\frac{1}{2r\sin^2\theta}\Delta C_{\phi\phi}^{(1)}s^\phi=-\frac{1}{2r}\Delta C_{\theta\theta}^{(1)}s^\phi~.
\end{align}
Thus the memory produces opposite fractional distortions along the two transverse directions,
\begin{align}
\frac{\Delta s^\theta}{s^\theta}=-\frac{\Delta s^\phi}{s^\phi}=\frac{1}{2r}\Delta C_{\theta\theta}^{(1)}~,
\end{align}
whenever both coordinate initial separations are nonzero.

\begin{figure}[h]
\centering
\includegraphics[width=\textwidth]{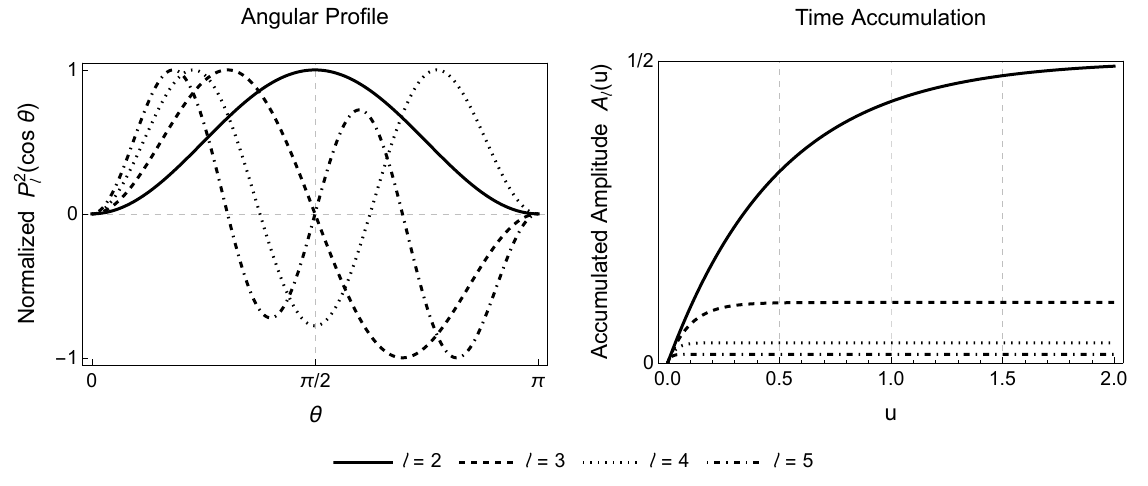}
\caption{
Angular profile and time accumulation of the linearized Robinson--Trautman displacement memory. Left: normalized angular profiles on the celestial sphere, proportional to $P_\ell^2(\cos\theta)$. Right: accumulated memory amplitude $A_\ell(u)=(1-e^{-\omega_\ell(u-u_i)})/\omega_\ell$, up to an overall constant. The solid, dashed, dotted, and dot-dashed curves correspond respectively to $\ell=2,3,4,5$. Higher multipoles relax faster but accumulate a smaller total memory, so for comparable initial amplitudes the dominant contribution comes from the first radiative mode $\ell=2$.
}
\label{Fig:RTLinearDispMemory}
\end{figure}

The angular dependence and time accumulation of these distortions are displayed in \autoref{Fig:RTLinearDispMemory}. The left panel shows the normalized angular profile on the celestial sphere, controlled by $P_\ell^2(\cos\theta)$, where the polar angle $\theta$ labels points on the asymptotic two-sphere. This illustrates that the memory acts as an angle-dependent transverse deformation, with different Robinson--Trautman multipoles producing different distortion patterns. The right panel shows the accumulated memory amplitude
\begin{align}
A_\ell(u):=\frac{1}{\omega_\ell}\left(1-e^{-\omega_\ell(u-u_i)}\right)~,
\end{align}
up to an overall mode amplitude. This quantity approaches $1/\omega_\ell$ at late times. Since
\begin{align}
\omega_\ell =\frac{(\ell-1)\ell(\ell+1)(\ell+2)}{12m}~,
\end{align}
higher multipoles relax faster and accumulate a smaller total memory. Geometrically, the Robinson--Trautman equation acts as a fourth-order angular diffusion equation on the celestial sphere: higher harmonics describe sharper angular distortions and are therefore more strongly damped. Thus, for comparable initial amplitudes $\varepsilon_\ell$, the dominant contribution comes from the first genuinely radiative mode, namely the quadrupolar mode $\ell=2$.

At the same order, the Bondi mass aspect is obtained from the linearized Bondi constraint,
\begin{align}
\partial_u m_{\rm B}^{(1)}=\frac14 D_A D_B N^{AB}_{(1)}~,
\qquad
N_{AB}^{(1)}=\partial_u C_{AB}^{(1)}~.
\end{align}
Choosing the integration constant so that the final stationary state has mass aspect $m$, one finds
\begin{align}
m_{\rm B}=m+m_{\rm B}^{(1)}+{\cal O}(\varepsilon_\ell^2)~,
\qquad
m_{\rm B}^{(1)}=\frac14 D_A D_B C^{AB}_{(1)}~.
\end{align}
Using \eqref{LinearShearO1} together with $D^2P_\ell(\cos\theta)=-\ell(\ell+1)P_\ell(\cos\theta)$, this gives
\begin{align}
m_{\rm B}=m\left(1+\frac{3}{2}\,
\varepsilon_\ell e^{-\omega_\ell u} P_\ell(\cos\theta)\right)+{\cal O}(\varepsilon_\ell^2)~.
\end{align}
Thus the linearized Robinson--Trautman perturbation produces an angle-dependent Bondi mass aspect, but, due to the orthogonality of Legendre polynomials, its integral over the celestial sphere vanishes for $\ell\ge2$,
\begin{align}
\int_{S^2}\d\Omega_\circ\, m_{\rm B}^{(1)}=0~,
\end{align}
so that the total Bondi mass is still proportional to $m+{\cal O}(\varepsilon_\ell^2)$. At linear order, the perturbation therefore changes the local angular profile of the Bondi mass aspect but leaves the total Bondi mass unchanged. This feature will also appear in the $\Lambda\neq0$ analysis of \autoref{Sec:RTLambda}. In order to study the flux and monotonicity of the mass in that case, one must go to second order in perturbation theory, where the radiative contribution enters quadratically in the first-order data.

The angular momentum aspect is also nontrivial locally. At linear order, however, the axisymmetric Robinson--Trautman mode produces an aspect of pure electric type,
\begin{align}
N_A^{(1)}=D_A {\cal J}^{(1)}_\ell(u,\theta)~,\qquad {\cal J}^{(1)}_\ell(u,\theta)=\frac{3m\,\varepsilon_\ell}{2\omega_\ell}
e^{-\omega_\ell u}
P_\ell(\cos\theta)~.
\end{align}
In standard spherical coordinates,
\begin{align}
N_\theta^{(1)}
&=
\frac{3m\varepsilon_\ell}{2\omega_\ell}
e^{-\omega_\ell u}
\partial_\theta P_\ell(\cos\theta)
=
\frac{3m(l+1)}{2\omega_\ell\sin\theta}
\left[P_{\ell+1}(\cos\theta)-\cos\theta P_\ell(\cos\theta)\right]\varepsilon_\ell e^{-\omega_\ell u}~,
\\
N_\phi^{(1)}
&=0~.
\end{align}

Therefore the associated Lorentz charge, obtained by contraction with a conformal Killing vector $Y^A$ of the unit sphere,\footnote{A BMS generator is parametrized by a supertranslation $f$ and a vector field $Y^A$ on the celestial sphere,
\begin{align}
\xi=f\,\partial_u+Y^A\partial_A+\cdots~,
\end{align}
where \(Y^A\) is a conformal Killing vector of the unit sphere for the global Lorentz subgroup,
\begin{align}
D_A Y_B+D_B Y_A=\gamma_{AB}D_CY^C~.
\end{align}
The six globally well-defined solutions split into three rotations, satisfying $D_A Y^A=0$, and three boosts, for which $D_A Y^A$ is an $\ell=1$ spherical harmonic.} is
\begin{align}
Q_Y^{(1)} \propto \int_{S^2}\d\Omega_\circ\,Y^A N_A^{(1)} = \int_{S^2}\d\Omega_\circ\,Y^A D_A{\cal J}^{(1)}_\ell=-\int_{S^2}\d\Omega_\circ\,{\cal J}^{(1)}_\ell D_A Y^A~.
\end{align}
For rotations one has $D_A Y^A=0$, while for boosts $D_A Y^A$ is an $\ell=1$ spherical harmonic. Since ${\cal J}^{(1)}_\ell$ is proportional to an $\ell\ge2$ harmonic, orthogonality on $S^2$ gives
\begin{align}
Q_Y^{(1)}=0~.
\end{align}
Thus the linearized Robinson--Trautman mode can generate a local angular momentum aspect in Bondi gauge, but it carries no global Lorentz charge at this order. For example, in the Barnich--Troessaert prescription\footnote{For discussions of BMS charges, angular momentum at null infinity, and related ambiguities, see for example \cite{Ashtekar:1981bq,Wald:1999wa,Flanagan:2015pxa, Chandrasekaran:2018aop,Compere:2018aar, Compere:2018ylh,Compere:2019gft,Harlow:2019yfa,Odak:2021axr,Freidel:2021cjp,Chandrasekaran:2021vyu,Odak:2022ndm, Odak:2023pga,Speziale:2025lkm}.} the integrable charge associated with a BMS generator $s=(f,Y^A)$ takes the form \cite{Barnich:2011mi}
\begin{align}
Q_{\rm BT}=\frac{1}{16\pi G}
\int_{S^2}\d\Omega_\circ
\left[4fM+Y^A\left(2N_A+\frac{1}{16}D_A(C_{BC}C^{BC})\right)\right]~.
\end{align}
For a Lorentz generator one sets $f=0$. At linear order around Schwarzschild, the shear-squared improvement is of order ${\cal O}(\varepsilon_\ell^2)$, so that
\begin{align}
\left.Q_{\rm BT}^{(1)}\right|_{f=0}=\frac{1}{8\pi G}\int_{S^2}\d\Omega_\circ\,Y^A N_A^{(1)}=-\frac{1}{8\pi G}\int_{S^2}\d\Omega_\circ\, {\cal J}^{(1)}_\ell D_A Y^A~.
\end{align}
This expression vanishes for all global Lorentz generators
\begin{align}
    \left.Q_{\rm BT}^{(1)}\right|_{f=0}= 0~.
\end{align}

Thus the Robinson--Trautman-to-Bondi map may produce a nontrivial local angular-momentum aspect in Bondi gauge, but this aspect carries no global Lorentz charge at linear order.

Let us summarize the outcome of the linearized Bondi analysis. A single Robinson--Trautman mode produces a purely electric Bondi shear and hence a standard displacement-memory effect at null infinity. The memory is controlled by the integrated relaxation of the mode, with higher multipoles suppressed by their faster Robinson--Trautman decay rates. The total Bondi mass is unchanged at linear order, while the locally nonzero angular momentum aspect has vanishing projection onto all global Lorentz charges. Thus the linearized solution leaves a nontrivial local imprint on the celestial sphere without generating global angular momentum.
We will use this asymptotic result as a reference point for the finite-distance null-surface analysis, where the same radiative process will be described in terms of the intrinsic and extrinsic geometry of Carrollian screens.

We close this section with a comment on the relation with celestial descriptions
of asymptotically flat gravity. The Bondi variables obtained above are precisely
the radiative data entering the celestial generators at $\scri^+$. In particular,
the leading BMS current and the celestial stress tensor can be written in terms
of retarded-time moments of the Bondi news \cite{strominger2014bms, kapec20172d},\footnote{For the relation between celestial and Carrollian descriptions of asymptotically
flat holography, see \cite{Donnay:2022aba,Donnay:2022wvx}.}
\begin{align}
    P_z
    =
    \frac{1}{4G}D^z\int\d u\,N_{zz}~,
    \qquad
    T_{zz}
    =
    \frac{i}{8\pi G}
    \int\d^2w\,\frac{1}{z-w}
    D_w^2D_{\bar w}
    \int\d u\,u\,N_{\bar w\bar w}~.
\end{align}
Similarly, the positive-helicity celestial graviton can be obtained from the
retarded-time transform
\begin{align}
    {\cal O}^{+}_{\Delta,2}(z,\bar z)
    =
    \frac{\Gamma(\Delta-2)}{4\pi i^{\Delta+1}}
    \int\d u\,
    u^{-\Delta+2}\partial_uN_{zz}(u,z,\bar z)~.
\end{align}
The Robinson--Trautman solution provides a simple classical profile for these objects. For the Newman--Foster mode discussed above, the linearized Bondi shear \eqref{LinearShear02} gives
\begin{align}\label{LinearNews01}
N_{AB}^{(1)} = -\varepsilon_\ell e^{-\omega_\ell u} \left( D_AD_B-\frac12\gamma_{AB}D^2 \right)P_\ell(\cos\theta)~.
\end{align}
Passing to complex coordinates on the sphere, we write the axisymmetric scalar profile as $P_\ell(\mu)$ with
\begin{align}
\mu(z,\bar z):=\cos\theta(z,\bar z)=\frac{1-\frac12z\bar z}{1+\frac12z\bar z}~,
\end{align}
such that the news tensor takes the form
\begin{align}
    N_{zz}^{(1)}
    =
    -\varepsilon_\ell e^{-\omega_\ell u}
    D_z^2P_\ell(\mu)~.
\end{align}
Choosing the initial cut at $u=0$ and working with the decaying profile for
$u>0$, the leading retarded-time moment is
\begin{align}
    \int_0^\infty\d u\,N_{zz}^{(1)}
    =
    -\frac{\varepsilon_\ell}{\omega_\ell}
    D_z^2P_\ell(\cos\theta)
    =
    \Delta C_{zz}^{(1)}~.
\end{align}
Therefore the leading celestial current associated with the displacement-memory
mode is
\begin{align}
P_z=-\frac{\varepsilon_\ell}{4G\omega_\ell} D^zD_z^2P_\ell(\mu)~.
\end{align}
The first retarded-time moment gives
\begin{align}
\int_0^\infty\d u\,u\,N_{\bar z\bar z}^{(1)}=-\frac{\varepsilon_\ell}{\omega_\ell^2}D_{\bar z}^2P_\ell(\mu)~,
\end{align}
so that the corresponding celestial stress-tensor profile is
\begin{align}
T_{zz}=-\frac{i\varepsilon_\ell}{8\pi G\,\omega_\ell^2}\int\d^2w\,\frac{1}{z-w}D_w^2D_{\bar w}D_{\bar w}^2P_\ell(\mu_w)~.
\end{align}
Finally, substituting the same Robinson--Trautman profile into the celestial
graviton transform \cite{donnay2022goldilocks} gives
\begin{align}
{\cal O}^{+}_{\Delta,2}=\frac{\Gamma(\Delta-2)\Gamma(3-\Delta)}{4\pi i^{\Delta+1}}\,\varepsilon_\ell\,\omega_\ell^{\Delta-2}D_z^2P_\ell(\mu)~,
\end{align}
up to possible contact terms at the initial cut.
Thus the exponential relaxation of a linearized Robinson--Trautman mode gives closed-form classical profiles for the celestial radiative operator and for the soft currents associated with memory. More generally,
\begin{align}
\int_0^\infty\d u\,u^nN_{zz}^{(1)}=-\frac{n!\,\varepsilon_\ell}{\omega_\ell^{n+1}}D_z^2P_\ell(\mu)~,
\end{align}
so higher retarded-time moments are controlled by inverse powers of the Robinson--Trautman decay rate. 
This suggests that Robinson--Trautman spacetimes provide a useful classical testbed for the celestial representation of gravitational memory. The whole tower of retarded-time moments is fixed by the displacement-memory profile and the Robinson--Trautman decay rate. This structure is reminiscent of the light-ray towers appearing in celestial descriptions of asymptotic symmetries. Recent work has shown that appropriate classes of light-ray operators generate the wedge algebra of $w_{1+\infty}$ \cite{himwich2025light}, while generalized energy detectors at null infinity define celestial-primary observables adapted to energy-flux measurements \cite{Gonzalez:2025ene,Moult:2025njc}. A closely related direction is to promote memory itself to a genuine quantum observable. Recent in-in analyses of gravitational memory treat the integrated Bondi news as a soft operator at $\scri^+$ and study correlators of memory insertions, rather than only classical memory profiles \cite{Moult:2025njc}. From the finite-distance side, this is also naturally connected with recent approaches to the quantization of null hypersurface geometry, where the cut metric, shear, and area element become quantum data on the null surface \cite{Ciambelli:2024swv,Ciambelli:2025flo}. In the present work the Robinson--Trautman solution is treated classically, so it should be regarded as providing explicit classical profiles, or saddle configurations, for such operators. It would be interesting to understand whether Robinson--Trautman relaxation modes define a simple classical or semiclassical sector of these light-ray and celestial algebras, how the memory observable computed here is embedded in the corresponding tower of celestial charges, and whether ensembles or quantizations of Robinson--Trautman mode amplitudes lead to nontrivial memory correlators.

\section{Finite Null Screens and Late-Time Carrollian Relaxation}\label{Sec:CarrFluidRT}
We now turn from the asymptotic Bondi description to a quasilocal description of Robinson--Trautman radiation on finite null hypersurfaces. Following the finite-distance null-surface framework of \cite{Ciambelli:2025mex}, we regard such hypersurfaces as Carrollian screens: they carry an intrinsic degenerate geometry, optical data, and a null Brown--York tensor whose conservation equations take the form of Carrollian fluid equations.  Our aim is to show that, for Robinson--Trautman spacetimes, these finite-screen fluid equations are governed by the same Robinson--Trautman equation that controls the bulk radiative relaxation.  We will also use the corresponding asymptotic limit, in which the screen is pushed toward $\mathscr I^+$, to recover the standard Bondi radiative data.  Our conventions for null embeddings, riggings, projectors, optical tensors and Carrollian data are summarized in \autoref{App:CarrollManifolds}; here we only recall the ingredients needed for the Robinson--Trautman application. 
For notational simplicity, throughout this section we drop the subscript $_R$ on the Robinson--Trautman coordinates and write $(u,r,x^A)$ instead of $(u_R,r_R,x_R^A)$. The distinction with Bondi coordinates will be restored when taking the large-radius Bondi limit.

The basic object is a null hypersurface embedded in the bulk,
\begin{align}
    {\cal N}:\qquad
    \check\Phi = r-\rho(u,x^A)\overset{\cal N}{=}0~,
\end{align}
where $\check\Phi$ is a defining function for the screen.  The name Carrollian screen reflects the fact that a null hypersurface carries an intrinsic Carrollian structure: a degenerate metric on ${\cal N}$, a preferred null evolution vector, and spatial data living on the two-dimensional cuts.  Unlike future null infinity, the screen is placed at finite-distance and therefore keeps track of quasilocal optical data.

For \eqref{RTgenLineElement}, a convenient normal one-form is 
\begin{align}
\ell_\mu \d x^\mu = - \d \check\Phi = -\d(r-\rho) \overset{\cal N}{=} -\d r+\partial_u\rho \d u+\partial_A\rho \d x^A~. \end{align} 
When the nullity condition 
\begin{align} 
g^{\mu\nu}\ell_\mu\ell_\nu =2\partial_u\rho+    \frac{2e^{-\Phi}}{\rho^2}\partial\rho\bar\partial\rho+\rho \partial_u\Phi-\Delta_R\Phi-\frac{2m}{\rho}\overset{\cal N}{=}0 
\end{align} 
is imposed, the metric dual $\ell^\mu=g^{\mu\nu}\ell_\nu$ is tangent to ${\cal N}$ and the same object is both normal and tangent, as is characteristic of null hypersurfaces. Thus the embedding function $\rho$ is not freely specifiable: once an initial cut is chosen, its evolution is fixed by the nullity condition and is driven by the same Robinson--Trautman field $\Phi$ that controls the bulk radiation.

To define the Carrollian splitting of the screen data we also choose an auxiliary null rigging vector $n^\mu$, normalized as 
\begin{align} 
\ell_\mu n^\mu=-1~, \qquad n^\mu n_\mu=0~. 
\end{align}
For the screens considered below a convenient choice is 
\begin{align} 
n^\mu\partial_\mu=\partial_r~, 
\end{align}
or equivalently $n_\mu\d x^\mu=-\d u$ in the Robinson--Trautman coordinates. The rigging is not intrinsic to the null hypersurface; it fixes a transverse direction and hence a Carrollian decomposition of the screen geometry into a null evolution direction and spatial data on the cuts.

The embedding map of the screen is
\begin{align}
X^\mu(u,z,\bar z)=\big(u,\rho(u,z,\bar z),z,\bar z\big)~,
\end{align}
where we have ordered the bulk coordinates as $x^\mu=(u,r,z,\bar z)$. 
The corresponding tangent basis is
\begin{align}
e^\mu{}_a =    \frac{\partial X^\mu}{\partial y^a}~,\qquad y^a=(u,z,\bar z)~.
\end{align}
Equivalently,
\begin{align}
e^\mu{}_u\partial_\mu&=\partial_u+\partial_u\rho\,\partial_r~,\\e^\mu{}_z\partial_\mu&=\partial+\partial\rho\,\partial_r~,\\e^\mu{}_{\bar z}\partial_\mu&=\bar\partial+\bar\partial\rho\,\partial_r~.
\end{align}
These vectors span $T{\cal N}$ and will be used to pull back bulk tensors to the screen.  In particular, the metric induced on
${\cal N}$ is
\begin{align}
q_{ab}\d y^a\d y^b=g_{\mu\nu}e^\mu{}_a e^\nu{}_b \d y^a\d y^b \overset{{\cal N}}{=} \frac{2e^{-\Phi}}{\rho^2}\partial\rho\,\bar\partial\rho\,\d u^2-2\partial\rho\,\d u\,\d z-2\bar\partial\rho\,\d u\,\d\bar z+2\rho^2e^\Phi\d z\d\bar z~,
\end{align}
which is degenerate, as expected for a
null hypersurface.  Its non-degenerate spatial part on the cuts $u=\mathrm{const.}$
is
\begin{align}
q_{AB}\d x^A\d x^B\overset{\cal N}{=}2\rho^2e^\Phi\d z\d\bar z~,
\end{align}
and the kernel of $q_{ab}$ is generated by the Carrollian vector
\begin{align}
\ell^\mu\partial_\mu=\partial_u+V^z\partial+V^{\bar z}\bar\partial~,\qquad V^z = \frac{e^{-\Phi}}{\rho^2}\bar\partial\rho~, \qquad V^{\bar z}=\frac{e^{-\Phi}}{\rho^2}\partial\rho~.
\end{align}
Indeed, one verifies that $q_{ab}\ell^b=0$. Lowering the spatial index with $q_{AB}$ gives \begin{align} 
V_A=\mathscr{D}_A\rho~, 
\end{align} 
where $\mathscr D_A$ is the Levi--Civita connection of $q_{AB}$,
\begin{align}
    \mathscr D_A q_{BC}=0~,
    \qquad
    \mathscr D^2:=q^{AB}\mathscr D_A\mathscr D_B~.
\end{align}

The soldering form adapted to the rigging $n^\mu=\partial_r$ is fixed by $e^a{}_\mu e^\mu{}_b=\delta^a{}_b$ and $e^a{}_\mu n^\mu=0$.  For the embedding above one finds
\begin{align}
e^a{}_\mu=\begin{pmatrix}
        1 & 0 & 0 & 0 \\
        0 & 0 & 1 & 0 \\
        0 & 0 & 0 & 1
\end{pmatrix}~,
\end{align}
with rows $a=(u,z,\bar z)$ and columns $\mu=(u,r,z,\bar z)$.  It obeys
$e^\mu{}_a e^a{}_\nu=\delta^\mu{}_\nu+n^\mu\ell_\nu$, namely it projects
bulk tensors onto the hypersurface along the chosen rigging direction.

The inverse Carrollian metric is defined only on the spatial cuts. With the clock one-form induced by the rigging, $n_a\d y^a=-\d u$, it satisfies 
\begin{align} q^{ac}q_{cb} = \pi^a{}_b~, \qquad q^{ab}n_b=0~, 
\end{align} 
where $\pi^a{}_b=\delta^a{}_b+\ell^a n_b$ is the intrinsic spatial projector. For the Robinson--Trautman screen the nonzero components are 
\begin{align}
q^{z\bar z}=q^{\bar z z}=\frac{e^{-\Phi}}{\rho^2}~.
\end{align}

We can now give the optical data of the screen in the conventions of \autoref{App:CarrollManifolds}. The optical tensor is 
\begin{align} 
\theta_{ab} = \frac12{\cal L}_{\ell}q_{ab}~. 
\end{align} 
Since $q_{ab}\ell^b=0$, it is horizontal, 
\begin{align} 
\theta_{ab}\ell^b=0~. 
\end{align}
The expansion and shear are 
\begin{align} 
\theta = q^{ab}\theta_{ab}~, \qquad \sigma_{ab} = \theta_{ab} - \frac12\theta q_{ab}~, \qquad q^{ab}\sigma_{ab}=0~. 
\end{align}
In the adapted coordinate basis $y^a=(u,z,\bar z)$, the tensors $\theta_{ab}$ and $\sigma_{ab}$ may have $u$-components, but these are fixed by horizontality and carry no independent data. 
The independent optical data are therefore contained in the spatial components $\theta_{AB}$ and $\sigma_{AB}$, while the full tensors are reconstructed from horizontality.
Using $\ell=\partial_u+V^A\partial_A$ and $V_A=\mathscr{D}_A\rho$, one finds 
\begin{align} 
\theta_{AB} = \frac12\partial_u q_{AB} + \mathscr{D}_A \mathscr{D}_B\rho~.
\end{align}
Consequently, 
\begin{align}
\sigma_{AB} = \left( \mathscr{D}_A \mathscr{D}_B - \frac12q_{AB}\mathscr{D}^2 \right)\rho~.
\end{align}
For the Robinson--Trautman screen, the expansion is 
\begin{align} 
\theta = \frac{2}{\rho^2} \left[ m+\Delta_R\rho + \frac12\rho\,\Delta_R\Phi - \frac{e^{-\Phi}}{\rho} \partial\rho\,\bar\partial\rho \right]~,\qquad \Delta_R = e^{-\Phi}\partial\bar\partial~.
\end{align} 
The inaffinity is defined by 
\begin{align} 
\ell^\nu\nabla_\nu\ell^\mu = \kappa\ell^\mu~. 
\end{align}
For the normalization $\ell^u=1$, we find 
\begin{align} 
\kappa = -\frac12\partial_u\Phi - \frac{m}{\rho^2} + \frac{2e^{-\Phi}}{\rho^3} \partial\rho\,\bar\partial\rho~. 
\end{align} 
The full normal-connection one-form is
\begin{align} 
\Omega_a = -n_\mu e^\nu{}_a\nabla_\nu\ell^\mu~.
\end{align} 
For the Robinson--Trautman solution we find
\begin{align} 
\Omega_a \d y^a = -\left(\frac{m}{\rho^2}+\frac12\partial_u \Phi\right) \d u + \mathscr{D}_A\log\rho \d x^A~, 
\end{align}
while the \Hajicek one-form is its horizontal projection $\omega_a = \pi_a{}^b \Omega_b$ with components
\begin{align}
    \omega_a \d y^a=
        -\frac{2e^{-\Phi}}{\rho^3}\partial\rho\bar\partial\rho \d u+{\mathscr{D}}_A \log\rho \d x^A~.
\end{align}
The first term is fixed by the horizontality condition $\omega_a\ell^a=0$,
while the independent spatial components are
$\omega_A=\mathscr D_A\log\rho$.

Finally, we define 
\begin{align} 
\mu = \kappa+\frac12\theta~. 
\end{align}
Equivalently, for the Robinson--Trautman screen, 
\begin{align} 
\mu = -\frac12\partial_u\Phi + \frac{\Delta_R\rho}{\rho^2} + \frac{\Delta_R\Phi}{2\rho} + \frac{e^{-\Phi}}{\rho^3} \partial\rho \bar\partial\rho~. 
\end{align}

Thus the finite screen carries the optical data $(\mu,\omega_A,\sigma_{AB})$, together with the expansion $\theta$ and inaffinity $\kappa$ from which $\mu$ is built.  Equivalently, $\omega_a$ and $\sigma_{ab}$ denote the corresponding horizontal intrinsic tensors, whose independent components are $\omega_A$ and $\sigma_{AB}$.  The optical data above can be packaged into the null Weingarten map
\begin{align}
W_a{}^b=e^\mu{}_{a}e^b{}_\nu\nabla_\mu\ell^\nu~.
\end{align}
This is a mixed intrinsic tensor measuring the variation of the null generator along the screen.  In the Carrollian decomposition reviewed in \autoref{App:CarrollManifolds}, it contains the expansion, shear, inaffinity and \Hajicek data of the null surface.  Its trace is the ordinary trace of this mixed tensor,
\begin{align}
W=W_a{}^a=\theta+\kappa~.
\end{align}
This trace is not computed by contracting with the degenerate metric
$q_{ab}$.

The corresponding null Brown--York tensor is \cite{Chandrasekaran:2021hxc}
\begin{align}
T_a{}^b=\frac{1}{8\pi G}\left(W_a{}^b-\delta_a{}^b W\right)~.
\end{align}
Using the Carrollian decomposition of the Weingarten map, this can be written
in the fluid form
\begin{align}
T_a{}^b=\ell^b\tau_a+\tau_a{}^b~,
\end{align}
with
\begin{align}
\tau_a=\frac{1}{8\pi G}\left(\omega_a+\theta n_a\right)~,\qquad\tau_a{}^b=\frac{1}{8\pi G}\left(\sigma_a{}^b-\mu\,\pi_a{}^b\right)~.
\end{align}
Thus $T_a{}^b$ admits the interpretation of a Carrollian fluid stress tensor living on the finite screen.  The null generator $\ell^a$ plays the role of the Carrollian time direction, $\omega_a$ is the transverse momentum or heat-current datum, $\mu$ is the isotropic pressure, and $\sigma_a{}^b$ is the tracefree viscous stress.
For completeness, we display the independent nonzero components of the null Brown--York tensor for the Robinson--Trautman screen
\begin{align}
T_u{}^u
&=
-\frac{1}{4\pi G\rho^2}
\left(
\Delta_R\rho+\frac12\rho\,\Delta_R\Phi+m
\right)~,
\\[5pt]
T_u{}^z &=\frac{e^{-2\Phi}}{8\pi G\rho^5}\left[3\partial\rho(\bar\partial\rho)^2-\rho\partial\rho\bar\partial^2\rho+\rho\partial\rho\bar\partial\rho\bar\partial\Phi\right]-\frac{e^{-\Phi}}{16\pi G\rho^2}\bar\partial\rho\dot\Phi-\frac{e^{-\Phi}\bar\partial\rho}{8\pi G\rho^4}\left(\Delta_R\rho+\frac12\rho\Delta_R\Phi+2m\right)~,
\\[5pt]
T_z{}^u &=\frac{1}{8\pi G}
\partial \log\rho~,
\\[5pt]
T_z{}^z&=\frac{1}{16\pi G}\left[\dot\Phi -\frac{1}{\rho}\Delta_R\Phi -\frac{2}{\rho^2}\Delta_R\rho\right]~,
\\[5pt]
T_z{}^{\bar z}&=-\frac{e^{-\Phi}}{8\pi G\rho^3}\left(\partial\rho\right)^2-\frac{e^{-\Phi}}{8\pi G\rho^2}\left(\partial\rho \partial\Phi-\partial^2\rho\right)~.
\end{align}
The remaining components are obtained by the formal exchange $z\leftrightarrow\bar z$, namely by replacing $\partial\leftrightarrow\bar\partial$ in the expressions above. These expressions are the component form of the Carrollian fluid decomposition of the null Brown--York tensor in terms of the optical data $(\theta,\sigma_a{}^b,\omega_a,\mu)$ given above.

The projected Einstein equations become the conservation equations of the null Brown--York tensor,
\begin{align}\label{divthere}
D_bT_a{}^b=-\frac{1}{8\pi G}e^\mu{}_aG_{\mu\nu}\ell^\nu\overset{\cal N}{=}0~.
\end{align}
For the Robinson--Trautman screen, the components of \eqref{divthere} split into the Raychaudhuri equation, obtained from the projection along the Carrollian time direction, and the Damour equation, obtained from the spatial projections. 
A direct computation gives
\begin{align}
e^\mu{}_A G_{\mu\nu}\ell^\nu =0
\end{align}
identically once the nullity condition for the screen embedding is imposed. Thus the Damour equation is automatically satisfied for the Robinson--Trautman screen.
The remaining projection gives the nontrivial focusing equation. After using the nullity condition and its derivatives to eliminate the embedding dependence, one finds
\begin{align}
e^\mu{}_uG_{\mu\nu}\ell^\nu
=
\frac{1}{\rho^2}
\left(
\Delta_R^2\Phi+3m\partial_u\Phi
\right)~.
\label{RTScreenEinsteinProjection}
\end{align}
{ Hence the Raychaudhuri equation is satisfied precisely when the Robinson--Trautman equation holds, while the Damour equation follows from the screen nullity condition. 
The Carrollian conservation equations therefore provide a quasilocal description of the bulk radiative dynamics, with no additional independent degrees of freedom. 
In particular, the same fourth-order parabolic equation governing the Robinson--Trautman relaxation controls the evolution of the induced Carrollian fluid, relating the time-dependent expansion, shear, and momentum data of the screen to the gravitational radiation in the bulk. }
Following the asymptotic-screen construction of \cite{Ciambelli:2025mex}, one could in principle push the finite Carrollian screens to $\scri^+$ by introducing a large radial parameter and expanding the screen embedding order by order. In Robinson--Trautman coordinates this is a well-defined procedure, since $r_R$ is an affine parameter along the preferred outgoing null congruence and the limit $r_R\to\infty$ reaches future null infinity. The subtlety is that this limit is taken in the Robinson--Trautman/Newman--Unti conformal frame. The celestial metric is
\begin{align}
q^{\rm RT}_{AB}\d x_R^A\d x_R^B
=
2e^\Phi \d z_R \d\bar z_R
=
\frac{2}{P^2}\d z_R \d\bar z_R~,
\end{align}
which is generally time dependent in a radiative Robinson--Trautman spacetime. Therefore the associated asymptotic screen data do not coincide directly with the standard Bondi data defined with respect to a fixed round-sphere representative. For this reason, we do not use the Robinson--Trautman-frame asymptotic screen expansion to identify the usual displacement memory. Instead, the comparison with standard asymptotic observables is made through the Bondi map of \autoref{Sec:BondiMemoryRT}.

\subsection{Late-Time Behavior of the Carroll Fluid}\label{SubSec:LateTimeNearHorizon}
We separate the following late-time analysis from the construction of the exact Robinson--Trautman event horizon. The latter is a global, teleological object and its differentiability properties are infamously subtle \cite{Chrusciel:1991vxx,Chrusciel:1992rv,Chrusciel:2020fql}. A global Gaussian-null description of the event horizon may therefore obscure the local relaxation physics. We take a more modest route and study a family of null screens which approach the Schwarzschild horizon at late retarded time, asking how the induced Carrollian fluid data relax as $u\to+\infty$.

In the spirit of the near-horizon analysis of \cite{Podolsky:2009an, Bakas:2014kfa} and \cite{Redondo-Yuste:2022czg,Husnugil:2025edm}, we work perturbatively around the final Schwarzschild configuration. For the linearized Robinson--Trautman mode of \autoref{Sec:LinearRT}, we consider a screen approaching the final Schwarzschild horizon and take
\begin{align}
\Phi=\Phi_\circ+\varepsilon_\ell e^{-\omega_\ell u}P_\ell(\cos\theta)+{\cal O}(\varepsilon_\ell^2)~, \qquad \rho=2m\left[1+\varepsilon_\ell h_\ell(u)P_\ell(\cos\theta)\right]+{\cal O}(\varepsilon_\ell^2)~.
\label{LateTimeScreenAnsatz}
\end{align}
The only new unknown is the screen deformation $h_\ell(u)$.
Substituting \eqref{LateTimeScreenAnsatz} into the nullity condition gives
\begin{align}
4m\dot h_\ell+h_\ell+\left[-2m\omega_\ell+\frac12\ell(\ell+1)-1\right]e^{-\omega_\ell u}=0~.
\label{LateTimeScreenNullityMode}
\end{align}
Thus the near-horizon screen deformation is not an independent degree of freedom. It is fixed by the nullity condition and sourced by the decaying Robinson--Trautman mode.
Solving the nullity condition separates the screen deformation into a homogeneous screen transient and a piece driven by the Robinson--Trautman mode,
\begin{align}
h_\ell(u)=C_\ell e^{-u/(4m)}+B_\ell e^{-\omega_\ell u}~,\qquad B_\ell=\frac{-2m\omega_\ell+\frac12\ell(\ell+1)-1}{4m\omega_\ell-1}~.
\label{LateTimeScreenEmbeddingSolution}
\end{align}
The first term is the homogeneous solution of the nullity condition. It is already present for a perturbation of the screen in the final Schwarzschild geometry and is fixed by the initial choice of null screen. By contrast, the second term is the particular solution sourced by the Robinson--Trautman mode. Since we are interested in the screen response induced by the radiative relaxation, we set $C_\ell=0$ and keep
\begin{align}
h_\ell(u)=B_\ell e^{-\omega_\ell u}~.
\label{LateTimeScreenParticularSolution}
\end{align}
With this choice the screen approaches the Schwarzschild horizon,
\begin{align}
\rho(u,\theta)\longrightarrow 2m \qquad \text{as} \qquad u\to+\infty~.
\end{align}

We now evaluate the optical and Carrollian fluid data derived above on this late-time screen. Since both $h_\ell(u)$ and the Robinson--Trautman perturbation decay exponentially, the following expressions make explicit how the non-stationary screen data relax at late times. For compactness, we denote
\begin{align}
P_\ell:=P_\ell(\cos\theta)~,\qquad L_\ell:=\ell(\ell+1)~.
\end{align}
Using the optical data of \autoref{Sec:CarrFluidRT} and the nullity condition \eqref{LateTimeScreenNullityMode}, the expansion becomes
\begin{align}
\theta=\frac{\varepsilon_\ell}{2m}\left[-(L_\ell+1)h_\ell(u)+\left(1-\frac12L_\ell\right)e^{-\omega_\ell u}\right]P_\ell+{\cal O}(\varepsilon_\ell^2)~.
\label{LateTimeExpansionAxisym}
\end{align}
The expansion is therefore purely transient. It is sourced both by the Robinson--Trautman deformation of the angular metric and by the response of the null embedding.

The inaffinity approaches the Schwarzschild value with a decaying angular correction,
\begin{align}
\kappa=-\frac{1}{4m}+\frac{\varepsilon_\ell}{2}\left[\frac{h_\ell(u)}{m}+\omega_\ell e^{-\omega_\ell u}\right]P_\ell+{\cal O}(\varepsilon_\ell^2)~,
\label{LateTimeKappaAxisym}
\end{align}
while the independent spatial components of the \Hajicek one-form are controlled only by the angular variation of the screen embedding,
\begin{align}
\omega_A=\varepsilon_\ell h_\ell(u)D_AP_\ell+{\cal O}(\varepsilon_\ell^2)~.
\label{LateTimeHajicekAxisym}
\end{align}
The $u$-component is fixed by horizontality and starts only at quadratic order,
\begin{align}
\omega_u={\cal O}(\varepsilon_\ell^2)~.
\end{align}
The pressure variable $\mu=\kappa+\theta/2$ is
\begin{align}
\mu=-\frac{1}{4m}+\frac{\varepsilon_\ell}{4m}\left[(1-L_\ell)h_\ell(u)+\left(1-\frac12L_\ell+2m\omega_\ell\right)e^{-\omega_\ell u}\right]P_\ell+{\cal O}(\varepsilon_\ell^2)~.
\label{LateTimeMuAxisym}
\end{align}
Thus $\mu$ also relaxes to the Schwarzschild value fixed by our normalization of the null generator.

Finally, the tracefree optical shear is determined by the tracefree Hessian of the screen deformation. With one index raised by the screen metric,
\begin{align}
\sigma_A{}^B=\frac{\varepsilon_\ell h_\ell(u)}{2m}\left(D_AD^B-\frac12\delta_A{}^BD^2\right)P_\ell+{\cal O}(\varepsilon_\ell^2)~.
\label{LateTimeShearMixedAxisym}
\end{align}

After fixing the homogeneous screen transient by $C_\ell=0$, the embedding response is $h_\ell(u)=B_\ell e^{-\omega_\ell u}$. Hence all angular corrections above are proportional to $e^{-\omega_\ell u}$. Therefore,
\begin{align}
\theta\to0~,\qquad \omega_A\to0~,\qquad \sigma_A{}^B\to0~,\qquad \kappa\to-\frac{1}{4m}~,\qquad \mu\to-\frac{1}{4m}~.
\label{LateTimeOpticalLimits}
\end{align}
The late-time screen is non-expanding, carries no transverse momentum datum, and has no tracefree viscous stress. In terms of the null Brown--York fluid variables,
\begin{align}
\tau_a=\frac{1}{8\pi G}\left(\omega_a+\theta n_a\right)~,\qquad \tau_a{}^b=\frac{1}{8\pi G}\left(\sigma_a{}^b-\mu\pi_a{}^b\right)~,
\end{align}
the late-time limit is
\begin{align}
\tau_a\to0~,\qquad \tau_a{}^b\to\frac{1}{32\pi Gm}\pi_a{}^b~.
\label{LateTimeBrownYorkFluidLimit}
\end{align}
The full null Brown--York tensor does not vanish in the late-time limit. Rather, its non-perfect and angle-dependent parts decay, while the isotropic Schwarzschild screen stress remains,
\begin{align}
T_a{}^b\to \frac{1}{32\pi Gm}\pi_a{}^b~,\qquad
\Delta T_a{}^b:=T_a{}^b-T_a{}^b\big|_{\rm Schw}\to0~.
\end{align}

The same relaxation can be visualized directly from the intrinsic metric on the cuts. To linear order, the spatial metric induced on the late-time screen is
\begin{align}
q_{AB}=4m^2\left[1+\varepsilon_\ell\left(e^{-\omega_\ell u}+2h_\ell(u)\right)P_\ell\right]\gamma_{AB}+{\cal O}(\varepsilon_\ell^2)~.
\label{LateTimeCutMetricAxisym}
\end{align}
Equivalently, the conformal deformation of the cut can be represented by the effective radius
\begin{align}
r_\star(u,\theta)=2m\left[1+\frac{\varepsilon_\ell}{2}\left(1+2B_\ell\right)e^{-\omega_\ell u}P_\ell(\cos\theta)\right]+{\cal O}(\varepsilon_\ell^2)~.
\label{LateTimeEffectiveRadiusParticular}
\end{align}
Thus the intrinsic cut geometry is a transient conformal deformation of the round Schwarzschild horizon sphere.
The transverse momentum, or heat-current, datum can be displayed through the norm of the \Hajicek one-form. At leading order,
\begin{align}
|\omega|_{\cal N}^2:=q^{AB}\omega_A\omega_B=\frac{\varepsilon_\ell^2 h_\ell(u)^2}{4m^2}\,D^AP_\ell D_AP_\ell+{\cal O}(\varepsilon_\ell^3)=\frac{\varepsilon_\ell^2 h_\ell(u)^2}{4m^2}\left|\partial_\theta P_\ell(\cos\theta)\right|^2+{\cal O}(\varepsilon_\ell^4)~.
\label{LateTimeHajicekNormAxisym}
\end{align}
Both the shape deformation \eqref{LateTimeEffectiveRadiusParticular} and the \Hajicek norm \eqref{LateTimeHajicekNormAxisym} decay exponentially with the Robinson--Trautman rate $\omega_\ell$.
This provides a useful visualization of the Carrollian relaxation.

\begin{figure}[h]
\centering
\includegraphics[width=\textwidth]{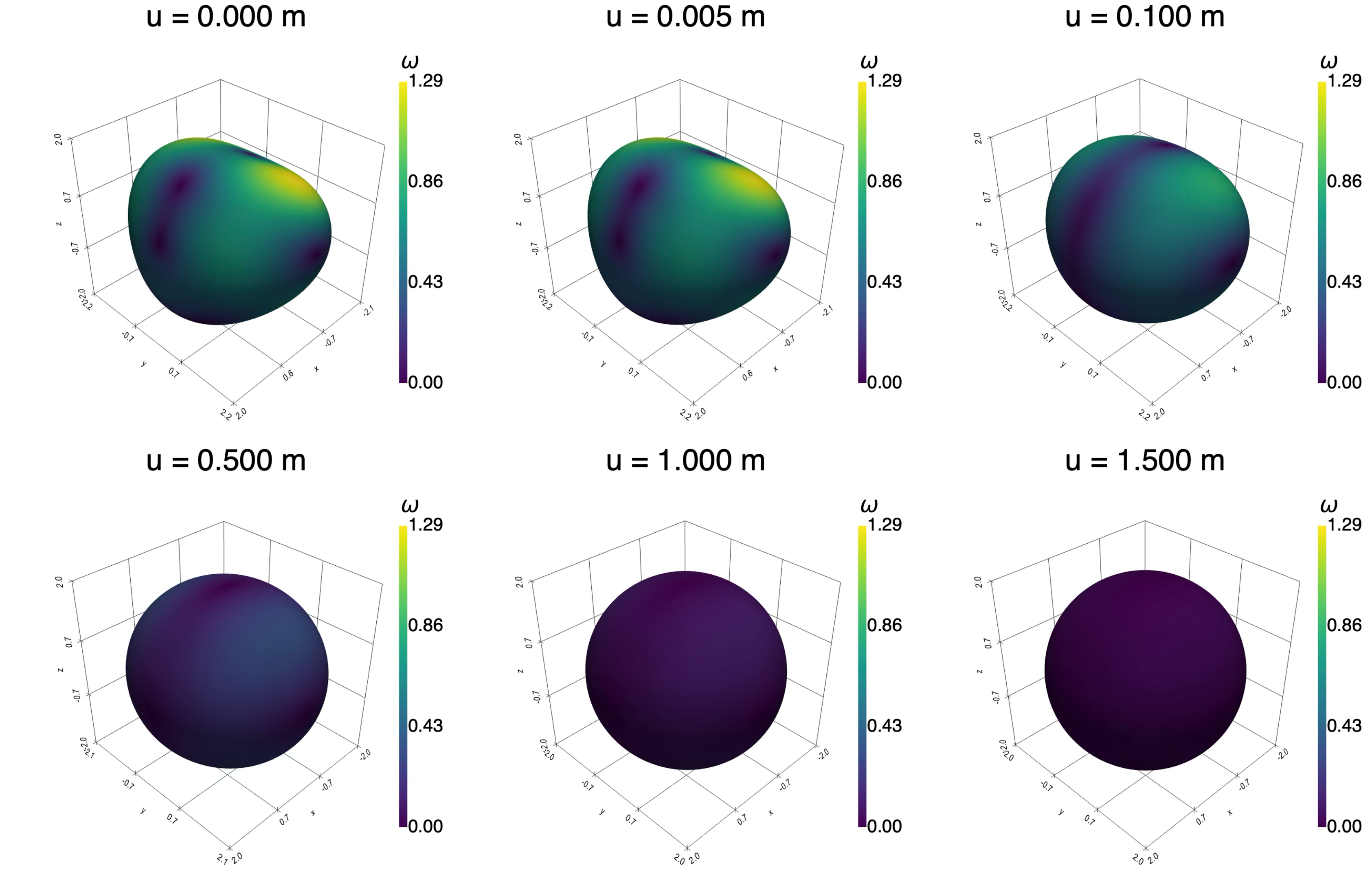}
\caption{Late-time relaxation of a Robinson--Trautman Carrollian screen toward the Schwarzschild horizon. The surface is represented by the effective radius $r_\star(u,x^A)$, while the color scale shows the norm $|\omega|_{\cal N}$ of the \Hajicek one-form. The initial perturbation is chosen as a representative superposition of spherical harmonics with nonvanishing coefficients ${\Phi}^{0}_{20}=1$, ${\Phi}^{0}_{2,\pm2}=0.5$, ${\Phi}^{0}_{3,1} = -{\Phi}^{0}_{3,-1}=0.3$, and ${\Phi}^{0}_{3,\pm2}=0.2$, while all remaining modes vanish. Since each multipole evolves independently in linearized Robinson--Trautman theory, every mode decays exponentially with its corresponding relaxation rate $\omega_\ell$, and the screen progressively approaches the round Schwarzschild horizon while the \Hajicek momentum density disappears. At late retarded times the geometry becomes the stationary Schwarzschild Carrollian screen, carrying only the isotropic null Brown--York stress.}
\label{Fig:LateTimeScreenRelaxation}
\end{figure}

To visualize the late-time relaxation in a more generic setting, the surfaces shown in \autoref{Fig:LateTimeScreenRelaxation} are generated from a linearized Robinson--Trautman perturbation expanded in spherical harmonics rather than a single axisymmetric Legendre mode. Since the Robinson--Trautman equation is linear at this order, each spherical harmonic evolves independently with its corresponding decay rate $\omega_\ell$, so the analysis above extends mode by mode upon replacing $P_\ell(\cos\theta)$ by $Y_{\ell m}(\theta,\phi)$. The figure therefore provides a representative visualization of the generic non-axisymmetric relaxation toward the round Schwarzschild horizon. As expected, both the intrinsic deformation of the cuts and the \Hajicek momentum decay exponentially, leaving only the isotropic Schwarzschild Brown--York stress at late retarded times.

{ The late-time analysis thus provides an explicit description of the relaxation of the Carrollian fluid induced on a finite null screen. The Robinson--Trautman radiative modes determine the time dependence of the screen geometry, optical data, and null Brown--York tensor, whose non-isotropic contributions decay exponentially toward the stationary Schwarzschild configuration. Although the final screen carries no non-perfect Carrollian data, its evolution records the transient geometric response to the gravitational radiation. This accumulated response will be characterized as finite-screen memory in the following section.}

%This late-time analysis shows explicitly that the Robinson--Trautman mode leaves no stationary non-perfect Carrollian data on the final screen. What survives is only the isotropic Schwarzschild Brown--York stress, while the angular deformation, momentum datum and viscous stress are exponentially damped.
%
%
%
%
%
%
%
\section{Finite-Screen Memory}
\label{Sec:FiniteScreenMemory}

We now use the finite Robinson--Trautman null screens constructed above as quasilocal detectors of radiation. The relevant data are the intrinsic Carrollian metric $q_{AB}$ on the cuts, the null generator $\ell=\partial_u+V^A\partial_A$, and the optical tensor
\begin{align}
\theta_{AB}=\frac12{\cal L}_\ell q_{AB}
=\frac12\theta q_{AB}+\sigma_{AB}~.
\label{ScreenOpticalDefinitionMemory}
\end{align}
All these quantities were evaluated explicitly in \autoref{Sec:CarrFluidRT}. Here we use them to define the corresponding finite-distance memory observable.

The basic point is that finite-screen memory is the residual change of the intrinsic screen geometry between two cuts. Since the null generators need not remain at fixed angular coordinate, there are two natural comparisons. If one follows the generators of the Carrollian vector $\ell$, with
\begin{align}
\frac{\d X^A(u;y)}{\d u}=V^A(u,X(u;y))~,\qquad
X^A(u_i;y)=y^A~.
\end{align}
Denoting by $S_u:=\{u=\text{constant}\}\cap{\cal N}$ the two-dimensional cut of the screen, $X_u:S_{u_i}\to S_u$ is the flow map of the Carrollian generator as it sends the initial generator label $y^A$ on $S_{u_i}$ to its angular position $X^A(u;y)$ on $S_u$. We denote by $X_u^*$ the corresponding pullback to the initial cut. 

The finite difference between the two cuts is then
\begin{align}
\Delta_\ell Q_{AB}(y)
=
\int_{u_i}^{u_f}\d u\,
X_u^*
\left(
\theta q_{AB}+2\sigma_{AB}
\right)~,
\label{GeneratorMemoryFiniteScreen}
\end{align}
where $Q_{AB}=X_u^*q_{AB}$ is the generator-adapted cut metric. This is the most intrinsic comparison: it measures the change in transverse geometry seen by the same neighboring null generators.

In the fixed screen coordinates used in the Robinson--Trautman construction, the same statement becomes
\begin{align}
\Delta_{\cal N}q_{AB}
:=
q_{AB}(u_f,x)-q_{AB}(u_i,x)
=
\int_{u_i}^{u_f}\d u\,
\left[
\theta q_{AB}+2\sigma_{AB}
-({\cal L}_Vq)_{AB}
\right]~.
\label{FixedFrameMemoryFiniteScreen}
\end{align}
The last term subtracts the angular drift of the generators relative to the fixed screen frame. For the screens considered here $V_A=\mathscr D_A\rho$, so that
\begin{align}
({\cal L}_Vq)_{AB}=2\mathscr D_A\mathscr D_B\rho~,
\end{align}
and therefore
\begin{align}
\Delta_{\cal N}q_{AB}
=
\int_{u_i}^{u_f}\d u\,
\left[
\theta q_{AB}+2\sigma_{AB}
-2\mathscr D_A\mathscr D_B\rho
\right]~.
\label{FixedFrameGradientMemoryFiniteScreen}
\end{align}
This formula is the finite-screen precursor of the Bondi displacement memory: its leading tracefree large-radius component reduces to the usual Bondi result, while at finite-distance it also contains area, focusing and embedding data. More explicitly, it contains an area/focusing part $\theta q_{AB}$, a tracefree shape part $2\sigma_{AB}$, and an embedding or drift part $-2\mathscr D_A\mathscr D_B\rho$.

Taking the trace\footnote{Here the trace is understood at the level of the evolution equation, using $\frac12 q^{AB}\partial_u q_{AB}=\partial_u\log\sqrt q$.} gives the scalar, or area-density, memory
\begin{align}
\Delta_{\cal N}\log\sqrt q
=
\int_{u_i}^{u_f}\d u\,
\left(
\theta-\mathscr D^2\rho
\right)~.
\label{ScalarMemoryFiniteScreen}
\end{align}
For Robinson--Trautman screens the cut metric is $q_{AB}\d x^A\d x^B=2\rho^2e^\Phi\d z\d\bar z$, and hence, up to the fixed coordinate density,
\begin{align}
\log\sqrt q=2\log\rho+\Phi~.
\end{align}
Thus the scalar memory can be written directly as
\begin{align}
\Delta_{\cal N}\log(\rho^2e^\Phi)
=
\left[
2\log\rho+\Phi
\right]_{u_i}^{u_f}
=
\int_{u_i}^{u_f}\d u\,
\left(
\theta-\mathscr D^2\rho
\right)~.
\label{RTScalarMemoryFiniteScreen}
\end{align}
This expression displays the two sources of the finite-distance response. The Robinson--Trautman field $\Phi$ changes the transverse geometry of the cuts, while the embedding function $\rho$ records the response of the null screen required by the nullity condition.

The tracefree part of \eqref{FixedFrameGradientMemoryFiniteScreen} defines the shape memory,
\begin{align}
\left(\Delta_{\cal N}q_{AB}\right)^{\rm TF}
=
2\int_{u_i}^{u_f}\d u\,
\left[
\sigma_{AB}
-
\left(\mathscr D_A\mathscr D_B\rho\right)^{\rm TF}
\right]~.
\label{ShapeMemoryFiniteScreen}
\end{align}
For the Robinson--Trautman screens considered here, the distinction between the two cut-identification prescriptions becomes essential. In fixed Robinson--Trautman angular coordinates, the tracefree deformation vanishes because the optical shear is exactly cancelled by the angular drift of the generators. The generator-adapted transition, however, is generally nonzero, as is explicit for the linearized horizon-settling screens. This illustrates the importance of the Carrollian generator in defining a geometric comparison between cuts, rather than identifying points solely by their angular coordinates. The cancellation in the fixed Robinson--Trautman frame therefore does not imply that the intrinsic generator-adapted memory vanishes.

Operationally, if two neighboring screen generators have initial separation $\eta^A$, then
\begin{align}
s^2(u,y)=Q_{AB}(u,y)\eta^A\eta^B~,
\end{align}
and the generator-adapted displacement is
\begin{align}
\Delta_\ell s^2(y)
=
\int_{u_i}^{u_f}\d u\,
X_u^*
\left(
\theta q_{AB}+2\sigma_{AB}
\right)\eta^A\eta^B~.
\label{SeparationMemoryFiniteScreen}
\end{align}
Thus the finite-screen memory is the residual change in the transverse separation of neighboring Carroll generators after the Robinson--Trautman wave has crossed the screen.

The vector sector is encoded in the horizontal \Hajicek one-form. For the Robinson--Trautman screens of \autoref{Sec:CarrFluidRT}, its independent spatial components are $\omega_A=\mathscr D_A\log\rho$. The corresponding vector memory is therefore given by $\Delta_\ell \omega_A$. Equivalently, in the fixed screen frame it measures the finite change of $\mathscr D_A\log\rho$. This is the momentum, or \Hajicek, component of finite-screen memory.

The dynamical interpretation follows from the projected Einstein equations on ${\cal N}$. Raychaudhuri controls the trace sector and gives the focusing balance law for \eqref{ScalarMemoryFiniteScreen}; the Sachs equation controls the tracefree shape sector in \eqref{ShapeMemoryFiniteScreen}; and the Damour equation controls the vector memory. Equivalently, in the null Brown--York language reviewed in \autoref{App:CarrollManifolds}, these equations are the Carrollian conservation equations of the screen stress tensor. For the Robinson--Trautman screens considered here, the optical data are determined by the same RT field $\Phi$ and screen embedding $\rho$, whose consistency is fixed by the RT equation and the nullity condition. Thus the bulk radiative relaxation controls the quasilocal Carrollian memory balance laws.

The relation with the standard displacement memory is obtained by pushing the screen to large radius. This limit is conceptually important because it separates the universal radiative information from the quasilocal optical response of the chosen screen. At finite-distance, $\Delta_{\cal N}q_{AB}$ contains focusing, embedding, Coulombic, and near-zone contributions. At null infinity, these data reorganize into an asymptotic expansion, and the usual Bondi memory is isolated as the leading renormalized tracefree part.

To see this, and to make contact with the standard displacement memory, we now switch to a large screen in Bondi gauge. We denote its radius by $r_{\cal N}$ in order to distinguish it from the finite Robinson--Trautman embedding function $\rho(u,x^A)$. Using the asymptotic screen expansion of \cite{Ciambelli:2025mex} in Bondi coordinates, the induced metric on the cuts takes the form
\begin{align}
q^{\cal N}_{AB}
=
r_{\cal N}^2\gamma_{AB}
+r_{\cal N}C_{AB}
+{\cal O}(r_{\cal N}^0)~,
\label{LargeScreenMetricExpansion}
\end{align}
where $\gamma_{AB}$ is the unit-sphere metric and $C_{AB}$ is the Bondi shear. 
Operationally, these subleading terms are the geometric data that must be controlled, modeled, or subtracted when comparing finite-radius measurements with the idealized Bondi memory. 
The angular drift of the Carrollian generators is subleading,
\begin{align}
V^A={\cal O}(r_{\cal N}^{-2})~.
\label{LargeScreenDriftExpansion}
\end{align}
Substituting these asymptotic expansions into the fixed-frame memory formula \eqref{FixedFrameMemoryFiniteScreen} gives a simple large-radius hierarchy. The term $\theta q_{AB}$ contributes to the trace at leading order and therefore does not affect the tracefree radiative component. The drift contribution is also subleading in the radiative hierarchy: although $q_{AB}^{\cal N}\sim r_{\cal N}^2$, the falloff \eqref{LargeScreenDriftExpansion} implies
\begin{align}
({\cal L}_Vq)_{AB}={\cal O}(r_{\cal N}^0)~.
\end{align}
By contrast, the tracefree optical shear contains the Bondi news at order $r_{\cal N}$,
\begin{align}
\sigma_{AB}
=
\frac12 r_{\cal N}N_{AB}
+{\cal O}(r_{\cal N}^0)~.
\end{align}
Therefore the tracefree part of the finite-screen memory behaves as
\begin{align}
\left(\Delta_{\cal N}q_{AB}\right)^{\rm TF}
=
r_{\cal N}
\int_{u_i}^{u_f}\d u\,N_{AB}
+
{\cal O}(r_{\cal N}^0)~.
\end{align}
Dividing by the screen radius and taking the limit gives
\begin{align}
\Delta C_{AB}
=
\lim_{r_{\cal N}\to\infty}
\frac{1}{r_{\cal N}}
\left(\Delta_{\cal N}q_{AB}\right)^{\rm TF}
=
\int_{u_i}^{u_f}\d u\,N_{AB}~.
\label{FiniteScreenToBondiMemory}
\end{align}
This is precisely the usual Bondi displacement memory.  

{ Equation \eqref{FiniteScreenToBondiMemory} establishes the connection between the finite-screen response and the standard displacement memory in the asymptotic Bondi frame. At finite-distance, the geometric transition depends on the prescribed screen and the identification of its cuts, retaining focusing, angular drift, embedding dependence, and near-zone information. After transforming to Bondi gauge, the leading tracefree component of the large-radius deformation isolates the universal radiative contribution and reproduces the Bondi memory. Thus the finite-screen construction provides a quasilocal geometric description whose standard asymptotic limit is recovered only after the appropriate choice of frame and radiative projection}

%Equation \eqref{FiniteScreenToBondiMemory} is the bridge between the finite and asymptotic descriptions. At finite distance, $\Delta_{\cal N}q_{AB}$ contains focusing, angular drift, embedding dependence, Coulombic contributions, and other near-zone information. These pieces are mixed at finite radius and are not separately universal. The Bondi memory is recovered only after extracting the leading tracefree coefficient in the large-radius Bondi limit. Thus the finite-screen memory is not a new independent asymptotic observable; it is a quasilocal geometric response whose universal wave-zone component is the standard displacement memory.

For Robinson--Trautman spacetimes this statement has a particularly simple interpretation. In the natural Robinson--Trautman frame, the finite screen metric is conformal to the angular metric and the memory is encoded in the change of $\rho^2e^\Phi$. After transforming to Bondi gauge and sending the screen to large radius, the conformal deformation is redistributed into the Bondi shear $C_{AB}$. The tracefree part of the renormalized screen deformation then reproduces exactly the displacement memory computed in \autoref{Sec:LinearRT}. Thus the finite Carrollian screen provides a quasilocal completion of the asymptotic memory observable.

For the linearized Robinson--Trautman mode of \autoref{Sec:LinearRT}, let
\begin{align}
\Phi=\Phi_\circ+\varepsilon_\ell e^{-\omega_\ell u}P_\ell(\cos\theta)
+{\cal O}(\varepsilon_\ell^2)~,
\qquad
\rho=\rho_0(u)+\varepsilon_\ell h_\ell(u)P_\ell(\cos\theta)
+{\cal O}(\varepsilon_\ell^2)~.
\label{LinearRTScreenAnsatz}
\end{align}
The background embedding obeys
\begin{align}
2\dot\rho_0+1-\frac{2m}{\rho_0}=0~.
\end{align}
Then \eqref{RTScalarMemoryFiniteScreen} gives
\begin{align}
\Delta_{\cal N}\log(\rho^2e^\Phi)
=
\varepsilon_\ell P_\ell(\cos\theta)
\left[
\frac{2h_\ell(u)}{\rho_0(u)}
+e^{-\omega_\ell u}
\right]_{u_i}^{u_f}
+{\cal O}(\varepsilon_\ell^2)~.
\label{LinearRTFiniteScreenMemory}
\end{align}
This is the scalar finite-distance analogue of the linearized Robinson--Trautman displacement memory. It contains both the explicit RT deformation and the response of the chosen null screen. The tensorial tracefree component is obtained from \eqref{ShapeMemoryFiniteScreen}. After transforming to the Bondi frame, its leading large-radius piece reproduces \eqref{LinearRTMemoryTensor}.

For a horizon-settling screen, $\rho_0=2m$ and $h_\ell(u_f)\to0$ as $u_f\to+\infty$. Since the Robinson--Trautman mode also decays, the memory from a finite initial cut to the final Schwarzschild screen is
\begin{align}
\Delta_{\cal N}\log(\rho^2e^\Phi)
=
-\varepsilon_\ell P_\ell(\cos\theta)
\left[
\frac{h_\ell(u_i)}{m}
+e^{-\omega_\ell u_i}
\right]
+{\cal O}(\varepsilon_\ell^2)~.
\label{LinearRTHorizonSettlingMemory}
\end{align}
This is a transition memory: it records the change from an initially distorted Robinson--Trautman screen to the final round Schwarzschild screen. It should not be interpreted as permanent nonspherical horizon hair. At linear order the final screen is round; irreversible changes in area or mass begin at order ${\cal O}(\varepsilon_\ell^2)$, since the gravitational-wave energy flux is quadratic in the shear.

The finite-screen memory defined above also fits naturally within the broader landscape of memory effects associated with null hypersurfaces. Besides the standard displacement memory at $\scri^+$, memory observables have been proposed for black-hole horizons, generic null boundaries, and bulk null congruences \cite{Hamada:2017gdg,Donnay:2018ckb,Rahman:2019bmk,Bart:2019gnf,Adami:2021nnf,Mao:2025yne}. These constructions differ in their precise observables and symmetry interpretations, but they share the same geometric mechanism: radiation or matter flux through a null hypersurface can leave a residual change in the data carried by that hypersurface.
It is also close in spirit to the broader class of persistent gravitational-wave observables, where displacement memory is one component of a more general finite-time response of detector configurations \cite{Grant:2021hga}.

{ The null-boundary phase analysis of \cite{Adami:2021nnf} provides a general framework for gravitational memory associated with null hypersurfaces, including expansion and spin memory. The Robinson--Trautman construction considered here instead follows the explicit geometric evolution of prescribed finite null screens, determining their accumulated scalar, tensorial, and momentum responses through the Carrollian data. The corresponding null Brown--York tensor describes the induced fluid relaxation, while the large-radius analysis identifies the standard Bondi displacement memory within the quasilocal geometric response.}

This viewpoint is reinforced by the symmetry structure of generic null surfaces. The covariant phase-space analysis of \cite{Chandrasekaran:2018aop} shows that a null boundary admits boundary-preserving diffeomorphisms, including supertranslation-like transformations intrinsic to the null surface. The resulting algebra is not simply the standard BMS algebra at infinity, but a null-boundary algebra adapted to the chosen hypersurface: diffeomorphisms of the spatial cross-sections act on an infinite-dimensional supertranslation sector along the null generators, with additional structure depending on the boundary conditions and on the allowed variations of the null data. The associated charges and fluxes can be localized on cuts of the null boundary.

{ The finite Robinson--Trautman screens are treated as quasilocal geometric detectors rather than as asymptotic boundaries. Their memory is defined by the residual change of intrinsic Carrollian data between two cuts, without imposing a finite-screen charge-memory relation. Establishing such a relation within the general null-boundary covariant phase space remains an interesting open question.}

%The finite Robinson--Trautman screens considered here should be understood in this broader spirit, but with a different emphasis. We do not treat ${\cal N}$ as an asymptotic boundary of spacetime, nor do we formulate a full charge-memory relation for the screen. Instead, the screen is used as a quasilocal geometric detector. The memory observable is the residual change of its intrinsic Carrollian data between two cuts. In this sense, finite-screen memory is a finite-distance geometric analogue of displacement memory rather than a new independent asymptotic observable.

This interpretation is consistent with the near-screen symmetry analysis summarized in \autoref{App:NearScreen}. There we show that a Robinson--Trautman null screen can be brought locally to Gaussian-null form and that the diffeomorphisms preserving the corresponding near-screen boundary conditions realize the expected null-surface structure. The same Carrollian null geometry that supports these near-screen symmetries is the geometry whose residual transition defines the finite-screen memory. It would be interesting to understand whether the finite-screen observable defined here admits a charge-memory interpretation within the covariant phase-space framework for finite-distance null hypersurfaces recently developed in \cite{Ruzziconi:2025fuy}.

A complementary bulk perspective was developed in \cite{Bart:2019gnf,Mao:2025yne}, where memory is associated with null-geodesic deviation or with transitions between families of null hypersurfaces in the bulk. Our construction is close in spirit, but the observable is different. We do not define memory as the transition between two null foliations. Rather, after choosing a finite null screen ${\cal N}$, we follow its Carrollian generators and measure the residual change of the intrinsic screen geometry.

For Robinson--Trautman spacetimes, this gives a simple final picture. The same parabolic relaxation that produces the Bondi displacement memory at $\scri^+$ also produces a quasilocal Carrollian memory on finite null screens. At finite-distance this memory retains focusing, shear, embedding, and near-zone information, while its universal wave-zone component is the standard displacement memory recovered in the large-radius limit.
\section{Cosmological Constant, Loss of Nullity at the Boundary, and Fluxes}
\label{Sec:RTLambda}

We finally comment on the effect of a nonzero cosmological constant. When $\Lambda\neq0$, the conformal boundary is no longer null: it is spacelike for de Sitter and timelike for anti-de Sitter. Therefore, the standard displacement-memory interpretation at $\mathscr I^+$, based on Bondi cuts and the time integral of the news, is not directly available. This is the sense in which the null-infinity memory construction loses its usual asymptotic arena.

{ We finally consider Robinson--Trautman spacetimes with nonzero cosmological constant where the conformal boundary is now either spacelike or timelike for dS and AdS, respectively. Therefore, the standard Bondi displacement memory construction at null infinity is no longer directly applicable. Nonetheless, finite null screens in the bulk retain their Carrollian structure, allowing us to extend the quasilocal memory construction of the previous sections. We first study their geometric and fluid response, before examining holographic charge evolution in the asymptotically AdS branch. }

%This obstruction is specific to the asymptotic Bondi-memory interpretation and does not remove the finite-screen construction. For $\Lambda\neq0$, finite-distance null hypersurfaces may still be chosen in the bulk, and one may study the residual deformation of their Carrollian data. The relevant optical quantities and the associated screen memory are collected in \autoref{App:SubSec:LBMSOptic}, where the modifications induced by the cosmological constant are displayed explicitly. In the finite-screen Carrollian fluid, $\Lambda$ appears as an isotropic background-curvature pressure in the scalar sector of the null Brown--York tensor. It changes the screen embedding and the scalar memory, but it is not a new dissipative channel; dissipation remains tied to the Robinson--Trautman relaxation and the optical shear.

\subsection{Finite-Distance Carrollian Screens and Memory}
\label{App:SubSec:LBMSOptic}

{ We extend the construction of \autoref{Sec:CarrFluidRT} to nonzero cosmological constant, determining how $\Lambda$ modifies the screen embedding, optical data, and null Brown--York tensor. We then use these results to evaluate the corresponding finite-distance geometric memory. }

%Here we repeat the finite-screen construction of \autoref{Sec:CarrFluidRT} for Robinson--Trautman spacetimes with nonzero cosmological constant. The point is not to define a new asymptotic memory observable, since for $\Lambda\neq0$ the conformal boundary is no longer null, but rather to show that finite null screens in the bulk still carry a well-defined Carrollian geometry and optical data.

We consider the Robinson--Trautman solution with cosmological constant given in \eqref{eq:RT_quadratic_metric}, and introduce a finite null screen
\begin{align}
{\cal N}:\qquad r=\rho(u,x^A)~,
\end{align}
with normal one-form
\begin{align}
\ell_\mu\d x^\mu=-\d(r-\rho)=-\d r+\dot\rho\,\d u+\partial_A\rho\,\d x^A~.
\end{align}
The nullity condition is
\begin{align}
2\dot\rho+\bar K+\rho\dot\Phi-\frac{2m}{\rho}
-\frac{\Lambda}{3}\rho^2
+\frac{e^{-\Phi}}{\rho^2}\gamma^{AB}\partial_A\rho\,\partial_B\rho=0~.
\label{eq:LambdaRTScreenNullity}
\end{align}
Thus, as in the asymptotically flat case, the embedding function $\rho$ is not
arbitrary. Once an initial cut is chosen, its evolution is fixed by the nullity
condition.

The induced spatial metric on the cuts of the screen is
\begin{align}
q_{AB}=\rho^2e^\Phi\gamma_{AB}~,
\qquad
q^{AB}=\rho^{-2}e^{-\Phi}\gamma^{AB}~,
\end{align}
and the Carrollian generator reads
\begin{align}
\ell^a\partial_a=\partial_u+V^A\partial_A~,
\qquad
V^A=q^{AB}\partial_B\rho
=
\frac{e^{-\Phi}}{\rho^2}\gamma^{AB}\partial_B\rho~,
\end{align}
or equivalently
\begin{align}
V_A=q_{AB}V^B=\partial_A\rho~.
\end{align}

Let $\mathscr D_A$ be the Levi--Civita connection of $q_{AB}$. The optical
tensor is
\begin{align}
\theta_{AB}
=
\frac12{\cal L}_\ell q_{AB}
=
\frac12\partial_uq_{AB}+\mathscr D_A\mathscr D_B\rho~.
\end{align}
Its trace and tracefree parts are
\begin{align}
\theta
=
\frac{2\dot\rho}{\rho}+\dot\Phi+\mathscr D^2\rho~,
\qquad
\sigma_{AB}
=
\left(\mathscr D_A\mathscr D_B-\frac12q_{AB}\mathscr D^2\right)\rho~.
\label{eq:LambdaRTOpticalBasic}
\end{align}
Using the nullity condition, the expansion can also be written as
\begin{align}
\theta
=
\mathscr D^2\rho
-\frac{\bar K}{\rho}
+\frac{2m}{\rho^2}
+\frac{\Lambda}{3}\rho
-\frac{1}{\rho}\mathscr D_A\rho\,\mathscr D^A\rho~.
\label{eq:LambdaRTExpansionExact}
\end{align}
With the rigging $n^\mu\partial_\mu=\partial_r$, the inaffinity is
\begin{align}
\kappa
=
-\frac12\dot\Phi-\frac{m}{\rho^2}
+\frac{\Lambda}{3}\rho
+\frac{1}{\rho}\mathscr D_A\rho\,\mathscr D^A\rho~.
\label{eq:LambdaRTKappaExact}
\end{align}
The normal-connection one-form is
\begin{align}
\Omega_a\d y^a
=
\left(
-\frac12\dot\Phi-\frac{m}{\rho^2}
+\frac{\Lambda}{3}\rho
\right)\d u
+\mathscr D_A\log\rho\,\d x^A~.
\end{align}
Its horizontal projection gives the \Hajicek one-form,
\begin{align}
\omega_A=\mathscr D_A\log\rho~,
\qquad
\omega_u=-V^A\omega_A
=
-\frac{1}{\rho}\mathscr D_A\rho\,\mathscr D^A\rho~.
\end{align}
Finally,
\begin{align}
\mu=\kappa+\frac12\theta
=
-\frac12\dot\Phi
+\frac12\mathscr D^2\rho
-\frac{\bar K}{2\rho}
+\frac{\Lambda}{2}\rho
+\frac{1}{2\rho}\mathscr D_A\rho\,\mathscr D^A\rho~.
\label{eq:LambdaRTMuExact}
\end{align}

Using the null Brown--York/Carrollian-fluid dictionary of
\autoref{Sec:CarrFluidRT}, the nonzero components of the Carrollian one-form
$\tau_a$ are
\begin{align}
\tau_u
&=
\frac{1}{8\pi G}
\left(
\omega_u-\theta
\right)
=
\frac{1}{8\pi G}
\left[
-\mathscr D^2\rho
+\frac{\bar K}{\rho}
-\frac{2m}{\rho^2}
-\frac{\Lambda}{3}\rho
\right]~,
\\
\tau_A
&=
\frac{1}{8\pi G}\omega_A
=
\frac{1}{8\pi G}\mathscr D_A\log\rho~.
\label{eq:LambdaRTTauOneForm}
\end{align}
Similarly, the independent spatial components of $\tau_a{}^b$ are
\begin{align}
\tau_A{}^B
=
\frac{1}{8\pi G}
\left[
\mathscr D_A\mathscr D^B\rho
-\frac12\delta_A{}^B\mathscr D^2\rho
-\mu\,\delta_A{}^B
\right]~.
\label{eq:LambdaRTTauSpatialStress}
\end{align}
Using \eqref{eq:LambdaRTMuExact}, this becomes
\begin{align}
\tau_A{}^B
=
\frac{1}{8\pi G}
\left[
\mathscr D_A\mathscr D^B\rho
-\frac12\delta_A{}^B\mathscr D^2\rho
+
\delta_A{}^B
\left(
\frac12\dot\Phi
-\frac12\mathscr D^2\rho
+\frac{\bar K}{2\rho}
-\frac{\Lambda}{2}\rho
-\frac{1}{2\rho}\mathscr D_C\rho\,\mathscr D^C\rho
\right)
\right]~.
\label{eq:LambdaRTTauSpatialStressExplicit}
\end{align}
The remaining components are fixed by horizontality and need not be displayed. 
Therefore the nonzero components of the mixed null Brown--York tensor
$T_a{}^b=\ell^b\tau_a+\tau_a{}^b$ are
\begin{align}
T_u{}^u
&=
\tau_u~,
&
T_A{}^u
&=
\tau_A~,
\\
T_u{}^B
&=
V^B\tau_u
-
V^A\tau_A{}^B~,
&
T_A{}^B
&=
V^B\tau_A+\tau_A{}^B~.
\label{eq:LambdaRTNullBYComponents}
\end{align}

These expressions show that $\Lambda$ enters explicitly in the scalar sector of the Carrollian fluid: it shifts $\tau_u$ by $-\Lambda\rho/(24\pi G)$ and the isotropic part of $\tau_A{}^B$ by $-\Lambda\rho\,\delta_A{}^B/(16\pi G)$. The momentum component $\tau_A$ has no explicit $\Lambda$-term, although it is modified indirectly through the embedding $\rho$ determined by \eqref{eq:LambdaRTScreenNullity}.
Using the second-order solution \eqref{eq:Phi_second_order_final}, we find that the radiative corrections to the late-time screen decay and the null Brown--York tensor relaxes to its stationary value. In particular,
\begin{align}
T_u{}^u\to0~,\qquad T_u{}^A\to0~,\qquad T_A{}^u\to0~,
\end{align}
while the only nonzero components are the isotropic spatial stresses
\begin{align}
T_A{}^B \to \frac{1-\Lambda\rho_h^2}{16\pi G\rho_h}\delta_A{}^B~, 
\end{align}
with the corresponding final screen determined by
\begin{align}
    1-\frac{2m}{\rho_h}-\frac{\Lambda}{3}\rho_h^2=0~.
    \label{LambdaFinalScreen}
\end{align}
This is the null Brown--York tensor of the corresponding Schwarzschild--(A)dS black-hole horizon.

%%%%
%%%%
%%%%
The finite-screen memory is defined as the residual change of the intrinsic spatial metric between two cuts of the same null screen. In the fixed screen frame, this is given in \eqref{ScalarMemoryFiniteScreen}. Using the explicit expansion \eqref{eq:LambdaRTExpansionExact}, we find
\begin{align}
\Delta_{\cal N}\log\sqrt{q}
=
\int_{u_i}^{u_f}\d u
\left[
-\frac{\bar K}{\rho}
+\frac{2m}{\rho^2}
+\frac{\Lambda}{3}\rho
-\frac{1}{\rho}\mathscr D_A\rho\,\mathscr D^A\rho
\right]~.
\label{eq:LambdaRTScalarScreenMemoryExplicit}
\end{align}

{ Equation \eqref{eq:LambdaRTScalarScreenMemoryExplicit} shows that the finite-screen memory retains its geometric definition for nonzero cosmological constant. The effect of $\Lambda$ enters explicitly through the scalar response and indirectly through the evolution of the screen embedding. For radiative Robinson--Trautman perturbations, the screen data relax toward the corresponding Schwarzschild-(A)dS configuration, with vanishing shear and momentum data for screens settling to a stationary black hole horizon. The accumulated memory therefore remains well defined independently of whether the conformal boundary is null. 
}

\subsection{Holographic Charge Evolution and Fluxes}

%In the remainder of this section, we focus on a different consequence of $\Lambda\neq0$: the evolution of a holographic energy charge in the asymptotically AdS branch of Robinson--Trautman geometries. General asymptotically locally (A)dS$_4$ solutions in Bondi gauge and their relation to Fefferman--Graham data were studied in \cite{Poole:2018koa,Compere:2019bua,compere2020lambda}. The Bondi-to-Fefferman--Graham dictionary determines the boundary metric and holographic stress tensor from the asymptotic Bondi data. We use this formalism in a conformal frame in which the spatial sections of the AdS boundary are round and time independent, while the Robinson--Trautman dynamics is encoded in a time-dependent boundary shift.

{ The finite-screen description above characterizes the geometric response to radiation through the intrinsic Carrollian data of a null hypersurface. In the asymptotically AdS branch, a complementary perspective is provided by the holographic stress tensor and its associated charges. The Bondi-to-Fefferman--Graham dictionary developed in \cite{Poole:2018koa,Compere:2019bua,compere2020lambda} determines the boundary metric and holographic stress tensor from the asymptotic Bondi data. We use this formalism in a conformal frame in which the spatial sections of the AdS boundary are round and time independent, while the Robinson--Trautman dynamics is encoded in a time-dependent boundary shift.}

%We now apply the Newman--Unti-to-Bondi map derived previously to the perturbative Robinson--Trautman solution. Although the solution is known to quadratic order as discussed in \autoref{SubSec:QuadraticRTSolution}, the present charge evolution only requires the linearized asymptotic map. For the present analysis, it is sufficient to determine this transformation only to first order in $\varepsilon_\ell$. Indeed, as shown below, the stationary background has vanishing boundary momentum density and boundary shift, so the first nontrivial evolution of the holographic charge is quadratic and arises entirely from the product of the linear momentum density with the linear time derivative of the shift,

{ To determine the holographic charge evolution, we apply the Newman--Unti-to-Bondi map to the perturbative Robinson--Trautman solution. Although the solution is known to quadratic order, as discussed in \autoref{SubSec:QuadraticRTSolution}, only the linearized asymptotic map is needed. Indeed, the stationary background has vanishing boundary momentum density and shift, so the leading charge evolution arises at quadratic order from the product of their first-order perturbations,}
\begin{align}
p_A=\varepsilon_\ell p_A^{(1)}+\mathcal O(\varepsilon_\ell^2)~, \qquad \partial_uU^A=\varepsilon_\ell\partial_uU^{A(1)}+\mathcal O(\varepsilon_\ell^2)~.
\end{align}
Consequently, neither the second-order coordinate transformation nor the second-order stress tensor is required at the perturbative order under control. 
The result therefore determines the evolution of the sphere-integrated quadratic charge, but not the complete local second-order mass aspect. In particular, different second-order representatives related by subleading gauge choices lead to the same integrated balance law at this perturbative order.

Denoting the original Robinson--Trautman coordinates by $(u_{\rm RT},r_{\rm RT},x_{\rm RT}^A)$ and the fixed-round Bondi coordinates by $(u,r,x^A)$, the asymptotic Newman--Unti-to-Bondi map is expanded order by order in $1/r$. For the quadratic charge evolution considered here, only the leading linearized coefficients are required:
\begin{align}
u_{\rm RT}&=u+\varepsilon_\ell T_\ell^{(1)}+\mathcal O(\varepsilon_\ell^2)~,\\
r_{\rm RT}&=r\left(1-\frac{\varepsilon_\ell}{2}K_\ell^{(1)}\right)+\varepsilon_\ell\frac{\lambda_\ell}{4\omega_\ell}K_\ell^{(1)}+\mathcal O(\varepsilon_\ell r^{-1},\varepsilon_\ell^2)~,\\
x_{\rm RT}^A&=x^A+\varepsilon_\ell X^{A(1)}(u,x^B,r)+\mathcal O(\varepsilon_\ell^2)~,
\end{align}
where 
\begin{align}
T_\ell^{(1)}=-\frac{1}{2\omega_\ell}K_\ell^{(1)}~, \qquad \partial_uT_\ell^{(1)}=\frac12K_\ell^{(1)}~, 
\end{align}
and the angular coefficients are fixed recursively by the Bondi gauge conditions but will not enter the integrated balance law below.
The leading radial rescaling removes the Robinson--Trautman Weyl factor from the angular metric, while the finite radial term is required by the Bondi gauge conditions at the next relevant order. 

In the resulting fixed-round Bondi frame, the boundary metric is
\begin{align}
\d s_{(0)}^2=g^{(0)}_{ab}\d x^a\d x^b=-\d u^2-\frac{\Lambda}{3}\gamma_{AB}\left(\d x^A-U^A\d u\right)\left(\d x^B-U^B\d u\right)~,
\label{eq:RT_fixed_round_boundary_metric}
\end{align}
with
\begin{align}
U^A=\varepsilon_\ell U^{A(1)}+\mathcal O(\varepsilon_\ell^2)~, \qquad U_A^{(1)}=-\frac{\Lambda}{3}D_AT_\ell^{(1)}=\frac{\Lambda}{6\omega_\ell}D_AK_\ell^{(1)}~.
\label{eq:RT_boundary_shift_linear}
\end{align}
Consequently,
\begin{align}
\partial_uU^{A(1)}=-\frac{\Lambda}{6}D^AK_\ell^{(1)}~.
\label{eq:RT_boundary_shift_derivative}
\end{align}
The same map induces the Bondi shear through
\begin{align}
\frac{\Lambda}{3}C_{AB}^{(1)}=2D_{\langle A}U_{B\rangle}^{(1)}~, \qquad C_{AB}^{(1)}=\frac{1}{\omega_\ell}D_{\langle A}D_{B\rangle}K_\ell^{(1)}~,
\label{eq:RT_linear_Bondi_shear_Lambda}
\end{align}
with $\langle ~,~\rangle$ denoting trace-free symmetrization. 
It is important that the fixed-round frame used here differs from the zero-shift frame often employed in the finite-$\Lambda$ charge literature. In this frame the spatial metric is time independent and the boundary work is encoded instead in the time-dependent shift.

The holographic stress tensor satisfies the boundary Ward identity
\begin{align}
\nabla_a T^{ab}=0~.
\label{eq:holographic_Ward_RT}
\end{align}
Given any boundary vector field $\xi^a$, the current
\begin{align}
J^a[\xi]=T^a{}_b\xi^b
\end{align}
therefore obeys\footnote{Indeed, the right-hand side follows directly from the symmetry of $T^{ab}$ and the holographic Ward identity, since $\nabla_a\left(T^a{}_b\xi^b\right)=T^{ab}\nabla_a\xi_b=\frac12T^{ab}\left(\nabla_a\xi_b+\nabla_b\xi_a\right)$.}
\begin{align}
\nabla_aJ^a[\xi]=\frac12T^{ab}\mathcal L_\xi g_{ab}^{(0)}~.
\label{eq:nonKilling_current_identity}
\end{align}
The associated charge is obtained by integrating the normal component of this current over the spatial boundary sections. Equation \eqref{eq:nonKilling_current_identity} shows that a conserved stress tensor does not necessarily produce a conserved charge when $\xi$ is not a Killing vector. Instead, the failure of $\xi$ to preserve the boundary metric acts as an external work term. 

A radiative Robinson--Trautman boundary metric does not possess an exact timelike Killing vector. Nevertheless, following the holographic charge convention of Poole, Skenderis and Taylor \cite{Poole:2018koa}, we choose
\begin{align}
\xi=-\partial_u~.
\end{align}
For stationary asymptotically AdS configurations, this reduces to the boundary time-translation Killing vector entering the holographic mass definition and gives the standard positive Schwarzschild--AdS mass. In the radiative Robinson--Trautman geometry, the same vector is not Killing because the boundary metric contains a time-dependent shift. We nevertheless use it as the natural continuation of the stationary mass generator; its associated holographic charge therefore obeys a flux-balance law rather than a conservation law. More generally, the absence of a preferred timelike boundary symmetry implies that mass-like charges are not unique, a point to which we return below.

Let
\begin{align}
n=\partial_u+U^A\partial_A
\end{align}
be the future-directed unit normal to the constant-$u$ boundary sections, and define the normal momentum density by
\begin{align}
p_A=T_{ab}n^a(\partial_A)^b~.
\label{eq:RT_boundary_momentum_definition}
\end{align}
For the metric \eqref{eq:RT_fixed_round_boundary_metric}, the integrated form of \eqref{eq:nonKilling_current_identity} becomes
\begin{align}
\frac{\d M_\xi}{\d u}=-\int_{S^2}\d\Omega_\circ\,p_A\partial_uU^A~.
\label{eq:RT_shift_work_balance}
\end{align}
Thus, the evolution of $M_\xi$ is the work performed by the time-dependent boundary shift against the holographic momentum density.

Specializing the Bondi-to-Fefferman--Graham dictionary of \cite{Poole:2018koa} to the linear Robinson--Trautman mode gives
\begin{align}
p_A=\varepsilon_\ell p_A^{(1)}+\mathcal O(\varepsilon_\ell^2)~, \qquad p_A^{(1)}=\frac{1}{8\pi G}\left[\frac{\lambda_\ell-2}{4}+\frac{m\Lambda}{2\omega_\ell}\right]D_AK_\ell^{(1)}~,
\label{eq:RT_linear_holographic_momentum}
\end{align}
which may also be checked directly from the boundary momentum Ward identity.

The stationary background has vanishing momentum density. It follows that the quadratic charge evolution depends only on the product of the linear momentum density and the linear shift,
\begin{align}
\frac{\d M_\xi^{(2)}}{\d u}=-\varepsilon_\ell^2\int_{S^2}\d\Omega_\circ\,p_A^{(1)}\partial_uU^{A(1)}~.
\label{eq:RT_quadratic_charge_balance}
\end{align}
In particular, the explicit second-order coefficients contained in the quadratic solution \eqref{eq:Phi_second_order_final}, including possible resonant contributions, do not enter \eqref{eq:RT_quadratic_charge_balance}. They would be required for a local reconstruction of the second-order stress tensor or for the charge evolution at higher perturbative orders, but not for the integrated quadratic balance law.

Substituting \eqref{eq:RT_boundary_shift_derivative} and \eqref{eq:RT_linear_holographic_momentum} into \eqref{eq:RT_quadratic_charge_balance} gives
\begin{align}
\frac{\d M_\xi^{(2)}}{\d u}=-\frac{1}{16\pi G}\left[\frac{\lambda_\ell-2}{4}+\frac{m\Lambda}{2\omega_\ell}\right]\varepsilon_\ell^2e^{-2\omega_\ell u}\int_{S^2}\d\Omega_\circ\,D_AP_\ell D^AP_\ell~.
\label{eq:RT_charge_before_harmonic_integral}
\end{align}
Using
\begin{align}
\int_{S^2}\d\Omega_\circ\,D_AP_\ell D^AP_\ell=\frac{4\pi\lambda_\ell}{2\ell+1}~,
\end{align}
and defining the mass-normalized charge by $M_{\xi,\ell}\equiv GM_\xi^{(2)}$, we obtain
\begin{align}
\frac{\d M_{\xi,\ell}}{\d u}=-\frac{\lambda_\ell}{4(2\ell+1)}\left[\frac{\lambda_\ell-2}{4}+\frac{m\Lambda}{2\omega_\ell}\right]\varepsilon_\ell^2e^{-2\omega_\ell u}+\mathcal O(\varepsilon_\ell^3)~.
\label{eq:RT_holographic_charge_evolution}
\end{align}

In the asymptotically flat limit,
\begin{align}
\left.\frac{\d M_{\xi,\ell}}{\d u}\right|_{\Lambda=0}=-\frac{\lambda_\ell(\lambda_\ell-2)}{16(2\ell+1)}\varepsilon_\ell^2e^{-2\omega_\ell u}+\mathcal O(\varepsilon_\ell^3)~,
\label{eq:RT_holographic_charge_flat_limit}
\end{align}
which agrees with the Bondi news-squared flux for the corresponding linear Robinson--Trautman mode after matching the shear normalization to the standard asymptotically flat Bondi convention of \eqref{LinearNews01}.

\begin{figure}[t]
    \centering
    \begin{subfigure}[t]{0.485\linewidth}
        \centering
        \includegraphics[width=\linewidth]{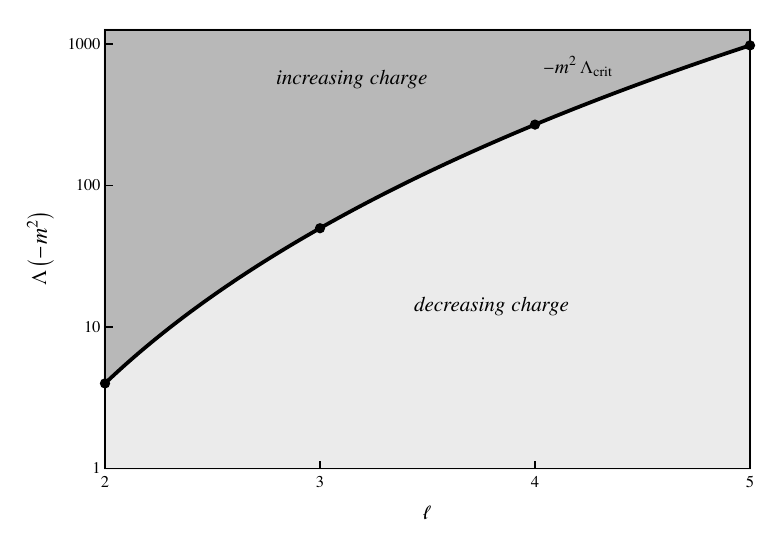}
        \label{fig:RT-monotonicity-phase}
    \end{subfigure}
    \hfill
    \begin{subfigure}[t]{0.485\linewidth}
        \centering
        \includegraphics[width=\linewidth]{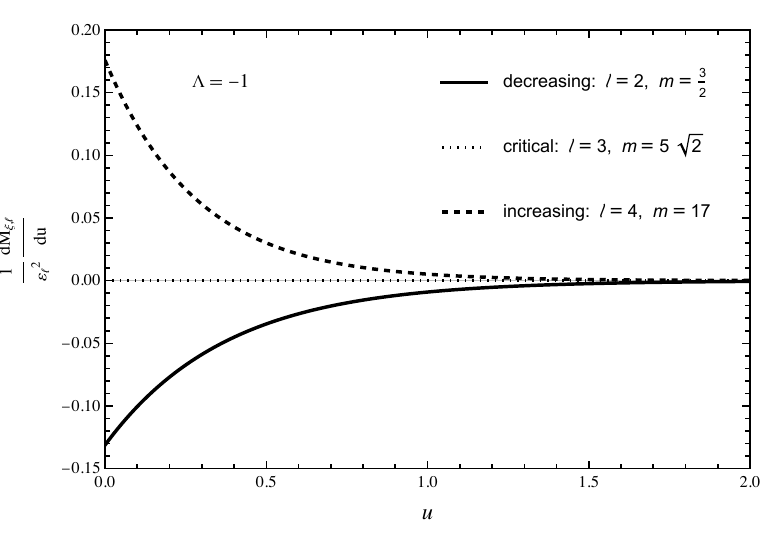}
        \label{fig:RT-flux-evolution}
    \end{subfigure}
    \caption{Mode-dependent monotonicity of the Robinson--Trautman holographic charge associated with $\xi=-\partial_u$. Left: the critical curve $-m^2\Lambda_{\mathrm{crit}}^{(\ell)} =\lambda_\ell(\lambda_\ell-2)^2/24$ separates regions with different monotonicity of the holographic charge. Right: the normalized evolution rate $\mathcal F_\ell(u)=\varepsilon_\ell^{-2}\partial_uM_{\xi,\ell}$ for representative configurations at fixed $\Lambda=-1$. Negative, vanishing, and positive values of $\mathcal F_\ell$ correspond, respectively, to decreasing, critical, and increasing charges.}
    \label{fig:RT-monotonicity}
\end{figure}

For the holographic charge associated with $\xi=-\partial_u$ in the fixed-round frame considered here, the coefficient controlling the monotonicity can change sign. The corresponding critical value is
\begin{align}
\Lambda_{\rm crit}^{(\ell)}=-\frac{\omega_\ell(\lambda_\ell-2)}{2m}=-\frac{\lambda_\ell(\lambda_\ell-2)^2}{24m^2}~.
\label{eq:RT_holographic_charge_critical_Lambda}
\end{align}
Since the remaining prefactor in \eqref{eq:RT_holographic_charge_evolution} is non-negative, the charge decreases for $\Lambda>\Lambda_{\rm crit}^{(\ell)}$, is stationary at $\Lambda=\Lambda_{\rm crit}^{(\ell)}$, and increases for $\Lambda<\Lambda_{\rm crit}^{(\ell)}$. This sign change characterizes the monotonicity of the particular holographic charge defined by the generator $\xi=-\partial_u$ and the fixed-round boundary frame used here; it should not be interpreted as a phase transition of the Robinson--Trautman geometry itself or as a universal property of all possible finite-$\Lambda$ charge definitions.
The mode-dependent change of monotonicity, together with representative decreasing, critical, and increasing fluxes, is summarized in Fig.~\ref{fig:RT-monotonicity}. For fixed $\ell$, the critical curve therefore separates different monotonicity regimes of the holographic charge rather than different branches of the Robinson--Trautman solution.

For the quadrupolar mode $\ell=2$,
\begin{align}
\frac{\d M_{\xi,2}}{\d u}=-\frac{3}{10}\left(1+\frac{m^2\Lambda}{4}\right)\varepsilon_2^2e^{-4u/m}+\mathcal O(\varepsilon_2^3)~,
\label{eq:RT_holographic_charge_quadrupole}
\end{align}
and
\begin{align}
\Lambda_{\rm crit}^{(2)}=-\frac{4}{m^2}~.
\end{align}

The choice of charge generator is physically motivated but is not unique because no exact timelike boundary symmetry exists during the radiative phase. More generally, one may consider vectors of the form \cite{Poole:2025cmv}
\begin{align}
\xi_f=-f(u,x^A)\partial_u~, \qquad f\rightarrow1 \quad \text{as} \quad u\rightarrow\infty~,
\end{align}
which define other weighted, mass-like holographic charges. Different choices of $f$ lead to different flux laws, and field-dependent choices require the corresponding field dependence to be included in a covariant phase-space treatment. We restrict here to the simplest choice $f=1$, which is directly connected to the final Schwarzschild--AdS time translation.

The loss of nullity of the conformal boundary prevents a direct extension of the standard Bondi displacement-memory construction to $\Lambda\neq0$, while finite-distance null-screen memory remains well defined in the bulk. In the AdS case, the holographic stress tensor additionally provides a boundary flux diagnostic. The charge associated with the continuation of the final Schwarzschild--AdS time translation evolves through the work performed by the time-dependent boundary shift. At quadratic order, this evolution is completely determined by the linear Robinson--Trautman data and reduces, after matching conventions, to the standard Bondi news-squared result in the flat limit.

For a complementary analysis of mass-like charges and their monotonicity properties in asymptotically de Sitter Robinson--Trautman spacetimes, see \cite{Poole:2025cmv}. This completes our analysis of the Robinson--Trautman solution at nonzero cosmological constant. Although the standard displacement-memory construction at null infinity is no longer available, the solution remains relevant both from the perspective of finite-distance screen memory and through the holographic charges that provide a complementary characterization of its radiative relaxation.
\section{Conclusions}

In this work we used Robinson--Trautman spacetimes as an exact laboratory for comparing asymptotic and finite-distance notions of gravitational memory. By transforming the asymptotically flat Robinson--Trautman family to Bondi gauge, we extracted the Bondi shear, news, mass aspect, and angular-momentum aspect directly in terms of the Robinson--Trautman field. In the linearized sector, a single Robinson--Trautman mode produces a purely electric Bondi shear and hence an ordinary displacement-memory effect. The corresponding memory records the relaxation from an initially radiative cut to the final Schwarzschild cut, with higher multipoles suppressed by their faster Robinson--Trautman decay rates.

We then moved away from null infinity and described Robinson--Trautman radiation on finite null hypersurfaces. Such hypersurfaces carry an intrinsic Carrollian geometry, with a degenerate metric on the cuts, a null generator, optical data, and a null Brown--York tensor. For the finite screens considered here, the nullity condition fixes the screen embedding in terms of the Robinson--Trautman field. The projected Einstein equations on the screen reduce to the corresponding Carrollian conservation equations, and their nontrivial content is controlled by the same Robinson--Trautman equation that governs the bulk radiative relaxation. In this sense, the parabolic Robinson--Trautman flow controls both the asymptotic Bondi data and the quasilocal Carrollian response of finite null screens.

%The finite-screen memory defined in this work is the residual change of the intrinsic Carrollian data of a prescribed null screen between two cuts. It includes the deformation of the transverse metric, the scalar area-density response, and the momentum data encoded in the \Hajicek one-form. This makes it a natural finite-distance analogue of displacement memory: just as Bondi memory measures the permanent change of the asymptotic shear between two cuts of $\scri^+$, finite-screen memory measures the permanent change of the intrinsic geometry of a chosen null screen. The analogy is, however, not an identity. 
{The finite-screen memory defined in this work is the residual change of the intrinsic Carrollian data of a prescribed null screen between two cuts. The Carrollian generator provides a natural prescription for identifying the cuts, allowing the accumulated geometric response to be expressed in terms of the optical expansion, shear, and momentum data. This includes the deformation of the transverse metric, the scalar area density response, and the momentum encoded in the \Hajicek one-form.}

For Robinson--Trautman screens this picture becomes especially transparent. In the natural Robinson--Trautman frame, the finite-screen scalar memory is encoded in the transition of the combination $\rho^2 e^\Phi$, where $\Phi$ is the Robinson--Trautman field and $\rho$ is the embedding of the null screen. The first contribution captures the deformation of the transverse angular geometry, while the second records the response of the chosen null hypersurface. For late-time screens approaching the final Schwarzschild horizon, the non-perfect Carrollian data decay exponentially: the expansion, tracefree shear, and transverse momentum vanish, while the isotropic Schwarzschild Brown--York stress remains. The corresponding memory is therefore a transition memory from an initially distorted Robinson--Trautman screen to the final round Schwarzschild screen, not permanent nonspherical horizon hair. Irreversible area or mass effects begin only at order ${\cal O}(\varepsilon_\ell^2)$, since the gravitational-wave energy flux is quadratic in the radiative amplitude.

The dynamical interpretation of finite-screen memory is also different from, but parallel to, the standard Bondi story. At null infinity, the displacement-memory equation follows from the Bondi constraints and relates the electric-parity memory to Coulombic data and to the news-squared flux. At finite-distance, the analogous role is played by the optical balance laws on the chosen null screen. Raychaudhuri controls the scalar focusing and area-density memory, Sachs controls the tracefree shape memory, and Damour controls the vector or momentum memory. In particular, the shear-squared term in Raychaudhuri is the finite-screen analogue of the news-squared contribution to nonlinear memory. Unlike the Bondi equation, however, these finite-screen balance laws are quasilocal and screen-dependent; they do not immediately reduce to a universal elliptic equation on the sphere until the large-radius tracefree Bondi limit is taken.

We also analyzed the role of a nonzero cosmological constant. The standard Bondi displacement-memory construction relies on the null character of $\mathscr I^+$ and therefore does not extend directly to $\Lambda\neq0$, where the conformal boundary is spacelike in de Sitter and timelike in anti-de Sitter. Finite null screens in the bulk, however, remain well defined and continue to carry Carrollian data. In this setting $\Lambda$ modifies the scalar sector of the null Brown--York tensor as an isotropic background curvature pressure, changing the screen embedding and scalar response without introducing a new dissipative channel. Separately, we constructed the Robinson--Trautman solution to second order in the radiative amplitude and studied the evolution of the associated holographic charge through the $\Lambda$-BMS/holographic dictionary \cite{Poole:2018koa,Compere:2019bua,compere2020lambda}. We find that the finite-$\Lambda$ charge associated with the continuation of the stationary time-translation generator does not obey a universal monotonicity property. This reflects the absence of a preferred timelike symmetry during the radiative phase and highlights the complementary role of holographic charges in characterizing Robinson--Trautman relaxation away from null infinity.

Several directions remain open. A first one is to relate the finite-screen observable to a genuine charge-memory relation. Recent covariant phase-space and quasi-local charge constructions for finite-distance null hypersurfaces \cite{Chandrasekaran:2018aop, Sheikh-Jabbari:2022mqi, Adami:2023wbe, Ruzziconi:2025fct,Ruzziconi:2025fuy,kmec2026quasi},\footnote{See for instance \cite{Grumiller:2019fmp, Adami:2020amw, Grumiller:2020vvv,Adami:2021kvx} for further developments on solution space and charges of null boundaries.} together with their associated asymptotic limit \cite{AsympCPS}, provide a natural framework for this question.

At null infinity, displacement, spin and center-of-mass memories are tied to BMS flux-balance laws and to the canonical soft sector at $\scri^+$ \cite{Strominger:2014pwa,Compere:2019gft}. Moreover, recent Carrollian-holographic analyses have shown that the BMS evolution equations can be understood as Ward identities of a Carrollian theory at null infinity, with radiation entering as a source for the boundary dynamics \cite{Hartong:2025jpp,Fiorucci:2025twa, Hartong:2026rbr}. This suggests a natural finite-distance analogue: the Raychaudhuri, Sachs and Damour equations on a finite null hypersurface should play the role of quasilocal flux-balance laws, or equivalently finite-screen Carrollian Ward identities. It would be interesting to determine whether the finite-screen memory considered here can be obtained from localized null-surface charges in this way.

A complementary direction is to formulate finite-screen memory directly in the language of Carrollian fluids. In the present construction, the null Brown--York tensor provides a natural hydrodynamic description of the screen dynamics, with the optical expansion, shear, and momentum densities controlling the response of the null fluid. It would be interesting to determine whether the memory observable $\Delta_{\cal N}q_{AB}$ admits a complete Carrollian fluid interpretation in terms of finite-distance Ward identities, constitutive relations, and entropy production. In particular, the Raychaudhuri equation suggests a possible connection between finite-screen memory and generalized notions of dynamical entropy for null hypersurfaces. While black-hole entropy changes are usually formulated in terms of event horizons and asymptotic charges,\footnote{See \cite{Rignon-Bret:2023fjq,Hollands:2024vbe,Visser:2024pwz,Ashtekar:2026jdz, Shajiee:2026coz} for recent proposals on defining entropy of dynamical black holes.} finite null screens provide a more general setting in which area evolution, focusing, and gravitational radiation can be studied without requiring a stationary horizon. Understanding whether the scalar memory variable $\Delta_{\cal N}\log\sqrt q$ admits an interpretation as a generalized null entropy transition could provide a bridge between Carrollian hydrodynamics, null-boundary thermodynamics, and gravitational memory.

A closely related question is the canonical status of the finite-screen memory variables. Recent work has related the soft sector of asymptotically flat gravity to the phase space of finite null surfaces, identifying the leading soft graviton with a finite radial fluctuation and the supertranslation Goldstone mode with its conjugate edge datum \cite{ciambelli2026asymptotically,ciambelli2026mapping}. This suggests that the memory observables studied here may admit a canonical interpretation as finite-distance representatives of the asymptotic soft modes. In the Robinson--Trautman example, one could test this proposal by comparing the finite-screen deformation $\Delta_{\cal N}q_{AB}$ with the Bondi memory $\Delta C_{AB}$ under the finite-to-asymptotic phase-space map. A useful way of formulating the problem is to regard both constructions as reductions of the general finite null-surface phase space. The causal-diamond analysis of \cite{ciambelli2026asymptotically,ciambelli2026mapping} selects a spherically symmetric sector, in which the relevant canonical pair is the finite radial fluctuation of the diamond and its conjugate edge datum. The Robinson--Trautman family selects a different reduced sector: an angular-dependent radiative sector whose time evolution is fixed by the Robinson--Trautman equation and whose finite-screen embedding is fixed by the nullity condition. Pulling back the general null-boundary symplectic form to this RT sector would determine the resulting reduced symplectic structure and clarify whether the finite-screen area-radius deformation and its conjugate null Brown--York or edge datum form the angular-dependent analogue of the causal-diamond pair, or whether the RT constraints make the sector a further constrained, possibly Lagrangian, submanifold of the full phase space.

This canonical question also has a natural quantum extension. Recent approaches to the quantization of null hypersurface geometry promote the cut metric, shear and area element to quantum data on the null surface \cite{Ciambelli:2024swv,Ciambelli:2025flo,Ciambelli:2026vxa}. From this viewpoint, the finite-screen memory observables considered here should become operator-valued transitions in the quantum geometry of a null hypersurface. In particular, the scalar memory $\Delta_{\cal N}\log\sqrt q$ is the natural observable in which quantum focusing effects and area fluctuations should appear, since its classical balance law is controlled by the Raychaudhuri constraint. One would then expect the leading tracefree large-radius component of this operator-valued transition to reproduce the usual soft or celestial memory operator at $\scri^+$. The Robinson--Trautman solutions studied here would provide explicit classical saddle configurations for such operators, and could therefore serve as a controlled sector in which to compare finite null-surface quantization with the asymptotic soft/celestial description of memory.

A second direction is to clarify the infrared and holographic interpretation of finite-screen memory. In the asymptotic case, memory is one corner of the infrared triangle relating asymptotic symmetries, soft theorems, and flux-balance laws. Recent work has shown that related structures can also be described in local detector frames, where memory modes arise as large residual diffeomorphisms and lead to corresponding Ward identities \cite{DeLuca:2024asq,DeLuca:2024cjl}. The near-screen analysis of \autoref{App:NearScreen} suggests that finite Robinson--Trautman screens carry the local null-surface symmetry structure expected from Gaussian-null boundary conditions. It would be worthwhile to understand whether the Carrollian memory data defined here admit an analogous interpretation in terms of finite-distance charges, Ward identities, or soft limits intrinsic to a null hypersurface.

A further aspect of this question concerns holography. One possible route is to understand the status of gravitational memory through flat and Carrollian limits of AdS/CFT and supergravities, following recent related developments in this direction \cite{Barnich:2012aw,Bagchi:2023fbj,Kraus:2024gso,Duary:2024kxl,Arenas-Henriquez:2025rpt,Lipstein:2025jfj,Fontanella:2025tbs,Poulias:2025eck,Bagchi:2026emg,Navarro:2026rna,Xu:2026spj,Bulunur:2026yav,Henneaux:2026dfc,Duary:2026hpb,Ballesteros:2026bqe}. In asymptotically flat gravity, displacement memory is part of the infrared triangle, and from the holographic viewpoint one would like to identify which CFT observables have a finite flat limit that becomes the Bondi news, the supertranslation charge, or the displacement-memory mode. The finite-screen construction suggests an intermediate step in this program: before the strict $\scri^+$ limit is taken, a finite null screen retains additional Coulombic, focusing, embedding, and momentum information. Understanding whether this quasilocal memory admits a holographic description, and how its large-radius projection matches the usual soft-memory observable of flat-space holography, could help separate the universal infrared part of memory from the screen-dependent finite-distance response.

A related direction concerns the analytical simplicity of Robinson--Trautman spacetimes. Dynamics of event horizons was studied previously in \cite{Redondo-Yuste:2022czg,Husnugil:2025edm} using perturbative methods. These techniques could be extended to the finite-distance null screens introduced in this work. Since Robinson--Trautman dynamics is analytically tractable, this geometry provides a controlled testing ground for understanding the robustness of perturbative approaches beyond event horizons and for investigating the relation between horizon dynamics, Carrollian fluid variables, and finite-screen memory.

More broadly, the lesson of the Robinson--Trautman example is that asymptotic memory can be viewed as the universal part of a richer finite-distance geometric response. Null infinity isolates the radiative tracefree component and removes the screen-dependent near-zone data. A finite Carrollian screen, by contrast, retains the full quasilocal optical response of the radiation. This makes finite-screen memory less universal than Bondi memory, but potentially more informative: it keeps track not only of the radiative shear, but also of focusing, embedding, angular drift, momentum, and thermodynamic data that are invisible in the strict asymptotic limit. Understanding how this finite-distance information fits into the infrared structure of gravity may provide a useful bridge between asymptotic symmetries, null-boundary phase spaces, Carrollian hydrodynamics, holography, and local measurements of gravitational radiation.

\acknowledgments  
We thank Gabriel Arenas-Henriquez, Glenn Barnich, Luca Ciambelli, Cristóbal Corral, Hern\'an Gonz\'alez, Michael Imseis, Luis Lehner, Thomas M\"adler, Rodrigo Olea, David Rivera-Betancour, Simone Speziale, and Per Sundell for comments and discussions on related topics.
The work of FD was supported by FONDECYT Grant No. 3260880. OL was partially supported by FONDECYT Grant No. 1230853. LS was supported by the Beca Doctorado Nacional ANID Grant No. 21221813 and FONDECYT Grant No. 1240043. This research was supported in part
by Perimeter Institute for Theoretical Physics. Research
at Perimeter Institute is supported in part by the Government of Canada through the Department of Innovation,
Science and Economic Development and by the Province
of Ontario through the Ministry of Colleges and Universities.

\begin{appendix}
%%%%%%%%%%%%%%%%%
%%%%%%%%%%%%%%%%%
%%%%%%%%%%%%%%%%%
%%%%%%%%%%%%%%%%%
\section{Bondi--Sachs Formalism}\label{App:BondiSachs}
In this appendix we collect the basic ingredients of the Bondi--Sachs description of asymptotically flat spacetimes in four dimensions. The Bondi, van der Burg, Metzner, and Sachs formalism \cite{Bondi:1960jsa,Bondi:1962px,Sachs:1961zz,Sachs:1962wk}, which we refer to as the Bondi--Sachs formalism, provides a framework to describe radiative spacetimes by analyzing the gravitational field in the asymptotic region near future null infinity.

Instead of foliating spacetime by spacelike hypersurfaces $t=\mathrm{constant}$, as in the ADM formulation \cite{Arnowitt:1962hi}, the Bondi--Sachs formalism uses outgoing null hypersurfaces of constant retarded time. We introduce coordinates
\begin{align}
x^\mu=(u,r,x^A)~,
\end{align}
where $u$ is retarded time, $r$ is an areal radial coordinate, and $x^A$ are coordinates on the celestial two-sphere. The surfaces $u=\mathrm{constant}$ are outgoing null hypersurfaces, and $\partial_r$ is tangent to their null generators. The coordinate $r$ labels points along these generators as an areal radius. The metric of the unit round sphere is denoted by $\gamma_{AB}$, with covariant derivative $D_A$, volume form $\epsilon_{AB}$, and Laplacian $D^2=D^A D_A$.

The Bondi gauge is defined by the conditions
\begin{align}
g_{rr}=0~, \qquad g_{rA}=0~, \qquad \partial_r\left(\frac{\det g_{AB}}{r^4}\right)=0~,
\end{align}
where the last is usually referred to as the Bondi determinant condition.

Equivalently, the angular part of the metric is normalized so that $r$ measures the area radius of the spherical cuts. A convenient form of the metric is\footnote{The radial coordinate is not, in general, an affine parameter unless $\partial_r\beta=0$.}
\begin{align}
\d s^2 =-\frac{V}{r}e^{2\beta}\d u^2 -2e^{2\beta}\d u\d r +g_{AB}\left(\d x^A-U^A\d u\right) \left(\d x^B-U^B\d u\right)~,
\label{BondiMetric}
\end{align}
where $V,\beta,U^A$, and $g_{AB}$ are functions of $x^\mu$ and admit asymptotic expansions at large $r$.

For an asymptotically flat spacetime, the angular metric takes the form
\begin{align}
g_{AB}=r^2q_{AB}+rC_{AB}+D_{AB}+\frac{1}{r}E_{AB}+\mathcal{O}(r^{-2})~,
\label{TransverseMetricExpansion}
\end{align}
together with 
\begin{align}
q^{AB}C_{AB}=0~,
\label{ShearTrace}
\end{align}
coming from the determinant condition. Furthermore, the determinant condition fixes only the area density of $q_{AB}$, requiring
\begin{align}
    \det q_{AB}=\det \gamma_{AB}~,
\end{align}
while leaving its conformal class unfixed. In the standard Bondi frame one uses the residual boundary freedom to choose
\begin{align}
    q_{AB}=\gamma_{AB}~.
\end{align}
In what follows we adopt this standard round-sphere Bondi frame.

The symmetric tracefree tensor $C_{AB}(u,x^C)$ is the Bondi shear. Its retarded-time derivative,
\begin{align}
N_{AB}:=\partial_u C_{AB}~,
\label{NewsDefinition}
\end{align}
is the Bondi news tensor. The news is the radiative datum of the asymptotically flat spacetime: $N_{AB}=0$ characterizes non-radiative regions of future null infinity. The importance of the news tensor is analogous to the role of the field
strength in Maxwell theory. The Bondi shear $C_{AB}$ is the asymptotic strain variable, while its retarded-time derivative $N_{AB}=\partial_u C_{AB}$ is the genuinely radiative quantity. Just as the electromagnetic radiation
flux is quadratic in the radiative components of $F_{\mu\nu}$, the gravitational energy flux through future null infinity is quadratic in the news tensor.
Moreover, the news appears naturally in the radiative phase space of asymptotically flat gravity. In the Ashtekar--Streubel description of null infinity \cite{Ashtekar:1981bq}, the symplectic structure on the gravitational radiative data takes the form
\begin{align}
\Omega_{\scri^+}(\delta_1,\delta_2) = \frac{1}{32\pi G}\int_{\scri^+}\d u\d^2\Omega\left( \delta_1 C^{AB} \delta_2 N_{AB}-\delta_2 C^{AB} \delta_1 N_{AB}\right)~.
\end{align}
Thus $C_{AB}$ is the radiative configuration variable, while $N_{AB}$ is its retarded-time derivative and momentum-like datum. The condition $N_{AB}=0$ therefore characterizes non-radiative regions of $\scri^+$, whereas nonzero news signals the passage of gravitational waves. This is why the Bondi news is the main ingredient in the asymptotic description of gravitational radiation and memory.
%%%%
%%%%

In vacuum Einstein gravity, the remaining metric functions have the asymptotic expansion
\begin{align}
\beta&=-\frac{1}{32r^2}C_{AB}C^{AB}
+\mathcal{O}(r^{-3})~,\\ V&=r-{2m_{\rm B}}+\mathcal{O}(r^{-1})~,
\\ U^A&=-\frac{1}{2r^2}D_B C^{AB}+\frac{1}{r^3}\left[-\frac{2}{3}N^A+\frac{1}{16}D^A(C_{BC}C^{BC})+\frac{1}{2}C^{AB}D^C C_{BC}\right]+\mathcal{O}(r^{-4})~,\\ D_{AB} &=\frac14 \gamma_{AB}C_{CD}C^{CD}~.
\label{BondiExpansions}
\end{align}
Here $m_{\rm B}(u,x^A)$ is the Bondi mass aspect and $N_A(u,x^A)$ is the angular-momentum aspect. Different conventions in the literature may shift signs or numerical factors in the definition of $N_A$ and $m_{\rm B}$. In what follows all quantities are defined with respect to the Bondi ansatz \eqref{BondiMetric}.
%%%%%%%%%%%%%%%%%%

The Bondi data also admit a direct interpretation in terms of the asymptotic curvature. A convenient way to organize this information is through the Newman--Penrose formalism \cite{Newman:1961qr, Penrose:1963iua}, in which one projects the Weyl tensor onto a complex null tetrad adapted to the outgoing null foliation as
\begin{align}\label{WPScalars}
    \Psi_0 ={}& - W_{\mu\nu\lambda\sigma}\ell^\mu m^\nu \ell^\lambda m^\sigma~, \\ \Psi_1 ={}& -W_{\mu\nu\lambda\sigma}l^\mu n^\nu l^\lambda m^\sigma~, \\ \Psi_2 ={}& -W_{\mu\nu\lambda\sigma}l^\mu m^\nu \bar{m}^\lambda n^\sigma~, \\ \Psi_3 ={}&-W_{\mu\nu\lambda\sigma}l^\mu n^\nu \bar{m}^\lambda n^\sigma~, \\ \Psi_4 ={}& -W_{\mu\nu\lambda\sigma}n^\mu \bar{m}^\nu n^\lambda \bar{m}^\sigma~.
\end{align}
The resulting five complex scalars $\Psi_n$, $n=0,\ldots,4$, encode the independent components of the free gravitational field. Near future null infinity they obey the peeling hierarchy,
\begin{align}
\Psi_n=\frac{\Psi_n^{(0)}(u,x^A)}{r^{5-n}}+\mathcal O(r^{n-6})~,\qquad n=0,\ldots,4~.
\end{align}
This hierarchy separates the different physical sectors of the gravitational field: $\Psi_4^{(0)}$ is the leading outgoing radiative component, $\Psi_3^{(0)}$ is related to angular derivatives of the radiation, $\Psi_2^{(0)}$ contains the Coulombic mass aspect, $\Psi_1^{(0)}$ contains angular-momentum data, and $\Psi_0^{(0)}$ encodes the leading component of the opposite, incoming-oriented curvature sector in this tetrad. Thus the Newman--Penrose scalars provide a curvature-space counterpart of the Bondi expansion.

To make this dictionary more explicit, we introduce a complex dyad $q^A$ on the celestial sphere, normalized with respect to the leading angular metric $q_{AB}$ as
\begin{align}
q^A q_{AB}\bar q^B=1~,
\qquad
q^A q_{AB}q^B=0~.
\end{align}
In local stereographic coordinates $z,\bar z$, one convenient representative is
\begin{align}
q^A\partial_A=\frac{1}{\sqrt{q_{z\bar z}}}\partial_{\bar z}~,\qquad\bar q^A\partial_A=\frac{1}{\sqrt{q_{z\bar z}}}\partial_z~,
\end{align}
but the following construction is independent of this coordinate choice. Writing $g_{AB}=r^2h_{AB}$, we parametrize the angular metric in terms of the spin-weighted variables
\begin{align}
{\cal K}&:=q^A h_{AB}\bar q^B~,\qquad{\cal J}:=q^A h_{AB}q^B~.
\end{align}
The Bondi determinant condition implies
\begin{align}
{\cal K}^2-{\cal J}\bar{\cal J}=1~,
\end{align}
and the angular metric may be reconstructed as
\begin{align}
h_{AB}={\cal K}q_{AB}+\frac12\left({\cal J}\bar q_A\bar q_B+\bar{\cal J}q_Aq_B\right)~.
\end{align}

With this tetrad, the Weyl scalars satisfy the peeling hierarchy once the Einstein equations and the Bondi expansions are imposed. With the curvature and tetrad conventions used in this appendix, the leading coefficients are
\begin{align}
\Psi_0&=\frac{1}{r^5}\left[3\left(E_{AB}-\frac{1}{16}C_{AB}C_{CD}C^{CD}\right)
\right]q^Aq^B+\mathcal O(r^{-6})~,\\\Psi_1&=\frac{1}{r^4}\left[\frac23q^AN_A\right]+\mathcal O(r^{-5})~,\\\Psi_2&=\frac{1}{r^3}\left[-\left(m_{\rm B}-\frac14 C_{AB}N^{AB}+\frac14 D_AD_BC^{AB}\right)-i\left(\frac{1}{4}\epsilon^{AC}D_AD^BC_{BC}\right)
\right]+\mathcal O(r^{-4})~,\\ \Psi_3&=\frac{1}{r^2}\left[-\frac12\bar q^A D^B N_{AB}\right]+\mathcal O(r^{-3})~,
\\ \Psi_4&=\frac{1}{r}\left[\bar{q}^A\bar{q}^B\partial_u N_{AB}\right]+\mathcal O(r^{-2})~.
\label{BondiWeylScalars}
\end{align}
The precise overall signs depend on the curvature convention and on the orientation of the complex null tetrad, but the physical content of the expansion is invariant. The scalar $\Psi_4$ contains the leading outgoing radiative curvature. In the present Bondi frame its leading coefficient is controlled by $\partial_u N_{AB}$, or equivalently by $\partial_u^2C_{AB}$, while the news $N_{AB}$ itself is the radiative datum whose square gives the gravitational energy flux through $\scri^+$. The scalar $\Psi_3$ is determined by angular derivatives of the news and therefore also belongs to the radiative sector. In the minimal smooth Bondi-Sachs phase space the tracefree integration tensor at order $1/r$ in $g_{AB}$ is set to zero, equivalently $\Psi_0^{(0)}=0$. This is the standard no-incoming-radiation sector and does not eliminate outgoing radiation, which is instead encoded in the Bondi news. Allowing $E_{AB}-\frac{1}{16}C_{AB}C_{CD}C^{CD}\neq0$ corresponds to enlarging the asymptotic phase space by an independent leading $\Psi_0$ datum (see for instance \cite{Geiller:2022vto,Geiller:2024amx}).

The Coulombic information is contained in $\Psi_2$. Its real part contains the Bondi mass aspect $m_{\rm B}$ together with the hard term $C_{AB}N^{AB}$ and the soft electric term $D_AD_BC^{AB}$. Its imaginary part 
\begin{align}
\operatorname{Im}\Psi_2^{0}= -\frac14\epsilon^{AC}D_AD^BC_{BC}~,
\end{align}
is proportional to the magnetic-parity scalar appearing in the decomposition of the Bondi shear. Thus $\operatorname{Im}\Psi_2^0$ measures a magnetic, or NUT-like, Coulombic component of the asymptotic gravitational field. Finally, $\Psi_1^{(0)}$ contains the angular-momentum aspect $N_A$, while $\Psi_0^{(0)}$ contains the next independent, incoming-type asymptotic curvature data. In this way the Newman--Penrose scalars provide a curvature-space interpretation of the Bondi quantities: $N_{AB}$ controls radiation, $m_{\rm B}$ controls the mass aspect, $N_A$ controls angular-momentum data, and the electric and magnetic parity sectors of $C_{AB}$ appear directly in the Coulombic scalar $\Psi_2^{(0)}$. For further details on the curvature interpretation of the Bondi data in terms of the Weyl--Penrose scalars, see for instance \cite{Geiller:2024bgf,Geiller:2022vto}. 

The Newman--Penrose scalars also provide a convenient way to state the Petrov--Penrose classification. The scalars $\Psi_i$ are the components of the Weyl tensor in a chosen null tetrad, whereas the Petrov type characterizes the algebraic degeneracy of the Weyl tensor through the multiplicity of its principal null directions. In an arbitrary Bondi-adapted tetrad the five scalars are generically nonzero, but by aligning one of the null tetrad vectors with a principal null direction some of them may vanish. In particular, repeated principal null directions are signaled by stronger vanishing conditions among the $\Psi_i$, and the different possible degeneracies give the standard Petrov types I, II, D, III, N, and O, as summarized in \autoref{Table:PetrovClass}).

\begin{table}[h]
\centering
\renewcommand{\arraystretch}{1.25}
\begin{tabular}{llll}
\hline
\textbf{Petrov type} &
\textbf{Principal null directions} &
\textbf{Adapted Weyl scalars} &
\textbf{Interpretation} \\
\hline
I &
$4$ distinct &
generic &
Algebraically general \\
II &
$1$ repeated pair &
$\Psi_0=\Psi_1=0$ &
One repeated null direction \\
D &
$2$ repeated pairs &
$\Psi_0=\Psi_1=\Psi_3=\Psi_4=0$ &
Coulomb/black-hole type \\
III &
$1$ triple direction &
$\Psi_0=\Psi_1=\Psi_2=0$ &
Degenerate radiative type \\
N &
$1$ quadruple direction &
$\Psi_0=\Psi_1=\Psi_2=\Psi_3=0$ &
Pure transverse radiation \\
O &
$W_{\mu\nu\rho\sigma}=0$ &
$\Psi_i=0\ \forall i$ &
Conformally flat \\
\hline
\end{tabular}
\caption{Petrov--Penrose classification in terms of principal null directions and the corresponding vanishing pattern of Newman--Penrose Weyl scalars in an adapted null tetrad.}
\label{Table:PetrovClass}
\end{table}

%%%%%%%%%%%%%%%%
The $uu$ component of Einstein's equations gives the \textit{Bondi mass-loss formula}
\begin{align}
\partial_u m_{\rm B}=\frac{1}{4}D_A D_B N^{AB}
-T_{uu}~,\qquad T_{uu}:= \frac{1}{8}N_{AB}N^{AB}+4\pi G\lim_{r\to\infty} r^2 T_{uu}^{\mathrm{matter}}~,
\label{BMEvolution}
\end{align}
where $T^{\mathrm{matter}}_{\mu\nu}$ denotes the stress tensor of the matter fields.\footnote{Near $\scri^+$, only the component of the matter stress tensor carrying energy flux along outgoing null directions contributes at leading order.}
In vacuum this reduces to
\begin{align}
\partial_u m_{\rm B}=\frac{1}{4}D_A D_B N^{AB}-\frac{1}{8}N_{AB}N^{AB}~.
\end{align}
The total Bondi mass is
\begin{align}
M_{\rm B}(u)=\int \d^2\Omega m_{\rm B}(u,x^A)~.
\end{align}
Since the integral of the total-derivative term vanishes on the sphere, one obtains the global mass-loss formula
\begin{align}
\frac{\d M_{\rm B}}{\d u}=-\frac{1}{32\pi G}\int d^2\Omega N_{AB}N^{AB}-\lim_{r\to\infty} r^2 \int \d^2\Omega~ T_{uu}^{\mathrm{matter}} \leq0~.
\label{BML}
\end{align}
Thus positive outgoing radiation decreases the Bondi mass, assuming the matter satisfies the appropriate positive-energy condition.

Having identified the Bondi shear and news as the radiative data at future null infinity, we now explain how these quantities are related to an observable effect. The physical content of $C_{AB}$ can be seen by considering a set of freely falling detectors located at large radius $r$. In the wave zone, their relative separation is governed by the geodesic-deviation equation. If $s^A(u)$ denotes the angular separation vector between two neighboring detectors, then the leading tidal equation takes the form
\begin{align}
\frac{\d^2 s^A}{\d u^2}=- R^A{}_{uBu}s^B~.
\end{align}
In Bondi gauge, the radiative component of the curvature is controlled by the retarded-time variation of the shear,
\begin{align}
R_{uAuB}=-\frac{r}{2}\partial_u^2 C_{AB}
+\mathcal{O}(r^0)~.
\end{align}
Thus the passage of radiation produces a transient tidal acceleration determined by the news tensor. When the burst has passed, the news vanishes again, but the separation of the detectors need not return to its original value. Integrating the geodesic-deviation equation between an initial and a final non-radiative cut of $\scri^+$, one obtains the permanent displacement
\begin{align}
\Delta s^A=\frac{1}{2r}\gamma^{AC}\Delta C_{BC} s^B+\mathcal{O}(r^{-2})~,
\end{align}
where
\begin{align}
\Delta C_{AB} :=C_{AB}(u_f,x^C)-C_{AB}(u_i,x^C)=\int_{u_i}^{u_f}\d u N_{AB}~.
\end{align}
This residual change in the relative separation of inertial detectors is the displacement memory effect. It shows that gravitational radiation is not characterized only by an oscillatory waveform: a burst of radiation can also leave a permanent imprint, encoded asymptotically in the change of the Bondi shear.
%%%%%

Once the Bondi gauge and the asymptotic falloff conditions are imposed, there remains a nontrivial residual gauge freedom. The corresponding asymptotic symmetry group is the Bondi--Metzner--Sachs (BMS) group.  In four dimensions, the standard BMS group contains the Lorentz group together with an infinite-dimensional Abelian subgroup of angle-dependent translations in retarded time, called \textit{supertranslations}. More precisely,
\begin{align}
    \mathrm{BMS}_4=SL(2,\mathbb C)\ltimes C^\infty(S^2)~,
\end{align}
where $SL(2,\mathbb C)$ is the double cover of the proper orthochronous Lorentz group and $C^\infty(S^2)$ denotes the space of supertranslation functions on the celestial sphere. Here we restrict to the standard globally well-defined BMS group. Enlarged versions allowing local conformal transformations, usually called \emph{superrotations}, or more general sphere diffeomorphisms will not be needed in this appendix.\footnote{For the extension by local conformal transformations, see \cite{Barnich:2010eb,Barnich:2011mi}. For the generalized BMS group involving $\mathrm{Diff}(S^2)$, see \cite{Campiglia:2014yka}. For a review of the physical implications of finite superrotations, see \cite{Strominger:2016wns}.} It is also useful to recall that the BMS group admits a Carrollian interpretation. As shown in \cite{Duval:2014uva}, the BMS group can be viewed as a conformal Carroll group acting on null infinity. The Carroll group arises as the ultra-relativistic contraction, in the In\"on\"u-Wigner sense, of the Poincar\'e group \cite{LevyLeblond:1965}, corresponding formally to the limit $c\to0$, in contrast with the Galilei group, which arises from the non-relativistic limit $c\to\infty$.\footnote{See also \cite{Hartong:2022lsy} for a modern review of non-relativistic gravity and its connections to holography and string theory.} Geometrically, a Carrollian structure is naturally associated with a null hypersurface: it consists of a degenerate spatial metric together with a preferred null direction. Since $\scri^+$ is itself a null hypersurface, its universal geometry is Carrollian, and the BMS transformations act as conformal automorphisms of this Carrollian structure.

The relation between displacement memory and supertranslations becomes manifest after decomposing the Bondi shear into electric- and magnetic-parity components. On the unit sphere, a symmetric tracefree tensor can be written as
\begin{align}
    C_{AB}=-2\left(D_A D_B-\frac12\gamma_{AB}D^2\right)C+\epsilon_{C(A}D_{B)}D^C\widetilde C~,
    \label{ShearHodgeDecomposition}
\end{align}
where $C$ is the electric-parity potential and $\widetilde C$ is the magnetic-parity potential. Under a supertranslation generated by $f(x^A)$, the shear transforms as
\begin{align}
    \delta_f C_{AB}=f\partial_u C_{AB}-2D_A D_B f+\gamma_{AB}D^2 f~.
    \label{ShearSupertranslation}
\end{align}
In a non-radiative region, where $N_{AB}=0$, this reduces to
\begin{align}
    \delta_f C_{AB}=-2\left(D_A D_B-\frac12\gamma_{AB}D^2\right)f~.
\end{align}
Thus a supertranslation shifts only the electric-parity part of the shear,
\begin{align}
    \delta_f C=f~,\qquad\delta_f\widetilde C=0~,
\end{align}
up to the $\ell=0,1$ harmonics, which lie in the kernel of the tracefree operator and correspond to ordinary translations. Therefore, when a burst of radiation interpolates between two non-radiative regions, the electric-parity displacement memory can be interpreted as the observable transition between two supertranslation-related Bondi vacua.

The memory tensor is invariant under a common supertranslation of the initial and final non-radiative cuts. Indeed, the same $f$-dependent vacuum shift appears in both $C_{AB}(u_i)$ and $C_{AB}(u_f)$, and hence cancels in the difference:
\begin{align}
    \Delta C'_{AB}=C'_{AB}(u_f)-C'_{AB}(u_i)=\Delta C_{AB}~.
\end{align}
Thus displacement memory is a well-defined observable, although the individual shear values on the initial and final cuts are themselves supertranslation-frame dependent. Equivalently, gravitational radiation induces a transition between supertranslation-related Bondi vacua, and the observable is the relative transition rather than the absolute vacuum label. The electric-parity displacement memory may be interpreted as a
supertranslation shift between the preferred Poincar\'e frames of the initial and final non-radiative regions. This shift also affects the comparison of angular momentum between the two regimes \cite{Madler:2017umy}.

The displacement memory admits two physically distinct sources. The first is the ordinary, or linear, memory, which is associated with changes in the Coulombic part of the gravitational field, such as the change in the distribution or velocity of massive bodies between the initial and final stationary regimes. The second is the null memory, sourced by energy flux reaching future null infinity. For a clear discussion of this distinction in simple scattering processes, see also \cite{Tolish:2014bka}. This flux may be carried by massless matter through the component $T^{\mathrm{matter}}_{uu}$, or by gravitational radiation itself. The latter contribution is nonlinear: gravitational waves carry energy, and their flux is proportional to the positive quantity $N_{AB}N^{AB}$. Therefore the total displacement memory can be viewed schematically as
\begin{align}
\Delta C_{AB}=\Delta C^{\mathrm{ordinary}}_{AB}+\Delta C^{\mathrm{null}}_{AB}~,
\end{align}
where the null part contains both matter flux and the purely gravitational Christodoulou contribution \cite{Christodoulou:1991cr,Favata:2010zu}. This decomposition makes clear that memory is not only a response to changes in the matter configuration of the source, but also to the energy radiated away by the gravitational field itself.
%%%
%%%

A closely related observable is the spin memory effect \cite{Pasterski:2015tva}, which probes the magnetic-parity sector of the radiative data. While displacement memory measures a permanent relative displacement of inertial detectors and is associated with the electric-parity change of the shear, spin memory is a relative time delay between counter-propagating light rays that travel around a closed contour on the celestial sphere. In terms of the Hodge decomposition \eqref{ShearHodgeDecomposition}, the electric potential $C$ controls the
ordinary displacement memory, whereas the magnetic potential $\widetilde C$ controls the curl, or $B$-mode, part of the shear.

A useful pair of scalar diagnostics for the two parity sectors is
\begin{align}
\mathcal E &:=D^A D^B C_{AB}=- D^2(D^2+2)C~,\\ \mathcal B &:=\epsilon^{AC}D_A D^B C_{BC}=\frac12 D^2(D^2+2)\widetilde C~.
\end{align}
To see why these are faithful tests, expand the potentials in spherical harmonics,
\begin{align}
C=\sum_{\ell m} C_{\ell m}Y_{\ell m}~,\qquad\widetilde C=\sum_{\ell m}\widetilde C_{\ell m}Y_{\ell m}~,
\end{align}
with
\begin{align}
D^2Y_{\ell m}=-\ell(\ell+1)Y_{\ell m}~.
\end{align}
Then
\begin{align}
D^2(D^2+2)Y_{\ell m}=\ell(\ell+1)\bigl(\ell(\ell+1)-2\bigr)Y_{\ell m}~.
\end{align}
This eigenvalue vanishes only for $\ell=0,1$. These modes lie in the kernel
of the tracefree operator
\begin{align}
D_A D_B-\frac12\gamma_{AB}D^2
\end{align}
and therefore do not contribute to $C_{AB}$. Hence, on the physical
$\ell\geq2$ shear sector,
\begin{align}
\mathcal E=0\quad\Longleftrightarrow\quad C=0~,\qquad\mathcal B=0\quad\Longleftrightarrow\quad\widetilde C=0~,
\end{align}
where equality of the potentials is understood modulo the irrelevant
$\ell=0,1$ modes. Thus $\mathcal E$ and $\mathcal B$ isolate,
respectively, the electric and magnetic radiative parts of the Bondi shear.

The spin memory observable is sensitive not simply to the endpoint difference
$\Delta C_{AB}$, but to the retarded-time integral of the magnetic shear
data. Schematically, for a closed curve $\mathcal C\subset S^2$ of length
$L$, the accumulated relative time delay takes the form
\begin{align}
    \Delta u_{\mathcal C}^{\rm spin}=\frac{1}{2\pi L}\oint_{\mathcal C}\epsilon_{AB} D_C\left(\int_{u_i}^{u_f}\d u\,C^{BC}\right)\d x^A~.
\end{align}
Equivalently, spin memory isolates the magnetic-parity part of the
time-integrated shear. It is sourced by angular-momentum flux through $\scri^+$, including the flux carried by gravitational radiation itself, and is naturally associated with the subleading soft graviton theorem and the superrotation sector of asymptotic symmetries \cite{Pasterski:2015tva,Cachazo:2014fwa}. For applications to compact binaries, see also \cite{Nichols:2017rqr}.

%%%%%%%%%%%%%%%%%%%%%
%%%%%%%%%%%%%%%%%%%%%
For physically reasonable scattering spacetimes one usually assumes that the
news tensor decays sufficiently fast toward the far past and far future of
$\scri^+$,
\begin{align}
    N_{AB}(u,x^C)\to 0 \qquad \text{as} \qquad u\to\pm\infty~,
\end{align}
so that the total radiated energy flux is finite,
\begin{align}
    \int_{-\infty}^{+\infty}du
    \int_{S^2}d^2\Omega\,N_{AB}N^{AB}<\infty~.
\end{align}
This type of behavior is realized in the Christodoulou--Klainerman class of asymptotically flat vacuum spacetimes, where the nonlinear stability of Minkowski space ensures a controlled radiative decay at null infinity \cite{Christodoulou:1993uv}. The news is therefore effectively localized in retarded time: it decays in the early and late non-radiative regimes, while its integrated flux gives the total energy carried away by gravitational waves. This integrated flux is precisely the source of the nonlinear, or Christodoulou, contribution to displacement memory.

%%%%%%%%%%%%%%%%%%%%
%%%%%%%%%%%%%%%%%%%%
%%%%%%%%%%%%%%%%%%%%
%%%%%%%%%%%%%%%%%%%%
%%%%%%%%%%%%%%%%%
%%%%%%%%%%%%%%%%%
%%%%%%%%%%%%%%%%%
%%%%%%%%%%%%%%%%%

\section{Carrollian Surfaces and Optical Data}\label{App:CarrollManifolds}
Let $\cal M$ be a $(d+2)$-dimensional Lorentzian manifold endowed with a metric tensor $g_{\mu\nu}$. Let
\begin{align}
X: {\cal N} \hookrightarrow {\cal M}
\end{align}
be an embedded null hypersurface of dimension $d+1$. Intrinsic coordinates on $\N$ are denoted by $y^a$, and the tangent map 
\begin{align}
    e^\mu{}_a = \frac{\partial X^\mu}{\partial y^a}~,
\end{align}
such that an intrinsic vector field $V^a$ on $\N$ is pushed forward to tangent bulk vector by
$V^\mu=e^\mu{}_a V^a$, and similarly a bulk one-form $w_\mu$ restricts to an intrinsic one-form by pullback $w_a = e^\mu{}_a w_\mu$. So there is no two-sided inverse $e^a{}_\mu$ satisfying both
\begin{align}
    e^a{}_\mu e^\mu{}_b = \delta^a{}_b~,\qquad e^\mu{}_a e^a{}_\nu = \delta^\mu{}_\nu~,
\end{align}
as the second equality can not hold because the image of $e^\mu{}_a$ is only $T\N \subset T{\cal M}$.

Locally, the hypersurface may be described as a level set $\N: \{\check\Phi = 0\}$~, such that a normal one-form $\ell_\mu$ is proportional to $-\d\check\Phi$ and annihilates all tangent vectors to the hypersurface,
\begin{align}
\ell_\mu V^\mu = 0~\text{for all}~V^\mu \in T{\cal N}~.
\end{align}
Since $\N$ is null, the metric dual $\ell^\mu=g^{\mu\nu}\ell_\nu \in T\N$ is tangent to $\N$. We denote this null generator by $\ell^\mu$ and write $\ell^\mu  = e^\mu{}_a\ell^a$. The null direction is both normal and tangent implying a degeneracy of the metric tensor and there is no canonical transverse direction. Therefore, in order to define projectors, extrinsic curvature, or connections, we need to supplement each point with a notion of transverse direction. In order to do so one defines a \textit{rigging} vector \cite{Mars:1993mj} normalized as
\begin{align}
    \ell^\mu n_\mu = -1~.
\end{align}
We further impose that 
\begin{align}
    n^\mu n_\mu = 0~,
\end{align}
so that $n^\mu=g^{\mu\nu}n_\nu$ is transverse and null, which will make the definition of projectors clean. 

The pullback of the rigging one-form defines the \textit{Carrollian clock}: it selects the one-form dual to the null Carrollian time direction on the degenerate hypersurface. The intrinsic one-form $n_a$ is given by the pullback
\begin{align}
    n_a := e^\mu{}_a n_\mu~,
\end{align}
and therefore
\begin{align}
    \ell^a n_a = \ell^ae^\mu{}_an_\mu = \ell^\mu n_\mu =-1~.
\end{align}
Thus, after a bulk rigging has been chosen, the intrinsic Carrollian clock is not an independent new datum; it is induced by pullback.

Once a rigging has been chosen, one can define a left-inverse, or \textit{soldering form}, $e^a{}_\mu$, adapted to the splitting of the ambient tangent space along $n^\mu$. It satisfies
\begin{align}
e^a{}_\mu e^\mu{}_b=\delta^a{}_b~,
\qquad
e^a{}_\mu n^\mu=0~.
\end{align}
Its failure to be a two-sided inverse is precisely measured by the ambient projector onto the hypersurface:\footnote{We use $\Pi^\mu{}_\nu$ for the ambient rigged projector acting on bulk vectors, and $\Pi_\mu{}^\nu$ for its dual action on one-forms. Since the rigged projector is oblique in null geometry, these index placements should be distinguished and should not be identified by raising or lowering with the spacetime metric. Intrinsic and ambient indices are related by the soldering maps $e^\mu{}_a$ and $e^a{}_\mu$.}
\begin{align}
e^\mu{}_a e^a{}_\nu=\Pi^\mu{}_\nu:=\delta^\mu{}_\nu+n^\mu\ell_\nu~.
\end{align}
This projector obeys
\begin{align}
\Pi^\mu{}_\rho\Pi^\rho{}_\nu=\Pi^\mu{}_\nu~, \qquad \Pi^\mu{}_\nu n^\nu=0~, \qquad \ell_\mu \Pi^\mu{}_\nu=0~.
\end{align}
For every tangent vector $V^\mu\in T{\cal N}$, equivalently $\ell_\mu V^\mu=0$, one has
\begin{align}
\Pi^\mu{}_\nu V^\nu=V^\mu~.
\end{align}
Thus $\Pi^\mu{}_\nu$ projects ambient vectors onto $T{\cal N}$ along the chosen rigging direction $n^\mu$.

The ambient projector $\Pi^\mu{}_\nu$ should be distinguished from the projectors that act on the spatial directions of the null surface. The reason is simple but important: projecting onto the null hypersurface still leaves the null generator $\ell^\mu$ inside the image. Thus $\Pi^\mu{}_\nu$ removes the transverse rigging direction $n^\mu$, but it does not remove the degenerate direction $\ell^\mu$. In order to isolate the $d$-dimensional spatial screen inside $T{\cal N}$, one must project once more, now intrinsically on the hypersurface. Then, the induced metric on ${\cal N}$ is obtained by pullback,
\begin{align}
q_{ab}=e^\mu{}_a e^\nu{}_b g_{\mu\nu}~.
\end{align}
Since ${\cal N}$ is null, this metric is degenerate. Its kernel is generated by $\ell^a$:
\begin{align}
q_{ab}\ell^b=0~.
\end{align}
The pair $(q_{ab},\ell^a)$ is the weak Carrollian structure induced on the null hypersurface. The additional choice of the clock one-form $n_a$, obtained from the rigging by pullback, defines a splitting of the tangent bundle
\begin{align}\label{TNSplit}
T{\cal N}=\mathrm{span}({\ell})\oplus H~,
\end{align}
where the horizontal or spatial screen distribution is
\begin{align}
H:=\{X^a\in T{\cal N}~|~ n_aX^a=0\}~.
\end{align}

The intrinsic spatial projector is therefore defined by
\begin{align}
\pi^a{}_b:=\delta^a{}_b+\ell^a n_b~.
\end{align}
It obeys
\begin{align}
\pi^a{}_b\pi^b{}_c=\pi^a{}_c~,\qquad
\pi^a{}_b\ell^b=0~,\qquad
n_a\pi^a{}_b=0~.
\end{align}
Thus $\pi^a{}_b$ projects intrinsic vectors on ${\cal N}$ onto the spatial screen $H$, along the null direction $\ell^a$. The projector $\pi^a{}_b$ acts on intrinsic vectors, while its dual
$\pi_a{}^b$ acts on intrinsic one-forms. Equivalently, any intrinsic vector field $V^a\in T{\cal N}$ decomposes uniquely as
\begin{align}
V^a=V_{\parallel}\ell^a+V_\perp^a~,
\qquad
V_{\parallel}:=-n_bV^b~,
\qquad
V_\perp^a:=\pi^a{}_bV^b~,
\end{align}
with $n_a V_\perp^a=0$. This is the intrinsic Carrollian analogue of decomposing a relativistic vector into a time direction plus spatial directions, except that the ``time'' direction is null and lies in the kernel of the degenerate metric.

The tensor $q_{ab}$ is already horizontal, in the sense that
\begin{align}
q_{ab}\ell^b=e^\mu{}_a e^\nu{}_b g_{\mu\nu}\ell^b=e^\mu{}_a g_{\mu\nu}\ell^\nu=0~,
\qquad
\pi_a{}^c\pi_b{}^d q_{cd}=q_{ab}~.
\end{align}
There is no full inverse metric on ${\cal N}$. However, once the clock $n_a$ has been chosen, one can define an inverse spatial metric $q^{ab}$ on the horizontal subspace by requiring
\begin{align}
q^{ac}q_{cb}=\pi^a{}_b~,
\qquad
q^{ab}n_b=0~.
\end{align}
This object raises indices only after projection to the screen. It should not be confused with an inverse of $q_{ab}$ on all of $T{\cal N}$.

The same spatial projection may also be described directly with bulk indices. Composing the projection onto $T{\cal N}$ with the intrinsic projection onto $H$ gives the ambient screen projector
\begin{align}
\pi^\mu{}_\nu
:=e^\mu{}_a\pi^a{}_b e^b{}_\nu
=\delta^\mu{}_\nu+\ell^\mu n_\nu+n^\mu\ell_\nu~.
\end{align}
This projector annihilates both null directions:
\begin{align}
\pi^\mu{}_\nu \ell^\nu=0~,
\qquad
\pi^\mu{}_\nu n^\nu=0~,
\qquad
\ell_\mu\pi^\mu{}_\nu=0~,
\qquad
n_\mu\pi^\mu{}_\nu=0~,
\end{align}
and satisfies
\begin{align}
\pi^\mu{}_\rho\pi^\rho{}_\nu=\pi^\mu{}_\nu~.
\end{align}
Therefore $\pi^\mu{}_\nu$ projects contravariant ambient indices onto the
$d$-dimensional spatial screen orthogonal to both $\ell^\mu$ and $n^\mu$.
For covariant ambient indices one uses the dual projector
\begin{align}
\pi_\mu{}^\nu=\delta_\mu{}^\nu+\ell_\mu n^\nu+n_\mu\ell^\nu~.
\end{align}

%%%%%%%%%%%%%%
%%%%%%%%%%%%%%
%%%%%%%%%%%%%%
This splitting is not canonical; it depends on the choice of rigging. The rigging is not fixed by the null hypersurface itself. Since the normal direction to a null hypersurface is tangent, the ambient metric does not provide a canonical transverse complement to $T\N$. A rigging is precisely the choice of such a complement. Different choices of rigging define different horizontal distributions $H\N = {\rm ker}(n)$ and hence different projectors $\pi^a{}_b$ while leaving the underlying weak Carrollian structure $(\N,q_{ab},\ell^a)$ unchanged.

It is useful to summarize the three projectors as follows. The first one,
\begin{align}
\Pi^\mu{}_\nu=\delta^\mu{}_\nu+n^\mu\ell_\nu~,
\end{align}
is the ambient rigged projector onto the full tangent space $T{\cal N}$. It removes the transverse direction $n^\mu$, but keeps the null generator $\ell^\mu$. The second one,
\begin{align}
\pi^a{}_b=\delta^a{}_b+\ell^a n_b~,
\end{align}
is the intrinsic Carrollian projector from $T{\cal N}$ to the spatial screen $H$. It removes the degenerate direction $\ell^a$. The third one,
\begin{align}
\pi^\mu{}_\nu=\delta^\mu{}_\nu+\ell^\mu n_\nu+n^\mu\ell_\nu~,
\end{align}
is the ambient screen projector. It removes both null directions and projects directly onto the codimension-two spatial screen. The corresponding dual projectors $\Pi_\mu{}^\nu$, $\pi_a{}^b$ and $\pi_\mu{}^\nu$ act on one-forms.

Accordingly, the geometric hierarchy is
\begin{align}
T{\cal M}
\xrightarrow{\ \Pi^\mu{}_\nu\ }
T{\cal N}
\xrightarrow{\ \pi^a{}_b\ }
H~,
\end{align}
or, equivalently, in one ambient step,
\begin{align}
T{\cal M}
\xrightarrow{\ \pi^\mu{}_\nu\ }
H~.
\end{align}
This hierarchy is the basic algebraic structure behind the Carrollian geometry of embedded null hypersurfaces. 

We now use the rigged projector to differentiate ambient tensors along the null hypersurface. Since the Levi-Civita connection $\nabla_\mu$ is an ambient connection on $T{\cal M}$, its action on a tangent tensor is not automatically intrinsic to ${\cal N}$. The rigging provides the missing projection. For an intrinsic vector field $V^a$, with pushforward $V^\mu=e^\mu{}_aV^a$, we define the rigged derivative by
\begin{align}
D_a V^b:= e^\mu{}_a e^b{}_\nu \nabla_\mu V^\nu~.
\end{align}
For covectors and higher-rank tensors, one projects every free ambient index with the soldering form or, equivalently, with the rigged projector before interpreting the result as intrinsic data on ${\cal N}$. More generally, every free bulk index must be projected with $\Pi^\mu{}_\nu$ before being interpreted as an intrinsic index on the null hypersurface. Because the ambient connection is torsionless and the tangent basis is induced from an embedding, the projected rigged connection is torsionless on ${\cal N}$, but it is not uniquely determined by the weak Carrollian data $(q_{ab},\ell^a)$ alone; it depends on the rigging through $e^b{}_\nu$.

The derivative of the null generator along the hypersurface defines the null Weingarten map. In intrinsic indices, it is
\begin{align}
W_a{}^b:=e^\mu{}_a e^b{}_\nu \nabla_\mu \ell^\nu~.
\end{align}
The lower index labels the tangent direction along ${\cal N}$ in which the
derivative is taken, while the upper index labels the resulting tangent
vector.  Thus $W_a{}^b$ is a one-form on ${\cal N}$ valued in tangent
vectors.
Equivalently, using bulk indices,
\begin{align}
W_\mu{}^\nu:=\Pi_\mu{}^\rho \Pi^\nu{}_\sigma\nabla_\rho\ell^\sigma~,
\end{align}
such that the intrinsic tensor is obtained by pullback and soldering,
\begin{align}
W_a{}^b=e^\mu{}_a e^b{}_\nu W_\mu{}^\nu~.
\end{align}

The Weingarten map measures how the null generator changes when transported along the hypersurface. Since $\ell^\mu$ is tangent to ${\cal N}$, and since
\begin{align}
\ell_\mu \nabla_\rho \ell^\mu=\frac12\nabla_\rho(\ell_\mu\ell^\mu)=0~,
\end{align}
the derivative $\nabla_\rho\ell^\mu$, for $\rho$ tangent to ${\cal N}$, is again tangent to ${\cal N}$. Thus the Weingarten map is an endomorphism of $T{\cal N}$,
\begin{align}
W:T{\cal N}\longrightarrow T{\cal N}~.
\end{align}

The intrinsic content of $W_a{}^b$ is obtained by decomposing it with respect to the Carrollian splitting \eqref{TNSplit}. The purely horizontal part is the null extrinsic curvature, or expansion tensor,
\begin{align}
\theta_{ab}
:=
e^\mu{}_a e^\nu{}_b \nabla_\mu \ell_\nu
\end{align}
which for a null hypersurface
\begin{align}
\theta_{ab}=\frac12 {\cal L}_\ell q_{ab}~.
\end{align}
Because ${\cal N}$ is a null hypersurface, its null generator is hypersurface-orthogonal. By Frobenius' theorem the associated null congruence is therefore twist-free, and the null extrinsic curvature is symmetric.

Its trace gives the expansion,
\begin{align}
\theta:=q^{ab}\theta_{ab}~,
\end{align}
and its traceless part gives the shear,
\begin{align}
\sigma_{ab}:=\theta_{ab}-\frac1d\theta q_{ab}~.
\end{align}

The component of $W_a{}^b$ along the null direction is encoded in a one-form $\Omega_a$, defined by
\begin{align}
\Omega_a:=-n_\mu e^\nu{}_a\nabla_\nu \ell^\mu~.
\end{align}
This is the connection one-form associated with the variation of $\ell^\mu$. Then, the Weingarten map decomposes as
\begin{align}
W_a{}^b=\theta_a{}^b+\Omega_a\ell^b~,
\end{align}
where
\begin{align}
\theta_a{}^b:=q^{bc}\theta_{ac}
\end{align}
is horizontal:
\begin{align}
\theta_a{}^b n_b=0~, \qquad \theta_a{}^b\ell^a=0~.
\end{align}
The one-form $\Omega_a$ itself splits into its component along the Carrollian clock and a horizontal momentum one-form:
\begin{align}
\Omega_a=-\kappa n_a+\omega_a~, \qquad \omega_a:=\pi_a{}^b\Omega_b~.
\end{align}
Here
\begin{align}
\kappa:=\Omega_a\ell^a
\end{align}
is the inaffinity, equivalently defined by
\begin{align}
\ell^\nu\nabla_\nu\ell^\mu=\kappa\ell^\mu~,
\end{align}
while $\omega_a$ is the \Hajicek, or momentum, one-form on the screen. Geometrically, $\Omega_a$ is the connection one-form on the $SO(1,1)$ normal-frame bundle associated with local boosts of the null frame
\begin{align}
\ell^\mu\rightarrow e^\lambda \ell^\mu~,  \qquad  n^\mu\rightarrow e^{-\lambda}n^\mu~,
\end{align}
and therefore may be viewed as an Ehresmann connection on this normal bundle. Its horizontal projection $\omega_a$ is the corresponding screen connection, i.e. the \Hajicek one-form.

Under this local boost of the null frame, the screen projector is invariant, while the optical data transform as
\begin{align}
\theta_{ab}
&\to e^\lambda \theta_{ab}~,\\
\Omega_a
&\to\Omega_a+\partial_a\lambda~,\\
\kappa
&\to e^\lambda\left(\kappa+\ell^a\partial_a\lambda\right)~,\\
\omega_a
&\to\omega_a+\pi_a{}^b\partial_b\lambda~,
\end{align}
where $\partial_a\lambda := e^\mu{}_a\partial_\mu\lambda$.
Thus the \Hajicek one-form transforms as the horizontal part of an Abelian connection on the screen under local boosts of the null frame.

More generally, the choice of rigging is affected by a shift ambiguity. Let $v^\mu$ be a horizontal vector satisfying
\begin{align}
v^\mu \ell_\mu=0,
\qquad
v^\mu n_\mu=0 .
\end{align}
Then the shifted rigging
\begin{align}
n'^\mu=n^\mu+v^\mu+\frac12 v^\rho v_\rho\,\ell^\mu
\end{align}
preserves both $n'^2=0$ and $n'\cdot \ell=-1$. Since the pullback of $\ell_\mu$ to ${\cal N}$ vanishes, the intrinsic clock shifts as
\begin{align}
n'_a=n_a+v_a .
\end{align}
The weak Carrollian structure $({\cal N},q_{ab},\ell^a)$ is unchanged, and hence
\begin{align}
\theta'_{ab}=\theta_{ab},
\qquad
\kappa'=\kappa .
\end{align}
However, the horizontal connection data depend on the choice of rigging. In particular,
\begin{align}
\omega'_a=\omega_a+\kappa v_a-v^b\theta_{ab}.
\end{align}
Thus the expansion and inaffinity are invariant under horizontal rigging shifts, while the \Hajicek one-form is rigging-dependent.

Although the decomposition of the Weingarten map depends on the rigging, the Weingarten map itself is invariant under horizontal rigging shifts. Indeed, under the shift of the clock one has
\begin{align}
q'^{ab}=q^{ab} +\ell^a v^b+\ell^b v^a+q_{cd}v^cv^d\ell^a\ell^b~,
\end{align}
and therefore
\begin{align}
\theta'_a{}^b=\theta_a{}^b+\ell^b v^c\theta_{ac}~.
\end{align}
On the other hand, the connection one-form transforms as
\begin{align}
\Omega'_a=\Omega_a-v^c\theta_{ac}~.
\end{align}
Thus the two shifts compensate inside
\begin{align}
W_a{}^b=\theta_a{}^b+\Omega_a\ell^b~,
\label{WabOptical}
\end{align}
and one finds
\begin{align}
W'_a{}^b=W_a{}^b~.
\end{align}
Equivalently, the Weingarten map is rigging-independent as an endomorphism of $T{\cal N}$, while its Carrollian decomposition into horizontal expansion, inaffinity, and \Hajicek data is rigging-dependent.

Thus, in any chosen rigging, the full null Weingarten map takes the Carrollian form
\begin{align}
W_a{}^b=\theta_a{}^b-\left(\kappa n_a-\omega_a\right)\ell^b~.
\end{align}
This formula packages the optical data of the null hypersurface: the expansion and shear are contained in $\theta_{ab}$, the inaffinity is $\kappa$, and the transverse momentum data are contained in $\omega_a$. It is the null analogue of the extrinsic curvature of a non-null hypersurface, but because $q_{ab}$ is degenerate the extrinsic data are naturally organized by the Carrollian projectors rather than by a non-degenerate induced metric.

The Weingarten map provides the geometric input for the null Brown--York tensor  \cite{Chandrasekaran:2018aop, Chandrasekaran:2020wwn, Chandrasekaran:2021hxc, Chandrasekaran:2021vyu} defined as 
\begin{align}
    T_a{}^b:= \frac{1}{8\pi G}\left(W_a{}^b-\delta_a{}^b W\right)~.
\end{align}
In the finite-distance formalism, the latter plays the role of a Carrollian stress tensor living on the screen.  More precisely, the projection of the Einstein equations along the null generator can be written as a conservation equation for this stress tensor.  With the index placement used here, this conservation law takes the schematic form
\begin{align}
D_b T_a{}^b=-\frac{1}{8\pi G} e^\mu{}_a G_{\mu\nu}\ell^\nu~,
\end{align}
and therefore reduces to
\begin{align}\label{DivT}
    D_b T_a{}^b=0
\end{align}
in vacuum.  Thus the dynamics of the null screen can be read as the dynamics of a Carrollian fluid. The null generator $\ell^a$ is the Carrollian time direction, while the optical data of the screen determine the energy, momentum and stress carried by this fluid.

Using the Carrollian decomposition of the Weingarten map \eqref{WabOptical}, one obtains the fluid-like split
\begin{align}
T_a{}^b=\ell^b\tau_a+\tau_a{}^b~,
\end{align}
where
\begin{align}
\tau_a =\frac{1}{8\pi G}\left(\omega_a+\theta n_a\right)~,\qquad\tau_a{}^b=\frac{1}{8\pi G}\left(\sigma_a{}^b-\mu\,\pi_a{}^b\right)~,
\end{align}
and
\begin{align}
\mu=\kappa+\left(1-\frac1d\right)\theta~,
\end{align}
which for the four-dimensional Robinson--Trautman applications in the main text,
$d=2$, and this reduces to $\mu=\kappa+\theta/2$. Then,
\begin{align}
    T_a{}^b=\frac{1}{8\pi G}\left[ \theta_a{}^b + (\omega_a -\kappa n_a)\ell^b - \delta_a{}^b(\theta + \kappa) \right]~.
\end{align}

In this form the analogy with hydrodynamics is transparent.  The one-form $\tau_a$ contains the longitudinal energy-density sector, controlled by the expansion $\theta$, together with the transverse momentum or heat-current datum $\omega_a$.  The tensor $\tau_a{}^b$ is the spatial stress tensor with $\mu$ playing the role of an isotropic pressure, while the tracefree shear $\sigma_a{}^b$ is the viscous stress.  The important difference from an ordinary relativistic fluid is that the velocity is replaced by the Carrollian evolution vector $\ell^a$, and the stress tensor lives on a degenerate null geometry rather than on a Lorentzian worldvolume.
For the Robinson--Trautman screens considered here, the optical data entering this Carrollian stress tensor are those displayed in \autoref{Sec:CarrFluidRT}. The finite-screen conservation equations \eqref{DivT} are then satisfied modulo the nullity condition for the embedding function $\rho$ and the Robinson--Trautman equation, together with its angular derivatives. In this sense, the Robinson--Trautman equation controls not only the bulk radiative relaxation, but also the Carrollian fluid dynamics induced on finite null screens.

Let us finally stress that this null Brown--York tensor should be viewed as a quasilocal Carrollian stress tensor associated with the chosen null screen. Its conservation equations are the null analogues of fluid equations where the projection along the generator gives the Raychaudhuri equation, while the spatial projection gives the Damour equation. This interpretation is closely related to the horizon-fluid viewpoint in fluid/gravity, where the black-hole horizon supports an induced Carrollian fluid whose dynamics is tied to the bulk Einstein equations and, in holographic settings, to the boundary relativistic fluid \cite{Ciambelli:2023mir}. In the main text we use this geometric dictionary for Robinson--Trautman null screens, where the projected conservation equations are controlled by the Robinson--Trautman equation.

For a broader discussion of Carrollian geometry, its applications to flat-space holography, and further references, see for instance \cite{Bagchi:2025vri,Ciambelli:2025unn,Nguyen:2025zhg,Ruzziconi:2025fuy}.

%%%%%%%%%%%%%%%%%
%%%%%%%%%%%%%%%%%
%%%%%%%%%%%%%%%%

\section{Near Screen Geometry}\label{App:NearScreen}

In this appendix we show that a finite Robinson--Trautman null screen can be brought locally to the same Gaussian-null form used in near-horizon symmetry analyses. The purpose is to identify the universal near-null-surface geometry and the corresponding boundary-condition-preserving diffeomorphisms.
We use the Carrollian screens of Robinson--Trautman described in \autoref{Sec:CarrFluidRT}, where we recall that the nullity condition of the screen is
\begin{align}
F_{\cal N}+2\dot\rho+q_{\cal N}^{AB}\partial_A\rho\,\partial_B\rho=0~,
\label{eq:NearScreenNullityRho}
\end{align}
where a subscript ${\cal N}$ means evaluation at $r=\rho(u,z^A)$. This equation is the reason why $F_{\cal N}=0$ alone does not define a null hypersurface for a generic time-dependent and angle-dependent screen.

To expand near the screen, we introduce
\begin{align}
\varrho:=r-\rho(u,z^A)~,
\end{align}
so that ${\cal N}$ is located at $\varrho=0$. Since
\begin{align}
\d r=\d\varrho+\dot\rho\d u+\partial_A\rho\d z^A~,
\end{align}
the metric near $\varrho=0$ becomes
\begin{align}
\d s^2={}-2\d u\d\varrho-\left(F_{\cal N}+2\dot\rho\right)\d u^2-2\partial_A\rho\d u\d z^A+q^{\cal N}_{AB}\d z^A\d z^B +\varrho\left[-F'_{\cal N}\d u^2+q'_{AB}\d z^A\d z^B\right]+{\cal O}(\varrho^2)~.
\end{align}
Using \eqref{eq:NearScreenNullityRho}, the zeroth-order term can be written as
\begin{align}
\d s^2_{(0)}=q^{\cal N}_{AB}\left(\d z^A-v^A\d u\right)\left(\d z^B-v^B\d u\right)~,
\end{align}
where
\begin{align}
v^A:=q_{\cal N}^{AB}\partial_B\rho~.
\end{align}
Thus the null generator of the screen is
\begin{align}
\ell=\partial_u+v^A\partial_A~.
\end{align}
The coordinates $(u,z^A)$ are therefore not generically comoving with the null generators. Then, we now introduce local coordinates $y^{\mathfrak a}$ on the cuts by solving the transport equation
\begin{align}
\frac{\partial z^A}{\partial u}=v^A~,\qquad z^A=z^A(u,y^{\mathfrak a})~.
\end{align}
Then
\begin{align}
\d z^A-v^A\d u=e^A{}_{\mathfrak a}\d y^{\mathfrak a}~,\qquad e^A{}_{\mathfrak a}:=\frac{\partial z^A}{\partial y^{\mathfrak a}}~,
\end{align}
and the induced metric on the screen becomes
\begin{align}
\d s^2_{(0)}=\tilde q_{\mathfrak a\mathfrak b}\d y^{\mathfrak a}\d y^{\mathfrak b}~,\qquad \tilde q_{\mathfrak a\mathfrak b}=q^{\cal N}_{AB}e^A{}_{\mathfrak a}e^B{}_{\mathfrak b}~.
\end{align}
This coordinate system is local on the screen and may fail globally when the null congruence develops caustics.
In this coordinates, the metric takes the near-screen form
\begin{align}
\d s^2={}&-2\d u\d\varrho+\tilde q_{\mathfrak a\mathfrak b}\d y^{\mathfrak a}\d y^{\mathfrak b} \nonumber\\
&+\varrho\left[\left(-F'_{\cal N}+\tilde q'_{\mathfrak a\mathfrak b}\tilde v^{\mathfrak a}\tilde v^{\mathfrak b}\right)\d u^2+2\tilde q'_{\mathfrak a\mathfrak b}\tilde v^{\mathfrak a}\d u\d y^{\mathfrak b}+\tilde q'_{\mathfrak a\mathfrak b}\d y^{\mathfrak a}\d y^{\mathfrak b}\right]+{\cal O}(\varrho^2)~,
\end{align}
where
\begin{align}
\tilde q'_{\mathfrak a\mathfrak b}:=q'_{AB}e^A{}_{\mathfrak a}e^B{}_{\mathfrak b}~,\qquad \tilde v^{\mathfrak a}:=e^{\mathfrak a}{}_A v^A~.
\end{align}
Comparing with the standard Gaussian-null expansion
\begin{align}
\d s^2=-2\varrho\kappa\d u^2-2\d u\d\varrho-2\varrho\theta_{\mathfrak a}\d u\d y^{\mathfrak a}+\left(\Omega_{\mathfrak a\mathfrak b}+\varrho\lambda_{\mathfrak a\mathfrak b}\right)\d y^{\mathfrak a}\d y^{\mathfrak b}+{\cal O}(\varrho^2)~,
\end{align}
we identify
\begin{align}
\kappa={}&\frac12\left(F'_{\cal N}-\tilde q'_{\mathfrak a\mathfrak b}\tilde v^{\mathfrak a}\tilde v^{\mathfrak b}\right)~,\\
\theta_{\mathfrak b}={}&-\tilde q'_{\mathfrak a\mathfrak b}\tilde v^{\mathfrak a}~,\\
\Omega_{\mathfrak a\mathfrak b}={}&\tilde q_{\mathfrak a\mathfrak b}~,\\
\lambda_{\mathfrak a\mathfrak b}={}&\tilde q'_{\mathfrak a\mathfrak b}~.
\end{align}
For Robinson--Trautman screens one has $q_{AB}=r^2h_{AB}(u,z^C)$, and hence
\begin{align}
\tilde q'_{\mathfrak a\mathfrak b}=\frac{2}{\rho}\tilde q_{\mathfrak a\mathfrak b}~.
\end{align}
The first radial correction to the spatial metric is therefore isotropic in the radial null direction, reflecting the shear-free character of the preferred Robinson--Trautman outgoing congruence.

The local form above is precisely the type of near-null-surface geometry used in the horizon analysis of \cite{Donnay:2018ckb}. Therefore the diffeomorphisms preserving the Gaussian-null falloffs act on the finite screen in the same way. In particular, the residual vector fields can be parametrized by a function $T(y^{\mathfrak a})$ and a vector field $Y^{\mathfrak a}(y^{\mathfrak b})$ on the cut. To leading order near $\varrho=0$, they take the schematic form
\begin{align}
\xi^u=T(y^{\mathfrak a})~,\qquad \xi^{\mathfrak a}=Y^{\mathfrak a}(y^{\mathfrak b})+{\cal O}(\varrho)~,\qquad \xi^\varrho=-\varrho\,\partial_u T+{\cal O}(\varrho^2)~.
\end{align}
For the stationary near-horizon boundary conditions of \cite{Donnay:2018ckb}, one usually takes $T$ and $Y^{\mathfrak a}$ to be independent of $u$. The resulting algebra is the semidirect sum of diffeomorphisms of the cut with angle-dependent translations along the null generators,
\begin{align}
\left[(T_1,Y_1),(T_2,Y_2)\right]=(T_{12},Y_{12})~,
\end{align}
with
\begin{align}
Y_{12}^{\mathfrak a}=Y_1^{\mathfrak b}\partial_{\mathfrak b}Y_2^{\mathfrak a}-Y_2^{\mathfrak b}\partial_{\mathfrak b}Y_1^{\mathfrak a}~,
\end{align}
and
\begin{align}
T_{12}=Y_1^{\mathfrak a}\partial_{\mathfrak a}T_2-Y_2^{\mathfrak a}\partial_{\mathfrak a}T_1~.
\end{align}
Thus the near-screen symmetry algebra is
\begin{align}
\mathrm{Diff}(S^2)\ltimes C^\infty(S^2)~,
\end{align}
or, after restricting $Y^{\mathfrak a}$ to conformal Killing vectors of a round cut,
\begin{align}
SL(2,\mathbb C)\ltimes C^\infty(S^2)~,
\end{align}
which is the usual global BMS$_4$ algebra. In this sense the finite Robinson--Trautman null screen carries the same local BMS-type symmetry structure as the black-hole horizons considered in \cite{Donnay:2015abr, Donnay:2016ejv,Donnay:2018ckb,Donnay:2019jiz}. Since the screen admits the same Gaussian-null neighborhood, the same boundary-condition-preserving diffeomorphisms act on its Carrollian data.

The construction above shows that any smooth null screen can be put locally in the same Gaussian-null form used in the near-horizon analysis of black holes. Consequently, the diffeomorphisms preserving the near-screen falloffs contain an angle-dependent translation along the null generators together with diffeomorphisms of the spatial cuts. For round cuts, the restriction to globally well-defined conformal Killing vectors gives the usual BMS-type supertranslation/Lorentz sector, while allowing arbitrary vector fields gives the extended version. This statement is local and kinematical and follows from the universal geometry of null hypersurfaces.

The associated charge interpretation is more delicate. To use the Barnich--Brandt \cite{Barnich:2001jy} or covariant phase-space charges of the horizon analysis, the null screen must be treated as a boundary or corner of the variational problem. For a generic radiative finite screen the surface charge variation contains a non-integrable contribution controlled by the time dependence of the area density and other Carrollian data. This non-integrable term is not a problem; it is the finite-screen analogue of flux through the null surface. Only after imposing additional boundary conditions, such as a non-expanding screen and fixed surface gravity in phase space, does one recover an integrable charge algebra of the isolated-horizon type. Thus the generic finite screen carries a BMS-like symmetry algebra, while its charges are generically accompanied by flux. See \cite{Ruzziconi:2025fuy} for a discussion of the covariant phase space of finite-distance null hypersurfaces.

%%%
%%%
%%%
%%%
%%%
%%%
%%%
%%%
\section{\texorpdfstring{$\Lambda$}{Lambda}-BMS Gauge and the Quadratic Solution}
\label{app:LambdaBMS_RT}

In this appendix we describe the construction used in the main text to put the Robinson--Trautman solution with cosmological constant in $\Lambda$-BMS gauge. The purpose is not to display the full asymptotic coordinate transformation, whose subleading coefficients are lengthy, but rather to make explicit the gauge conditions, the order-by-order procedure, and the quadratic Robinson--Trautman solution used in the analysis. We use the same symbols as in the asymptotically flat Bondi-Sachs expansion for the fields and coefficients appearing in the $\Lambda$-BMS gauge, namely $\beta$, $U^A$, $V$, $g_{AB}$ and the expansion data $q_{AB}$, $C_{AB}$, $D_{AB}$, $E_{AB}$, $m_{\rm B}$ and $N_A$. These should be understood as the corresponding finite-$\Lambda$ expansion coefficients, reducing to the standard Bondi-Sachs quantities when a smooth flat limit exists. By contrast, the holographic combinations denoted by $m_{\rm B}^{(\Lambda)}$ and $N_A^{(\Lambda)}$ are finite-$\Lambda$ variables adapted to the $\Lambda$-BMS/Fefferman--Graham dictionary and need not reduce directly to the asymptotically flat Bondi mass and angular-momentum aspects. This convention avoids introducing a separate notation for every finite-$\Lambda$ coefficient while keeping distinct the quantities whose flat limit is nontrivial.

\subsection{\texorpdfstring{$\Lambda$}{Lambda}-BMS gauge and Robinson--Trautman Waves}

We start from the $\Lambda$-BMS form of the metric \cite{Poole:2018koa,Compere:2019bua,compere2020lambda},
\begin{align}
    \d s^2=e^{2\beta}\frac{V}{r}\d u^2
    -2e^{2\beta}\d u\d r+g_{AB}\left(\d x^A-U^A\d u\right)
    \left(\d x^B-U^B\d u\right)~,
    \label{eq:LambdaBMS_metric}
\end{align}
where the metric functions depend on $(u,r,x^A)$. The radial coordinate is chosen so that the angular metric satisfies the determinant condition
\begin{align}
    \partial_r\det\left(\frac{g_{AB}}{r^2}\right)=0~.
    \label{eq:LambdaBMS_det_condition}
\end{align}
Equivalently,
\begin{align}
    \det g_{AB}=r^4\,\chi(u,x^A)~,
\end{align}
with $\chi(u,x^A)$ independent of $r$. We use the asymptotic expansion
\begin{align}
    g_{AB}=r^2q_{AB}+rC_{AB}+D_{AB}+\frac{1}{r}E_{AB}+\frac{1}{r^2}F_{AB}+\cdots~.
    \label{eq:gAB_LambdaBMS_expansion}
\end{align}
The determinant condition implies
\begin{align}
    g^{AB}\partial_rg_{AB}=\frac{4}{r}~, \qquad q^{AB}C_{AB}=0~,
    \label{eq:det_condition_tracefree_C}
\end{align}
so that $C_{AB}$ is tracefree with respect to the boundary metric $q_{AB}$.

Solving the Einstein equations asymptotically gives, to the orders needed here,
\begin{align}
    \beta={}&\beta_0(u,x^A)
    -\frac{1}{32r^2}C^{AB}C_{AB}+\cdots~,
    \label{eq:beta_LambdaBMS}\\ V={}&\frac{\Lambda}{3}e^{2\beta_0}r^3
    -r^2\left(D_AU_0^A+\hat \ell\right)
    \nonumber\\ &\quad
    -re^{2\beta_0}
    \left[\frac{1}{2}R[q]+\frac{\Lambda}{16}C_{AB}C^{AB}+2D_A\partial^A\beta_0+4\partial_A\beta_0\partial^A\beta_0\right]+{2m}+\cdots~,
    \label{eq:V_LambdaBMS}
    \\ U^A={}&U_0^A(u,x^B)+\frac{2}{r}e^{2\beta_0}\partial^A\beta_0-\frac{e^{2\beta_0}}{r^2}\left(\frac{1}{2}D_BC^{AB}+C^{AB}\partial_B\beta_0\right)+\cdots~, \label{eq:UA_LambdaBMS}
\end{align}
where $D_A$ is the Levi-Civita connection of $q_{AB}$ and
\begin{align}
\hat\ell:=\partial_u\log\sqrt{q}=\frac{1}{2}q^{AB}\partial_uq_{AB}~.
    \label{eq:lhat_definition}
\end{align}
The radial momentum constraint imposes
\begin{align}
    \frac{\Lambda}{3}C_{AB}=e^{-2\beta_0}
    \left[\left(\partial_u-D_CU_0^C-\hat \ell\right)q_{AB}+D_AU_B^0+D_BU_A^0\right]~.
    \label{eq:CAB_constraint_LambdaBMS}
\end{align}
This equation is one of the essential differences with the asymptotically flat Bondi problem. For $\Lambda\neq0$, $C_{AB}$ is not freely specifiable radiative data; rather, it is fixed by the time-dependence of the boundary metric and by the leading shift $U_0^A$. In the flat limit, this constraint degenerates and the usual Bondi shear becomes free data.

The induced boundary conformal metric is
\begin{align}
    \d s^2_{(0)}=\left(\frac{\Lambda}{3}e^{2\beta_0}+U_0^AU_A^0\right)\d u^2-2U_A^0\d u\d x^A+q_{AB}\d x^A\d x^B~.
    \label{eq:boundary_metric_LambdaBMS}
\end{align}

Using the boundary gauge freedom one may choose the representative
\begin{align}
    U_0^A=0,\qquad\beta_0=0~,\qquad q_{AB}=\gamma_{AB}~.   \label{eq:boundary_gauge_choice}
\end{align}
In this frame the flat limit of the boundary metric is null, and the usual Bondi structure is recovered.

The dictionary between the $\Lambda$-BMS expansion and Fefferman--Graham gauge identifies the holographic stress tensor components with the $\Lambda$-deformed Bondi data. In the notation used in the main text,
\begin{align}
    T_{ab}=\frac{1}{8\pi G}
    \begin{pmatrix}
        -2m_{\rm B}^{(\Lambda)}
        &
        -N_B^{(\Lambda)}
        \\
        -N_A^{(\Lambda)}
        &
        \frac{3}{\Lambda}q_{AB}m_{\rm B}^{(\Lambda)}
        +\frac32J_{AB}
    \end{pmatrix}~,
    \label{eq:LambdaBMS_stress_tensor}
\end{align}
where
\begin{align}
    m_{\rm B}^{(\Lambda)}={}&m_{\rm B}+\frac{1}{16}\left(\partial_u+\hat \ell\right)C_{AB}C^{AB}~,
    \label{eq:mB_Lambda_dictionary}
    \\N_A^{(\Lambda)}={}&N_A-\frac{3}{2\Lambda}
    D^B\left(N_{AB}-\frac{1}{2}\hat \ell C_{AB}\right)+\frac{3}{4}
    \partial_A\left(\frac{3}{8}C^{CD}C_{CD}-\frac{1}{\Lambda}R[q]\right)~,
    \label{eq:NA_Lambda_dictionary}
    \\
    J_{AB}={}&-\frac{3}{\Lambda^2}\left[\partial_u\left(N_{AB}-\frac{1}{2}\hat \ell C_{AB}\right)+\left(D_AD_B-\frac{1}{2}q_{AB}D^2\right)\hat \ell
    \right]
    \nonumber\\ &-\frac{1}{\Lambda}
    \left[D_{(A}D^CC_{B)C}-\frac{1}{2}q_{AB}D^CD^DC_{CD}-\frac{1}{2}R[q]C_{AB}
    \right]
    \nonumber\\&
    +\frac{5}{16}C_{AB}C_{CD}C^{CD}-E_{AB}~.
    \label{eq:JAB_Lambda_dictionary}
\end{align}
Here $N_{AB}:=\partial_uC_{AB}$. For $\Lambda\neq0$ this object is not independent radiative data, since $C_{AB}$ is constrained by \eqref{eq:CAB_constraint_LambdaBMS}. In the flat limit, where $C_{AB}$ becomes freely specifiable, $N_{AB}$ reduces to the usual Bondi news tensor. Although this expression is written in terms of the coefficient $m_{\rm B}$, the combination $m_{\rm B}^{(\Lambda)}$ is the energy-density variable naturally selected by the finite-$\Lambda$ holographic dictionary. It should therefore not be identified a priori with the standard Bondi mass aspect before taking the appropriate asymptotically flat limit.

It is tempting to diagnose the radiative content of $\Lambda\neq0$ solutions directly through Newman--Penrose Weyl scalars as in \eqref{BondiWeylScalars} for $\Lambda=0$. We will not pursue this route here. In the asymptotically flat case the standard peeling hierarchy and the null character of $\scri^+$ give a distinguished interpretation to the leading $r^{-1}$ part of $\Psi_4$ as outgoing radiation. For nonzero cosmological constant the situation is more subtle. The conformal boundary is spacelike for de Sitter and timelike for anti-de Sitter, and there is no unique null direction singled out by the boundary geometry. Consequently, the radiative Weyl components depend on the choice of null tetrad, or equivalently on the orientation of the null geodesics by which the boundary is approached. This directional dependence is familiar in analyses of radiation in asymptotically (A)dS spacetimes, for example in the AdS C-metric,\footnote{For further holographic implications of radiation of the AdS C-metric, see \cite{Arenas-Henriquez:2022www,Arenas-Henriquez:2023hur,Cisterna:2023qhh,Arenas-Henriquez:2024ypo}} where the radiation pattern is explicitly tied to the null direction of approach to the timelike conformal boundary \cite{Podolsky:2003gm,Mao:2019ahc}.\footnote{See \cite{Mao:2019ahc,Chu:2019ssw,Poole:2021avh,Bonga:2023eml,Ciambelli:2024kre,Fernandez-Alvarez:2025qqx,Arenas-Henriquez:2025rpt,Poole:2025cmv,Revof:2025mgz,Solanki:2025zyu,Diaz:2026zzp,Campoleoni:2026abr} for related work on gravitational radiation, memory, and asymptotic structure with nonzero cosmological constant.}
For this reason we do not use the Weyl scalars as our main diagnostic in this section. Instead, we keep the discussion in terms of the quantities naturally associated with the $\Lambda$-BMS/holographic expansion: the boundary stress tensor, the local energy density, and the corresponding flux. These quantities are adapted to the non-null conformal boundary and provide a cleaner way to track the relaxation of the Robinson--Trautman solution when $\Lambda\neq0$.

%%%%%%%%%%%%%%
We now specialize this general $\Lambda$-BMS construction to the Robinson--Trautman family with cosmological constant, 
\begin{align}
    \d s^2=-F\d u_R^2-2\d u_R\d r_R+\frac{2r_R^2}{P(u_R,z_R,\bar z_R)^2}\d z_R\d\bar z_R~,
\label{eq:RT_Lambda_metric_complex_P}
\end{align}
where
\begin{align}
F=-\frac{\Lambda}{3}r_R^2-2r_R\partial_{u_R}\log P+2\Delta\log P-\frac{2m}{r_R}~,
\label{eq:RT_Lambda_F_P}
\end{align}
and the coordinates $(u_R,r_R,z_R,\bar z_R)$ denote the Robinson--Trautman frame, while $(u,r,z,\bar z)$ denote the $\Lambda$-BMS frame. By repeating the process of the main body of the paper using the asymptotic diffeomorphism
\begin{align}
    u_R
    &=
    \sum_{n\geq0}U_{(n)}(u,z,\bar z)\,r^{-n}~,
    \nonumber\\
    r_R
    &=
    rR_L(u,z,\bar z)
    +\sum_{n\geq0}R_{(n)}(u,z,\bar z)\,r^{-n}~,
    \nonumber\\
    x_R^A
    &=
    x^A + \sum_{n\geq1}Z_{(n)}^A(u,z,\bar z)\,r^{-n}~.
    \label{eq:RT_to_LambdaBMS_diffeo}
\end{align}
The functions $R_L$, $U_{(n)}$, $R_{(n)}$ and $Z_{(n)}^A$ are fixed by imposing the $\Lambda$-BMS gauge conditions order by order in the large-$r$ expansion.

Unlike the asymptotically flat Bondi problem, one should not impose $g_{ur}=-1+O(r^{-1})$ as an independent leading condition. In the $\Lambda$-BMS gauge the leading behavior is instead controlled by $e^{2\beta_0}$ and by the boundary conformal representative. At leading order one finds
\begin{align}
g_{uu}&=\frac{\Lambda}{3}R_L^2\dot U_{(0)}^2\,r^2 +\cdots~, \nonumber\\ g_{ur}&=-\left(1+\frac{\Lambda}{3}R_LU_{(1)}\right)\dot U_{(0)}R_L+\cdots~,\nonumber\\ g_{uA} &=\frac{\Lambda}{3}R_L^2 \partial_AU_{(0)}\dot U_{(0)}\,r^2+\cdots~,
\label{eq:leading_metric_components_LambdaBMS}
\end{align}
where a dot denotes $\partial_u$. Matching these terms with the $\Lambda$-BMS metric gives
\begin{align}
U_A^0&=-\frac{\Lambda}{3}R_L^2\partial_AU_{(0)}\dot U_{(0)}~, \label{eq:U0A_from_RT_map}
\\ e^{2\beta_0} &=\left(1+\frac{\Lambda}{3}R_LU_{(1)}\right)\dot U_{(0)}R_L~.
\label{eq:beta0_from_RT_map}
\end{align}
The determinant condition then fixes the leading radial rescaling as
\begin{align}
R_L=\frac{P}{P_\circ}\left(1+\frac{2}{3}\Lambda P^2\partial U_{(0)}\bar\partial U_{(0)}\right)^{-1/4}~,
\label{eq:RL_LambdaBMS}
\end{align}
where $P_\circ=1+\frac{1}{2}z\bar z$ is the round-sphere conformal factor and $P$ is defined by the Robinson--Trautman angular metric.

At the next order the gauge conditions give
\begin{align}
    U_{(1)}
    &=
    -\frac{3}{\Lambda R_L}
    \left[
        1+
        \left(
            1+\frac{2}{3}\Lambda P^2
            \partial U_{(0)}\bar\partial U_{(0)}
        \right)^{-1/2}
    \right],
    \label{eq:T1_LambdaBMS}
    \\
    Z_{(1)}
    &=
    \frac{
        P^2\bar\partial U_{(0)}
    }{
        R_L
        \left(
            1+\frac{2}{3}\Lambda P^2
            \partial U_{(0)}\bar\partial U_{(0)}
        \right)^{1/2}
    }~,
    \label{eq:Z1_LambdaBMS}
    \\
    \bar Z_{(1)}
    &=
    \frac{
        P^2\partial U_{(0)}
    }{
        R_L
        \left(
            1+\frac{2}{3}\Lambda P^2
            \partial U_{(0)}\bar\partial U_{(0)}
        \right)^{1/2}
    }~.
    \label{eq:Zbar1_LambdaBMS}
\end{align}
As in section~\ref{Sec:BondiMemoryRT}, all Robinson--Trautman functions in these expressions are understood after pullback to the $\Lambda$-BMS frame. In particular, $P$ should be read as $\widehat P=P(U_{(0)},z,\bar{z})$, and derivatives acting on it are the pulled-back Robinson--Trautman derivatives. We suppress hats in this appendix to keep the formulae readable.

The remaining coefficients are determined recursively by imposing the $\Lambda$-BMS gauge conditions order by order in the large-$r$ expansion. Their explicit expressions rapidly become lengthy and do not add useful insight, so we do not display them here. For the purposes of the main text, it is enough that the procedure fixes the asymptotic map once the leading function $U_{(0)}$ and the boundary conformal representative have been specified. The expressions above should be understood as pullback expressions to the $\Lambda$-BMS frame.
In practice, this recursive construction is best implemented algebraically. However, the expressions become increasingly involved at subleading orders, making a fully general analysis of the physical consequences of $\Lambda\neq0$ rather impractical. For this reason, in the main text we restrict to the minimal setup needed to extract the energy flux of Robinson--Trautman waves with cosmological constant. Since the flux is quadratic in the radiative amplitude, this requires the Robinson--Trautman field to be constructed to second order in perturbation theory.

\subsection{Quadratic Robinson--Trautman solution}
\label{SubSec:QuadraticRTSolution}

For the second-order analysis we use the axisymmetric Robinson--Trautman with cosmological constant metric in spherical coordinates,
\begin{align}
\d s^2=-\left(\bar{K}+r\partial_u\Phi-\frac{2m}{r}-\frac{\Lambda}{3}r^2\right)\d u^2-2\d u\d r+r^2e^{\Phi}\left(\d\theta^2 + \sin^2\theta\d\phi\right)~.
\label{eq:RT_quadratic_metric}
\end{align}
In these conventions,
\begin{align}
\bar{K}=e^{-\Phi}\left(1-\frac12\Delta_\circ\Phi\right)~,\qquad \Delta_\circ=\partial_\theta^2+\cot\theta\,\partial_\theta~.
\label{eq:K_Delta_gamma}
\end{align}
The Robinson--Trautman equation is
\begin{align}
\partial_u\Phi-\frac{1}{6m}e^{-\Phi}\Delta_\circ \bar{K}=0~.
\label{eq:RT_equation_quadratic}
\end{align}
We expand the Robinson--Trautman field perturbatively around the Schwarzschild solution,
\begin{align}
\Phi=\varepsilon_\ell f+\varepsilon_\ell^2 h+{\cal O}(\varepsilon_\ell^3)~.
\label{eq:Phi_quadratic_expansion}
\end{align}
The first-order field $f$ describes the linear Robinson--Trautman relaxation, while $h$ captures the quadratic backreaction sourced by $f$. Expanding the curvature function gives
\begin{align}
e^{-\Phi}=1-\varepsilon_\ell f+\varepsilon_\ell^2\left(\frac12 f^2-h\right)+{\cal O}(\varepsilon_\ell^3)~,\qquad 1-\frac12\Delta_\circ\Phi=1-\frac{\varepsilon_\ell}{2}\Delta_\circ f-\frac{\varepsilon_\ell^2}{2}\Delta_\circ h+{\cal O}(\varepsilon_\ell^3)~.
\end{align}
Therefore
\begin{align}
\bar{K}=1+\varepsilon_\ell \bar{K}^{(1)}+\varepsilon_\ell^2\bar{K}^{(2)}+{\cal O}(\varepsilon_\ell^3)~,
\end{align}
with
\begin{align}
\bar{K}^{(1)}=-f-\frac12\Delta_\circ f=-\frac12(\Delta_\circ+2)f~,\qquad \bar{K}^{(2)}=\frac12 f^2+\frac12 f\Delta_\circ f-h-\frac12\Delta_\circ h~.
\label{eq:K1K2_quadratic}
\end{align}
At first order the Robinson--Trautman equation becomes the usual linear relaxation equation,
\begin{align}
\partial_u f+\frac{1}{12m}\Delta_\circ(\Delta_\circ+2)f=0~.
\label{eq:f_linear_RT}
\end{align}
At second order one obtains an inhomogeneous equation for $h$,
\begin{align}
\partial_u h+\frac{1}{12m}\Delta_\circ(\Delta_\circ+2)h=\frac{1}{12m}\left[\Delta_\circ\left(f^2+f\Delta_\circ f\right)+f\Delta_\circ(\Delta_\circ+2)f\right]~.
\label{eq:h_forced_RT}
\end{align}
Once the linear profile $f$ is fixed, this is a linear forced equation for the second-order field $h$. The source is completely determined by the quadratic combinations of the linear solution.

For a single linearized mode we take
\begin{align}
f(u,\theta)=e^{-\omega_\ell u}P_\ell(\cos\theta)~,\qquad \omega_\ell=\frac{(\ell-1)\ell(\ell+1)(\ell+2)}{12m}~,\qquad \ell\geq2~.
\label{eq:f_single_mode}
\end{align}
The Legendre polynomials diagonalize the round-sphere Laplacian,
\begin{align}
\Delta_\circ P_\ell(x)=-\lambda_\ell P_\ell(x)~,\qquad \lambda_\ell=\ell(\ell+1)~,
\label{eq:Legendre_eigenvalue}
\end{align}
and therefore
\begin{align}
\Delta_\circ(\Delta_\circ+2)P_\ell(x)=\lambda_\ell(\lambda_\ell-2)P_\ell(x)=(\ell-1)\ell(\ell+1)(\ell+2)P_\ell(x)~.
\label{eq:RT_operator_eigenvalue}
\end{align}
Thus each linear mode decays exponentially with rate $\omega_\ell$. At quadratic order, however, the square of a single Legendre mode no longer has definite angular momentum. It decomposes into even modes,
\begin{align}
P_\ell(x)^2=\sum_{k=0}^{\ell}a_{2k}^{(\ell)}P_{2k}(x)~,
\label{eq:Legendre_square_decomposition}
\end{align}
where the coefficients are Legendre linearization coefficients (or Gaunt coefficients) and can be written in terms of Wigner $3j$ symbols as
\begin{align}
a_{2k}^{(\ell)}=\frac{4k+1}{2}\int_{-1}^{1}\d x\,P_\ell(x)^2P_{2k}(x)=(4k+1)\begin{pmatrix}\ell & \ell & 2k\\ 0 & 0 & 0\end{pmatrix}^2~.
\label{eq:Legendre_linearization_coefficients}
\end{align}
The relevant quadratic combinations are
\begin{align}
f^2+f\Delta_\circ f=(1-\lambda_\ell)e^{-2\omega_\ell u}P_\ell(\cos\theta)^2~,\qquad f\Delta_\circ(\Delta_\circ+2)f=\lambda_\ell(\lambda_\ell-2)e^{-2\omega_\ell u}P_\ell(\cos\theta)^2~.
\end{align}
Therefore the source in \eqref{eq:h_forced_RT} has the same time dependence $e^{-2\omega_\ell u}$, but contains all even angular modes appearing in $P_\ell^2$:
\begin{align}
\frac{1}{12m}\left[\Delta_\circ\left(f^2+f\Delta_\circ f\right)+f\Delta_\circ(\Delta_\circ+2)f\right]=e^{-2\omega_\ell u}\sum_{k=0}^{\ell}S_{2k}^{(\ell)}P_{2k}(\cos\theta)~,
\label{eq:quadratic_source_decomposition}
\end{align}
with
\begin{align}
S_{2k}^{(\ell)}=\frac{a_{2k}^{(\ell)}}{12m}\left[(\lambda_\ell-1)\lambda_{2k}+\lambda_\ell(\lambda_\ell-2)\right]~.
\label{eq:S2k_quadratic_source}
\end{align}

We solve for $h$ by expanding it in the same set of even modes sourced by $P_\ell^2$,
\begin{align}
h(u,\theta)=\sum_{k=0}^{\ell}h_{2k}^{(\ell)}(u,\theta)~.
\label{eq:h_mode_decomposition}
\end{align}
For each even mode $2k$, the forced equation reduces to an ordinary differential equation in $u$,
\begin{align}
\partial_u h_{2k}^{(\ell)}+\omega_{2k}h_{2k}^{(\ell)}=S_{2k}^{(\ell)}e^{-2\omega_\ell u}P_{2k}(\cos\theta)~,\qquad \omega_{2k}=\frac{\lambda_{2k}(\lambda_{2k}-2)}{12m}~.
\label{eq:h2k_mode_equation}
\end{align}
For non-resonant modes, $\omega_{2k}\neq2\omega_\ell$, a particular solution with the same time dependence as the source is
\begin{align}
h_{2k}^{(\ell)}(u,\theta)=\frac{S_{2k}^{(\ell)}}{\omega_{2k}-2\omega_\ell}e^{-2\omega_\ell u}P_{2k}(\cos\theta)~.
\label{eq:h2k_nonresonant}
\end{align}
Equivalently,
\begin{align}
h(u,\theta)=e^{-2\omega_\ell u}\sum_{k=0}^{\ell}b_{2k}^{(\ell)}P_{2k}(\cos\theta)~,\qquad b_{2k}^{(\ell)}=\frac{S_{2k}^{(\ell)}}{\omega_{2k}-2\omega_\ell}~.
\label{eq:h_quadratic_ansatz}
\end{align}
In terms of the Laplacian eigenvalues, this gives
\begin{align}
b_{2k}^{(\ell)}=\frac{a_{2k}^{(\ell)}\left[(\lambda_\ell-1)\lambda_{2k}+\lambda_\ell(\lambda_\ell-2)\right]}{\lambda_{2k}(\lambda_{2k}-2)-2\lambda_\ell(\lambda_\ell-2)}~,
\label{eq:b2k_quadratic_coefficients}
\end{align}
provided that the denominator does not vanish.

If instead
\begin{align}
\omega_{2k}=2\omega_\ell~,
\label{eq:resonance_condition}
\end{align}
the decay rate of the forced solution coincides with the homogeneous decay rate of the $2k$ mode.\footnote{The first such resonance occurs for $\ell=5$ and $k=3$, namely for the sourced mode $2k=6$, since $\omega_6=2\omega_5$.} The usual exponential ansatz then fails and must be replaced by the resonant particular solution
\begin{align}
h_{2k}^{(\ell)}(u,\theta)=S_{2k}^{(\ell)}u\,e^{-2\omega_\ell u}P_{2k}(\cos\theta)~.
\label{eq:h2k_resonant}
\end{align}
Combining both cases, one may write the solution mode by mode as follows. For $\omega_{2k}\neq2\omega_\ell$,
\begin{align}
h_{2k}^{(\ell)}(u,\theta)=\frac{S_{2k}^{(\ell)}}{\omega_{2k}-2\omega_\ell}e^{-2\omega_\ell u}P_{2k}(\cos\theta)~,
\label{eq:resonant_h_solution_nonres}
\end{align}
while for $\omega_{2k}=2\omega_\ell$,
\begin{align}
h_{2k}^{(\ell)}(u,\theta)=S_{2k}^{(\ell)}u\,e^{-2\omega_\ell u}P_{2k}(\cos\theta)~.
\label{eq:resonant_h_solution_res}
\end{align}
The homogeneous contribution $c_{2k}e^{-\omega_{2k}u}P_{2k}(\cos\theta)$ has been set to zero, since it corresponds to adding an independent second-order homogeneous Robinson--Trautman perturbation. Its coefficient is fixed by the choice of second-order initial data and is not part of the particular quadratic response sourced by the linear mode.

The full second-order Robinson--Trautman field is therefore described by
\begin{align}
\Phi(u,\theta)=\varepsilon_\ell e^{-\omega_\ell u}P_\ell(\cos\theta)+\varepsilon_\ell^2\sum_{k=0}^{\ell}h_{2k}^{(\ell)}(u,\theta)+{\cal O}(\varepsilon_\ell^3)~.
\label{eq:Phi_second_order_final}
\end{align}
This expression solves the Robinson--Trautman equation through ${\cal O}(\varepsilon_\ell^2)$ and is the form used in the main text to compute the $\Lambda$-BMS data to quadratic order.

\end{appendix}
\bibliographystyle{JHEP}
\bibliography{bib.bib}

@article{Duary:2026hpb,
    author = "Duary, Sarthak",
    title = "{Li{\'e}nard--Wiechert fields in AdS and flat-space antipodal matching from geodesic-centered Coulombic data}",
    eprint = "2607.05395",
    archivePrefix = "arXiv",
    primaryClass = "hep-th",
    month = "7",
    year = "2026"
}

@article{Adami:2023wbe,
    author = "Adami, H. and Parvizi, A. and Sheikh-Jabbari, M. M. and Taghiloo, V. and Yavartanoo, H.",
    title = "{Carrollian structure of the null boundary solution space}",
    eprint = "2311.03515",
    archivePrefix = "arXiv",
    primaryClass = "hep-th",
    doi = "10.1007/JHEP02(2024)073",
    journal = "JHEP",
    volume = "02",
    pages = "073",
    year = "2024"
}

@article{Sheikh-Jabbari:2022mqi,
    author = "Sheikh-Jabbari, M. M.",
    title = "{On symplectic form for null boundary phase space}",
    eprint = "2209.05043",
    archivePrefix = "arXiv",
    primaryClass = "gr-qc",
    doi = "10.1007/s10714-022-02997-2",
    journal = "Gen. Rel. Grav.",
    volume = "54",
    number = "11",
    pages = "140",
    year = "2022"
}

@article{Adami:2021kvx,
    author = "Adami, H. and Sheikh-Jabbari, M. M. and Taghiloo, V. and Yavartanoo, H.",
    title = "{Null surface thermodynamics}",
    eprint = "2110.04224",
    archivePrefix = "arXiv",
    primaryClass = "hep-th",
    doi = "10.1103/PhysRevD.105.066004",
    journal = "Phys. Rev. D",
    volume = "105",
    number = "6",
    pages = "066004",
    year = "2022"
}

@article{Grumiller:2020vvv,
    author = "Grumiller, D. and Sheikh-Jabbari, M. M. and Zwikel, C.",
    title = "{Horizons 2020}",
    eprint = "2005.06936",
    archivePrefix = "arXiv",
    primaryClass = "hep-th",
    reportNumber = "TUW--20--20",
    doi = "10.1142/S0218271820430063",
    journal = "Int. J. Mod. Phys. D",
    volume = "29",
    number = "14",
    pages = "2043006",
    year = "2020"
}

@article{Adami:2020amw,
    author = "Adami, H. and Grumiller, D. and Sadeghian, S. and Sheikh-Jabbari, M. M. and Zwikel, C.",
    title = "{T-Witts from the horizon}",
    eprint = "2002.08346",
    archivePrefix = "arXiv",
    primaryClass = "hep-th",
    reportNumber = "IPM/P-2019/041, TUW-20-02",
    doi = "10.1007/JHEP04(2020)128",
    journal = "JHEP",
    volume = "04",
    pages = "128",
    year = "2020"
}

@article{Grumiller:2019fmp,
    author = "Grumiller, Daniel and P{\'e}rez, Alfredo and Sheikh-Jabbari, M. M. and Troncoso, Ricardo and Zwikel, C{\'e}line",
    title = "{Spacetime structure near generic horizons and soft hair}",
    eprint = "1908.09833",
    archivePrefix = "arXiv",
    primaryClass = "hep-th",
    reportNumber = "Preprint: CECS-PHY-18/01, IPM/P-2019/009, TUW-18-03",
    doi = "10.1103/PhysRevLett.124.041601",
    journal = "Phys. Rev. Lett.",
    volume = "124",
    number = "4",
    pages = "041601",
    year = "2020"
}

@unpublished{AsympCPS,
    author= {Adamo, Martina and Ciambelli, Luca and Diaz, Felipe and Imseis, Michael and Sanhueza, Leonardo and Speziale, Simone},
    title= {{To appear}},
}

@article{Chandrasekaran:2018aop,
    author = "Chandrasekaran, Venkatesa and Flanagan, {\'E}anna {\'E}. and Prabhu, Kartik",
    title = "{Symmetries and charges of general relativity at null boundaries}",
    eprint = "1807.11499",
    archivePrefix = "arXiv",
    primaryClass = "hep-th",
    doi = "10.1007/JHEP11(2018)125",
    journal = "JHEP",
    volume = "11",
    pages = "125",
    year = "2018",
    note = "[Erratum: JHEP 07, 224 (2023)]"
}

@article{Gonzalez:2025ene,
    author = "Gonz{\'a}lez, Hern{\'a}n A. and Salzer, Jakob",
    title = "{Energy Detectors and Asymptotic Symmetries}",
    eprint = "2510.27348",
    archivePrefix = "arXiv",
    primaryClass = "hep-th",
    month = "10",
    year = "2025"
}

@article{Navarro:2026rna,
    author = "Navarro, N{\'u}ria and Raclariu, Ana-Maria",
    title = "{On bulk reconstruction in Lorentzian AdS and its flat space limit}",
    eprint = "2605.16641",
    archivePrefix = "arXiv",
    primaryClass = "hep-th",
    month = "5",
    year = "2026"
}

@article{Moult:2025njc,
    author = "Moult, Ian and Narayanan, Sruthi A. and Pasterski, Sabrina",
    title = "{Memory Correlators and Ward Identities in the 'in-in' Formalism}",
    eprint = "2512.02825",
    archivePrefix = "arXiv",
    primaryClass = "hep-th",
    month = "12",
    year = "2025"
}

@article{himwich2025light,
  title={Light-ray Operators and the $w_{1+\infty}$ Algebra},
  author={Himwich, Elizabeth and Pate, Monica},
  journal={arXiv preprint arXiv:2512.18973},
  year={2025}
}

@article{Poulias:2025eck,
    author = "Poulias, Georgios and Vandoren, Stefan",
    title = "{On Carroll partition functions and flat space holography}",
    eprint = "2503.20615",
    archivePrefix = "arXiv",
    primaryClass = "hep-th",
    doi = "10.1007/JHEP06(2025)232",
    journal = "JHEP",
    volume = "06",
    pages = "232",
    year = "2025"
}

@article{Solanki:2025zyu,
    author = "Solanki, Divyesh N. and Chattopadhyay, Pratik and Bhattacharjee, Srijit",
    title = "{Perturbative soft graviton theorems in de Sitter spacetime}",
    eprint = "2502.03481",
    archivePrefix = "arXiv",
    primaryClass = "hep-th",
    doi = "10.1007/JHEP12(2025)008",
    journal = "JHEP",
    volume = "12",
    pages = "008",
    year = "2025"
}

@article{Chu:2019ssw,
    author = "Chu, Chong-Sun and Koyama, Yoji",
    title = "{Memory effect in anti{\textendash}de Sitter spacetime}",
    eprint = "1906.09361",
    archivePrefix = "arXiv",
    primaryClass = "hep-th",
    reportNumber = "NCTS-TH/1903",
    doi = "10.1103/PhysRevD.100.104034",
    journal = "Phys. Rev. D",
    volume = "100",
    number = "10",
    pages = "104034",
    year = "2019"
}

@article{Ruzziconi:2025fct,
    author = "Ruzziconi, Romain and Zwikel, C{\'e}line",
    title = "{Celestial symmetries of black hole horizons}",
    eprint = "2504.08027",
    archivePrefix = "arXiv",
    primaryClass = "hep-th",
    doi = "10.1103/gx7p-8k34",
    journal = "Phys. Rev. D",
    volume = "113",
    number = "4",
    pages = "L041504",
    year = "2026"
}

@misc{kmec2026quasi,
      title={Quasi-Local Celestial Charges and Multipoles}, 
      author={Adam Kmec and Lionel Mason and Romain Ruzziconi},
      year={2026},
      eprint={2604.13362},
      archivePrefix={arXiv},
      primaryClass={hep-th},
      url={https://arxiv.org/abs/2604.13362}, 
}

@article{Bakas:2015hdc,
    author = "Bakas, Ioannis and Skenderis, Kostas and Withers, Benjamin",
    title = "{Self-similar equilibration of strongly interacting systems from holography}",
    eprint = "1512.09151",
    archivePrefix = "arXiv",
    primaryClass = "hep-th",
    doi = "10.1103/PhysRevD.93.101902",
    journal = "Phys. Rev. D",
    volume = "93",
    number = "10",
    pages = "101902",
    year = "2016"
}

@article{Gath:2015nxa,
    author = "Gath, Jakob and Mukhopadhyay, Ayan and Petkou, Anastasios C. and Petropoulos, P. Marios and Siampos, Konstantinos",
    title = "{Petrov Classification and holographic reconstruction of spacetime}",
    eprint = "1506.04813",
    archivePrefix = "arXiv",
    primaryClass = "hep-th",
    reportNumber = "CPHT-RR048.0914, CCQCN-2015-91, CCTP-2015-13",
    doi = "10.1007/JHEP09(2015)005",
    journal = "JHEP",
    volume = "09",
    pages = "005",
    year = "2015"
}

@article{Castro:2025itb,
    author = "Castro, Alejandra and Mancilla, Robinson and Papadimitriou, Ioannis",
    title = "{Near-extremal dynamics away from the horizon}",
    eprint = "2507.01126",
    archivePrefix = "arXiv",
    primaryClass = "hep-th",
    doi = "10.1007/JHEP11(2025)083",
    journal = "JHEP",
    volume = "11",
    pages = "083",
    year = "2025"
}

@article{Revof:2025mgz,
    author = "Revof, Anthi Voulgari and Tiwari, Shubhanshu",
    title = "{de Sitter corrections to gravitational wave memory}",
    eprint = "2509.24339",
    archivePrefix = "arXiv",
    primaryClass = "gr-qc",
    doi = "10.1088/1361-6382/ae6af2",
    journal = "Class. Quant. Grav.",
    volume = "43",
    number = "10",
    pages = "105013",
    year = "2026"
}

@article{Hamada:2017gdg,
    author = "Hamada, Yuta and Seo, Min-Seok and Shiu, Gary",
    title = "{Memory in de Sitter space and Bondi-Metzner-Sachs-like supertranslations}",
    eprint = "1702.06928",
    archivePrefix = "arXiv",
    primaryClass = "hep-th",
    reportNumber = "MAD-TH-17-01",
    doi = "10.1103/PhysRevD.96.023509",
    journal = "Phys. Rev. D",
    volume = "96",
    number = "2",
    pages = "023509",
    year = "2017"
}

@article{Poole:2021avh,
    author = "Poole, Aaron and Skenderis, Kostas and Taylor, Marika",
    title = "{Charges, conserved quantities, and fluxes in de Sitter spacetime}",
    eprint = "2112.14210",
    archivePrefix = "arXiv",
    primaryClass = "hep-th",
    doi = "10.1103/PhysRevD.106.L061901",
    journal = "Phys. Rev. D",
    volume = "106",
    number = "6",
    pages = "L061901",
    year = "2022"
}

@article{Mao:2019ahc,
    author = "Mao, Pujian",
    title = "{Asymptotics with a cosmological constant: The solution space}",
    eprint = "1901.04010",
    archivePrefix = "arXiv",
    primaryClass = "gr-qc",
    reportNumber = "CJQS-2019-015",
    doi = "10.1103/PhysRevD.99.104024",
    journal = "Phys. Rev. D",
    volume = "99",
    number = "10",
    pages = "104024",
    year = "2019"
}

@article{Fontanella:2025tbs,
    author = "Fontanella, Andrea and Payne, Oliver",
    title = "{A Carroll Limit of AdS/CFT: A Triality with Flat Space Holography?}",
    eprint = "2508.10085",
    archivePrefix = "arXiv",
    primaryClass = "hep-th",
    doi = "10.1016/j.physletb.2026.140669",
    month = "8",
    year = "2025"
}

@article{Ciambelli:2024swv,
    author = "Ciambelli, Luca and Freidel, Laurent and Leigh, Robert G.",
    title = "{Quantum null geometry and gravity}",
    eprint = "2407.11132",
    archivePrefix = "arXiv",
    primaryClass = "hep-th",
    doi = "10.1007/JHEP12(2024)028",
    journal = "JHEP",
    volume = "12",
    pages = "028",
    year = "2024"
}

@article{Ciambelli:2025flo,
    author = "Ciambelli, Luca and He, Temple and Zurek, Kathryn M.",
    title = "{Quantum area fluctuations from gravitational phase space}",
    eprint = "2504.12282",
    archivePrefix = "arXiv",
    primaryClass = "hep-th",
    reportNumber = "CALT-TH 2025-009",
    doi = "10.1007/JHEP08(2025)199",
    journal = "JHEP",
    volume = "08",
    pages = "199",
    year = "2025"
}

@article{Xu:2026spj,
    author = "Xu, Yingnan and Chu, Shuangshuang",
    title = "{Large-$N$ Carrollian Thermodynamics from AdS Black-Hole Phase-Space Contractions}",
    eprint = "2606.26163",
    archivePrefix = "arXiv",
    primaryClass = "hep-th",
    month = "6",
    year = "2026"
}

@article{Donnay:2019jiz,
    author = "Donnay, Laura and Marteau, Charles",
    title = "{Carrollian Physics at the Black Hole Horizon}",
    eprint = "1903.09654",
    archivePrefix = "arXiv",
    primaryClass = "hep-th",
    doi = "10.1088/1361-6382/ab2fd5",
    journal = "Class. Quant. Grav.",
    volume = "36",
    number = "16",
    pages = "165002",
    year = "2019"
}

@article{ciambelli2026mapping,
  title={Mapping the Infrared Phase Space of Gravity to Finite Subregions},
  author={Ciambelli, Luca and He, Temple and Klinger, Marc S and Zurek, Kathryn M},
  journal={arXiv preprint arXiv:2606.12515},
  year={2026}
}

@article{ciambelli2026asymptotically,
  title={From asymptotically flat gravity to finite causal diamonds},
  author={Ciambelli, Luca and He, Temple and Zurek, Kathryn M},
  journal={Physical Review Letters},
  volume={136},
  number={19},
  pages={191501},
  year={2026},
  publisher={APS}
}

@article{Javadinezhad:2023mtp,
    author = "Javadinezhad, Reza and Porrati, Massimo",
    title = "{Three Puzzles with Covariance and Supertranslation Invariance of Angular Momentum Flux and Their Solutions}",
    eprint = "2312.02458",
    archivePrefix = "arXiv",
    primaryClass = "hep-th",
    doi = "10.1103/PhysRevLett.132.151604",
    journal = "Phys. Rev. Lett.",
    volume = "132",
    number = "15",
    pages = "151604",
    year = "2024"
}

@article{Donnay:2016ejv,
    author = "Donnay, Laura and Giribet, Gaston and Gonz{\'a}lez, Hern{\'a}n A. and Pino, Miguel",
    title = "{Extended Symmetries at the Black Hole Horizon}",
    eprint = "1607.05703",
    archivePrefix = "arXiv",
    primaryClass = "hep-th",
    doi = "10.1007/JHEP09(2016)100",
    journal = "JHEP",
    volume = "09",
    pages = "100",
    year = "2016"
}

@article{Barnich:2011mi,
    author = "Barnich, Glenn and Troessaert, Cedric",
    title = "{BMS charge algebra}",
    eprint = "1106.0213",
    archivePrefix = "arXiv",
    primaryClass = "hep-th",
    reportNumber = "ULB-TH-11-10",
    doi = "10.1007/JHEP12(2011)105",
    journal = "JHEP",
    volume = "12",
    pages = "105",
    year = "2011"
}

@article{Braginsky:1985vlg,
    author = "Braginsky, V. B. and Grishchuk, L. P.",
    title = "{Kinematic Resonance and Memory Effect in Free Mass Gravitational Antennas}",
    journal = "Sov. Phys. JETP",
    volume = "62",
    pages = "427--430",
    year = "1985"
}

@article{Braginsky:1987kwo,
    author = "Braginsky, Vladimir B. and Thorne, Kip S.",
    title = "{Gravitational-wave bursts with memory and experimental prospects}",
    doi = "10.1038/327123a0",
    journal = "Nature",
    volume = "327",
    pages = "123--125",
    year = "1987"
}

@article{Barnich:2001jy,
    author = "Barnich, Glenn and Brandt, Friedemann",
    title = "{Covariant theory of asymptotic symmetries, conservation laws and central charges}",
    eprint = "hep-th/0111246",
    archivePrefix = "arXiv",
    reportNumber = "ULB-TH-01-19, MPI-MIS-94-2001",
    doi = "10.1016/S0550-3213(02)00251-1",
    journal = "Nucl. Phys. B",
    volume = "633",
    pages = "3--82",
    year = "2002"
}

@article{Donnay:2015abr,
    author = "Donnay, Laura and Giribet, Gaston and Gonzalez, Hernan A. and Pino, Miguel",
    title = "{Supertranslations and Superrotations at the Black Hole Horizon}",
    eprint = "1511.08687",
    archivePrefix = "arXiv",
    primaryClass = "hep-th",
    doi = "10.1103/PhysRevLett.116.091101",
    journal = "Phys. Rev. Lett.",
    volume = "116",
    number = "9",
    pages = "091101",
    year = "2016"
}

@article{Donnay:2018ckb,
    author = "Donnay, Laura and Giribet, Gaston and Gonz{\'a}lez, Hern{\'a}n A. and Puhm, Andrea",
    title = "{Black hole memory effect}",
    eprint = "1809.07266",
    archivePrefix = "arXiv",
    primaryClass = "hep-th",
    reportNumber = "CPHT-RR021.042018",
    doi = "10.1103/PhysRevD.98.124016",
    journal = "Phys. Rev. D",
    volume = "98",
    number = "12",
    pages = "124016",
    year = "2018"
}

@article{Riva:2023xxm,
    author = "Riva, Massimiliano Maria and Vernizzi, Filippo and Wong, Leong Khim",
    title = "{Angular momentum balance in gravitational two-body scattering: Flux, memory, and supertranslation invariance}",
    eprint = "2302.09065",
    archivePrefix = "arXiv",
    primaryClass = "gr-qc",
    reportNumber = "DESY-23-019",
    doi = "10.1103/PhysRevD.108.104052",
    journal = "Phys. Rev. D",
    volume = "108",
    number = "10",
    pages = "104052",
    year = "2023"
}

@article{Christodoulou:1993uv,
    author = "Christodoulou, D. and Klainerman, S.",
    title = "{The Global nonlinear stability of the Minkowski space}",
    year = "1993"
}

@article{Bondi:1962px,
    author = "Bondi, H. and van der Burg, M. G. J. and Metzner, A. W. K.",
    title = "{Gravitational waves in general relativity. 7. Waves from axisymmetric isolated systems}",
    doi = "10.1098/rspa.1962.0161",
    journal = "Proc. Roy. Soc. Lond. A",
    volume = "269",
    pages = "21--52",
    year = "1962"
}

@article{Zeldovich:1974gvh,
    author = "Zel'dovich, Y. B. and Polnarev, A. G.",
    title = "{Radiation of gravitational waves by a cluster of superdense stars}",
    journal = "Sov. Astron.",
    volume = "18",
    pages = "17",
    year = "1974"
}

@article{Madler:2016xju,
    author = {M\"adler, Thomas and Winicour, Jeffrey},
    title = "{Bondi-Sachs Formalism}",
    eprint = "1609.01731",
    archivePrefix = "arXiv",
    primaryClass = "gr-qc",
    doi = "10.4249/scholarpedia.33528",
    journal = "Scholarpedia",
    volume = "11",
    pages = "33528",
    year = "2016"
}

@article{Newman:1961qr,
    author = "Newman, Ezra and Penrose, Roger",
    title = "{An Approach to gravitational radiation by a method of spin coefficients}",
    doi = "10.1063/1.1724257",
    journal = "J. Math. Phys.",
    volume = "3",
    pages = "566--578",
    year = "1962"
}

@article{Strominger:2014pwa,
    author = "Strominger, Andrew and Zhiboedov, Alexander",
    title = "{Gravitational Memory, BMS Supertranslations and Soft Theorems}",
    eprint = "1411.5745",
    archivePrefix = "arXiv",
    primaryClass = "hep-th",
    doi = "10.1007/JHEP01(2016)086",
    journal = "JHEP",
    volume = "01",
    pages = "086",
    year = "2016"
}

@article{Tolish:2014bka,
    author = "Tolish, Alexander and Wald, Robert M.",
    title = "{Retarded Fields of Null Particles and the Memory Effect}",
    eprint = "1401.5831",
    archivePrefix = "arXiv",
    primaryClass = "gr-qc",
    doi = "10.1103/PhysRevD.89.064008",
    journal = "Phys. Rev. D",
    volume = "89",
    number = "6",
    pages = "064008",
    year = "2014"
}

@article{Favata:2010zu,
    author = "Favata, Marc",
    editor = "Marka, Zsuzsa and Marka, Szabolcs",
    title = "{The gravitational-wave memory effect}",
    eprint = "1003.3486",
    archivePrefix = "arXiv",
    primaryClass = "gr-qc",
    doi = "10.1088/0264-9381/27/8/084036",
    journal = "Class. Quant. Grav.",
    volume = "27",
    pages = "084036",
    year = "2010"
}

@article{Thorne:1992sdb,
    author = "Thorne, Kip S.",
    title = "{Gravitational-wave bursts with memory: The Christodoulou effect}",
    doi = "10.1103/PhysRevD.45.520",
    journal = "Phys. Rev. D",
    volume = "45",
    number = "2",
    pages = "520--524",
    year = "1992"
}

@article{Wald:1999wa,
  author        = {Wald, Robert M. and Zoupas, Andreas},
  title         = {{A General Definition of ``Conserved Quantities'' in General Relativity and Other Theories of Gravity}},
  eprint        = {gr-qc/9911095},
  archivePrefix = {arXiv},
  doi           = {10.1103/PhysRevD.61.084027},
  journal       = {Phys. Rev. D},
  volume        = {61},
  pages         = {084027},
  year          = {2000}
}

@article{Chen:2021kug,
    author = "Chen, Po-Ning and Wang, Mu-Tao and Wang, Ye-Kai and Yau, Shing-Tung",
    title = "{BMS Charges Without Supertranslation Ambiguity}",
    eprint = "2107.05316",
    archivePrefix = "arXiv",
    primaryClass = "gr-qc",
    doi = "10.1007/s00220-022-04390-1",
    journal = "Commun. Math. Phys.",
    volume = "393",
    number = "3",
    pages = "1411--1449",
    year = "2022"
}

@article{Geiller:2024amx,
    author = "Geiller, Marc and Zwikel, C{\'e}line",
    title = "{The partial Bondi gauge: Gauge fixings and asymptotic charges}",
    eprint = "2401.09540",
    archivePrefix = "arXiv",
    primaryClass = "hep-th",
    doi = "10.21468/SciPostPhys.16.3.076",
    journal = "SciPost Phys.",
    volume = "16",
    number = "3",
    pages = "076",
    year = "2024"
}

@article{Hartong:2022lsy,
  author        = {Hartong, Jelle and Obers, Niels A. and Oling, Gerben},
  title         = {Review on Non-Relativistic Gravity},
  journal       = {Front. in Phys.},
  volume        = {11},
  pages         = {1116888},
  year          = {2023},
  eprint        = {2212.11309},
  archivePrefix = {arXiv},
  primaryClass  = {gr-qc},
  doi           = {10.3389/fphy.2023.1116888}
}

@article{LevyLeblond:1965,
  author  = {L{\'e}vy-Leblond, Jean-Marc},
  title   = {Une nouvelle limite non-relativiste du groupe de Poincar{\'e}},
  journal = {Annales de l'I.H.P. Physique th{\'e}orique},
  volume  = {3},
  number  = {1},
  pages   = {1--12},
  year    = {1965}
}

@article{Ashtekar:2026jdz,
    author = "Ashtekar, Abhay and Paraizo, Daniel E. and Shu, Jonathan",
    title = "{Thermodynamics of dynamical black holes beyond perturbation theory}",
    eprint = "2604.00170",
    archivePrefix = "arXiv",
    primaryClass = "gr-qc",
    month = "3",
    year = "2026"
}

@article{Hollands:2024vbe,
    author = "Hollands, Stefan and Wald, Robert M. and Zhang, Victor G.",
    title = "{Entropy of dynamical black holes}",
    eprint = "2402.00818",
    archivePrefix = "arXiv",
    primaryClass = "hep-th",
    doi = "10.1103/PhysRevD.110.024070",
    journal = "Phys. Rev. D",
    volume = "110",
    number = "2",
    pages = "024070",
    year = "2024"
}

@article{Rignon-Bret:2023fjq,
    author = "Rignon-Bret, Antoine",
    title = "{Second law from the Noether current on null hypersurfaces}",
    eprint = "2303.07262",
    archivePrefix = "arXiv",
    primaryClass = "gr-qc",
    doi = "10.1103/PhysRevD.108.044069",
    journal = "Phys. Rev. D",
    volume = "108",
    number = "4",
    pages = "044069",
    year = "2023"
}

@article{Visser:2024pwz,
    author = "Visser, Manus R. and Yan, Zihan",
    title = "{Properties of dynamical black hole entropy}",
    eprint = "2403.07140",
    archivePrefix = "arXiv",
    primaryClass = "hep-th",
    doi = "10.1007/JHEP10(2024)029",
    journal = "JHEP",
    volume = "10",
    pages = "029",
    year = "2024"
}

@article{Shajiee:2026coz,
    author = "Shajiee, V. R. and Sheikh-Jabbari, M. M. and Taghiloo, V.",
    title = "{Dynamical Entropy Is a Noether Charge}",
    eprint = "2607.14289",
    archivePrefix = "arXiv",
    primaryClass = "hep-th",
    month = "7",
    year = "2026"
}

@article{Campiglia:2014yka,
  author        = {Campiglia, Miguel and Laddha, Alok},
  title         = {Asymptotic symmetries and subleading soft graviton theorem},
  journal       = {Phys. Rev. D},
  volume        = {90},
  number        = {12},
  pages         = {124028},
  year          = {2014},
  eprint        = {1408.2228},
  archivePrefix = {arXiv},
  primaryClass  = {hep-th},
  doi           = {10.1103/PhysRevD.90.124028}
}

@article{Penrose:1963iua,
  author  = {Penrose, Roger},
  title   = {Asymptotic Properties of Fields and Space-Times},
  journal = {Phys. Rev. Lett.},
  volume  = {10},
  pages   = {66--68},
  year    = {1963},
  doi     = {10.1103/PhysRevLett.10.66}
}

@article{Madler:2017umy,
    author = {M{\"a}dler, Thomas and Winicour, Jeffrey},
    title = "{Radiation Memory, Boosted Schwarzschild Spacetimes and Supertranslations}",
    eprint = "1701.02556",
    archivePrefix = "arXiv",
    primaryClass = "gr-qc",
    doi = "10.1088/1361-6382/aa6ca8",
    journal = "Class. Quant. Grav.",
    volume = "34",
    number = "11",
    pages = "115009",
    year = "2017"
}

@article{Cachazo:2014fwa,
  author        = {Cachazo, Freddy and Strominger, Andrew},
  title         = {Evidence for a New Soft Graviton Theorem},
  eprint        = {1404.4091},
  archivePrefix = {arXiv},
  primaryClass  = {hep-th},
  year          = {2014}
}

@article{Christodoulou:1991cr,
    author = "Christodoulou, D.",
    title = "{Nonlinear nature of gravitation and gravitational wave experiments}",
    doi = "10.1103/PhysRevLett.67.1486",
    journal = "Phys. Rev. Lett.",
    volume = "67",
    pages = "1486--1489",
    year = "1991"
}

@article{Ruzziconi:2025fuy,
    author = "Ruzziconi, Romain and Zwikel, C{\'e}line",
    title = "{Celestial Lw1+{\ensuremath{\infty}} symmetries and subleading phase space of null hypersurfaces}",
    eprint = "2511.07525",
    archivePrefix = "arXiv",
    primaryClass = "hep-th",
    doi = "10.1103/hrbd-cmr7",
    journal = "Phys. Rev. D",
    volume = "113",
    number = "4",
    pages = "044067",
    year = "2026"
}

@article{Adami:2021nnf,
    author = "Adami, H. and Grumiller, D. and Sheikh-Jabbari, M. M. and Taghiloo, V. and Yavartanoo, H. and Zwikel, C.",
    title = "{Null boundary phase space: slicings, news {\&} memory}",
    eprint = "2110.04218",
    archivePrefix = "arXiv",
    primaryClass = "hep-th",
    doi = "10.1007/JHEP11(2021)155",
    journal = "JHEP",
    volume = "11",
    pages = "155",
    year = "2021"
}

@article{Sachs:1961zz,
    author = "Sachs, R. K.",
    title = "{Gravitational waves in general relativity. 6. The outgoing radiation condition}",
    doi = "10.1098/rspa.1961.0202",
    journal = "Proc. Roy. Soc. Lond. A",
    volume = "264",
    pages = "309--338",
    year = "1961"
}

@article{Redondo-Yuste:2022czg,
    author = "Redondo-Yuste, Jaime and Lehner, Luis",
    title = "{Non-linear black hole dynamics and Carrollian fluids}",
    eprint = "2212.06175",
    archivePrefix = "arXiv",
    primaryClass = "gr-qc",
    doi = "10.1007/JHEP02(2023)240",
    journal = "JHEP",
    volume = "02",
    pages = "240",
    year = "2023"
}

@article{Grant:2021hga,
    author = "Grant, Alexander M. and Nichols, David A.",
    title = "{Persistent gravitational wave observables: Curve deviation in asymptotically flat spacetimes}",
    eprint = "2109.03832",
    archivePrefix = "arXiv",
    primaryClass = "gr-qc",
    doi = "10.1103/PhysRevD.105.024056",
    journal = "Phys. Rev. D",
    volume = "105",
    number = "2",
    pages = "024056",
    year = "2022",
    note = "[Erratum: Phys.Rev.D 107, 109902 (2023)]"
}

@article{Bondi:1960jsa,
    author = "Bondi, H.",
    title = "{Gravitational Waves in General Relativity}",
    doi = "10.1038/186535a0",
    journal = "Nature",
    volume = "186",
    number = "4724",
    pages = "535--535",
    year = "1960"
}

@article{Sachs:1962wk,
    author = "Sachs, R. K.",
    title = "{Gravitational waves in general relativity. 8. Waves in asymptotically flat space-times}",
    doi = "10.1098/rspa.1962.0206",
    journal = "Proc. Roy. Soc. Lond. A",
    volume = "270",
    pages = "103--126",
    year = "1962"
}

@article{Strominger:2016wns,
    author = "Strominger, Andrew and Zhiboedov, Alexander",
    title = "{Superrotations and Black Hole Pair Creation}",
    eprint = "1610.00639",
    archivePrefix = "arXiv",
    primaryClass = "hep-th",
    doi = "10.1088/1361-6382/aa5b5f",
    journal = "Class. Quant. Grav.",
    volume = "34",
    number = "6",
    pages = "064002",
    year = "2017"
}

@article{BernardideFreitas:2014eoi,
    author = "Bernardi de Freitas, Gabriel and Reall, Harvey S.",
    title = "{Algebraically special solutions in AdS/CFT}",
    eprint = "1403.3537",
    archivePrefix = "arXiv",
    primaryClass = "hep-th",
    doi = "10.1007/JHEP06(2014)148",
    journal = "JHEP",
    volume = "06",
    pages = "148",
    year = "2014"
}

@article{Podolsky:2003gm,
    author = "Podolsky, Jiri and Ortaggio, Marcello and Krtous, Pavel",
    title = "{Radiation from accelerated black holes in an anti-de Sitter universe}",
    eprint = "gr-qc/0307108",
    archivePrefix = "arXiv",
    doi = "10.1103/PhysRevD.68.124004",
    journal = "Phys. Rev. D",
    volume = "68",
    pages = "124004",
    year = "2003"
}

@article{strominger2014bms,
  title={On BMS invariance of gravitational scattering},
  author={Strominger, Andrew},
  journal={Journal of High Energy Physics},
  volume={2014},
  number={7},
  pages={1--20},
  year={2014},
  publisher={Springer}
}

@article{kapec20172d,
  title={2D stress tensor for 4D gravity},
  author={Kapec, Daniel and Mitra, Prahar and Raclariu, Ana-Maria and Strominger, Andrew},
  journal={Physical Review Letters},
  volume={119},
  number={12},
  pages={121601},
  year={2017},
  publisher={APS}
}

@article{Ciambelli:2017wou,
    author = "Ciambelli, Luca and Petkou, Anastasios C. and Petropoulos, P. Marios and Siampos, Konstantinos",
    title = "{The Robinson-Trautman spacetime and its holographic fluid}",
    eprint = "1707.02995",
    archivePrefix = "arXiv",
    primaryClass = "hep-th",
    reportNumber = "CPHT-PC037.062017",
    doi = "10.22323/1.292.0076",
    journal = "PoS",
    volume = "CORFU2016",
    pages = "076",
    year = "2017"
}

@article{Compere:2019gft,
    author = "Comp{\`e}re, Geoffrey and Oliveri, Roberto and Seraj, Ali",
    title = "{The Poincar{\'e} and BMS flux-balance laws with application to binary systems}",
    eprint = "1912.03164",
    archivePrefix = "arXiv",
    primaryClass = "gr-qc",
    doi = "10.1007/JHEP10(2020)116",
    journal = "JHEP",
    volume = "10",
    pages = "116",
    year = "2020",
    note = "[Erratum: JHEP 06, 045 (2024)]"
}

@article{Adami:2024mtu,
    author = "Adami, H. and Parvizi, A. and Sheikh-Jabbari, M. M. and Taghiloo, V.",
    title = "{Heisenberg soft hair on Robinson-Trautman spacetimes}",
    eprint = "2402.17658",
    archivePrefix = "arXiv",
    primaryClass = "hep-th",
    doi = "10.1007/JHEP05(2024)191",
    journal = "JHEP",
    volume = "05",
    pages = "191",
    year = "2024"
}

@article{Poole:2018koa,
    author = "Poole, Aaron and Skenderis, Kostas and Taylor, Marika",
    title = "{(A)dS$\mathbf{_4}$ in Bondi gauge}",
    eprint = "1812.05369",
    archivePrefix = "arXiv",
    primaryClass = "hep-th",
    doi = "10.1088/1361-6382/ab117c",
    journal = "Class. Quant. Grav.",
    volume = "36",
    number = "9",
    pages = "095005",
    year = "2019"
}

@article{Geiller:2024bgf,
    author = "Geiller, Marc",
    title = "{Celestial $w_{1+\infty}$ charges and the subleading structure of asymptotically-flat spacetimes}",
    eprint = "2403.05195",
    archivePrefix = "arXiv",
    primaryClass = "hep-th",
    doi = "10.21468/SciPostPhys.18.1.023",
    journal = "SciPost Phys.",
    volume = "18",
    number = "1",
    pages = "023",
    year = "2025"
}

@article{Podolsky:2009an,
    author = "Podolsky, Jiri and Svitek, Otakar",
    title = "{Past horizons in Robinson-Trautman spacetimes with a cosmological constant}",
    eprint = "0911.5317",
    archivePrefix = "arXiv",
    primaryClass = "gr-qc",
    doi = "10.1103/PhysRevD.80.124042",
    journal = "Phys. Rev. D",
    volume = "80",
    pages = "124042",
    year = "2009"
}

@article{Ciambelli:2025mex,
    author = "Ciambelli, Luca",
    title = "{Asymptotic limit of null hypersurfaces}",
    eprint = "2501.17357",
    archivePrefix = "arXiv",
    primaryClass = "hep-th",
    doi = "10.1088/1361-6382/ae22b5",
    journal = "Class. Quant. Grav.",
    volume = "42",
    number = "23",
    pages = "235020",
    year = "2025"
}

@article{Ciambelli:2025unn,
    author = "Ciambelli, Luca and Jai-akson, Puttarak",
    title = "{Foundations of Carrollian Geometry}",
    eprint = "2510.21651",
    archivePrefix = "arXiv",
    primaryClass = "hep-th",
    reportNumber = "RIKEN-iTHEMS-Report-26, RIKEN-iTHEMS-Report-25",
    month = "10",
    year = "2025"
}

@article{FosterNewman,
    author = {Foster, J. and Newman, E. T.},
    title = {Note on the Robinson‐Trautman Solutions},
    journal = {Journal of Mathematical Physics},
    volume = {8},
    number = {2},
    pages = {189-194},
    year = {1967},
    month = {02},
    issn = {0022-2488},
    doi = {10.1063/1.1705185},
    url = {https://doi.org/10.1063/1.1705185},
}

@article{Chrusciel:1992tj,
    author = "Chrusciel, Piotr T. and Singleton, David B.",
    title = "{Nonsmoothness of event horizons of Robinson-Trautman black holes}",
    doi = "10.1007/BF02099531",
    journal = "Commun. Math. Phys.",
    volume = "147",
    pages = "137--162",
    year = "1992"
}

@article{Chrusciel:1992rv,
    author = "Chrusciel, P. T.",
    title = "{On the global structure of Robinson-Trautman space-times}",
    doi = "10.1098/rspa.1992.0019",
    journal = "Proc. Roy. Soc. Lond. A",
    volume = "436",
    pages = "299--316",
    year = "1992"
}

@article{Chrusciel:1991vxx,
    author = "Chrusciel, Piotr",
    title = "{Semiglobal existence and convergence of solutions of the Robinson-Trautman (two-dimensional Calabi) equation}",
    doi = "10.1007/BF02431882",
    journal = "Commun. Math. Phys.",
    volume = "137",
    pages = "289--313",
    year = "1991"
}

@article{Ciambelli:2023mir,
    author = "Ciambelli, Luca and Freidel, Laurent and Leigh, Robert G.",
    title = "{Null Raychaudhuri: canonical structure and the dressing time}",
    eprint = "2309.03932",
    archivePrefix = "arXiv",
    primaryClass = "hep-th",
    doi = "10.1007/JHEP01(2024)166",
    journal = "JHEP",
    volume = "01",
    pages = "166",
    year = "2024"
}

@article{lukacs_perjes_sebestyen_porter_1983, title={Lyapunov functional approach to radiative metrics}, abstractNote={Lyapunov's second method is applied to the spherical radiative Robinson-Trautman vacuum space-times to prove that they settle down asymptotically to the Schwarzschild space-time. This class of Robinson-Trautman metrics is characterized by the surface S being topologically a two-sphere where S is invariantly defined by the intersection of the hypersurfaces u=const and r=const. It is shown that ∫sub(S)Ksup(2)dsigma is a Lyapunov functional where K is the Gaussian curvature and dsigma is the invariant measure on S. The critical point occurs at K=0 or, equivalently, at delta2K=0, which condition is shown to characterize the Schwarzschild space-time. (author)}, author={Lukacs, B. and Perjes, Z. and Sebestyen, A. and Porter, J.}, year={1983}, month={Jun}, journal={General Relativity and Gravitation}, volume="16", pages="691--701" }

@article{Ciambelli:2026vxa,
    author = "Ciambelli, Luca and Klinger, Marc S.",
    title = "{Quantization of Gravity on Null Hypersurfaces}",
    eprint = "2607.07785",
    archivePrefix = "arXiv",
    primaryClass = "hep-th",
    month = "7",
    year = "2026"
}

@misc{chow1995apparenthorizonsvacuumrobinsontrautman,
      title={Apparent Horizons in Vacuum Robinson-Trautman Spacetimes}, 
      author={E. W. M. Chow and A. W. -C. Lun},
      year={1995},
      eprint={gr-qc/9503065},
      archivePrefix={arXiv},
      primaryClass={gr-qc},
      url={https://arxiv.org/abs/gr-qc/9503065}, 
}

@book{Stephani:2003tm,
    author = "Stephani, Hans and Kramer, D. and MacCallum, Malcolm A. H. and Hoenselaers, Cornelius and Herlt, Eduard",
    title = "{Exact solutions of Einstein's field equations}",
    doi = "10.1017/CBO9780511535185",
    isbn = "978-0-521-46702-5, 978-0-511-05917-9",
    publisher = "Cambridge Univ. Press",
    address = "Cambridge",
    series = "Cambridge Monographs on Mathematical Physics",
    year = "2003"
}

@article{Speziale:2025lkm,
    author = "Speziale, Simone",
    title = "{GGI lectures on boundary and asymptotic symmetries}",
    eprint = "2512.16810",
    archivePrefix = "arXiv",
    primaryClass = "hep-th",
    month = "12",
    year = "2025"
}

@article{Harlow:2019yfa,
    author = "Harlow, Daniel and Wu, Jie-Qiang",
    title = "{Covariant phase space with boundaries}",
    eprint = "1906.08616",
    archivePrefix = "arXiv",
    primaryClass = "hep-th",
    doi = "10.1007/JHEP10(2020)146",
    journal = "JHEP",
    volume = "10",
    pages = "146",
    year = "2020"
}

@article{Chandrasekaran:2020wwn,
    author = "Chandrasekaran, Venkatesa and Speranza, Antony J.",
    title = "{Anomalies in gravitational charge algebras of null boundaries and black hole entropy}",
    eprint = "2009.10739",
    archivePrefix = "arXiv",
    primaryClass = "hep-th",
    doi = "10.1007/JHEP01(2021)137",
    journal = "JHEP",
    volume = "01",
    pages = "137",
    year = "2021"
}

@book{Chrusciel:2020fql,
    author = "Chrusciel, Piotr",
    title = "{Geometry of Black Holes}",
    isbn = "978-0-19-887320-4, 978-0-19-885541-5",
    publisher = "Oxford University Press",
    month = "4",
    year = "2023"
}

@article{Arnowitt:1962hi,
    author = "Arnowitt, Richard L. and Deser, Stanley and Misner, Charles W.",
    title = "{The Dynamics of general relativity}",
    eprint = "gr-qc/0405109",
    archivePrefix = "arXiv",
    doi = "10.1007/s10714-008-0661-1",
    journal = "Gen. Rel. Grav.",
    volume = "40",
    pages = "1997--2027",
    year = "2008"
}

@article{Chandrasekaran:2021hxc,
    author = "Chandrasekaran, Venkatesa and Flanagan, Eanna E. and Shehzad, Ibrahim and Speranza, Antony J.",
    title = "{Brown-York charges at null boundaries}",
    eprint = "2109.11567",
    archivePrefix = "arXiv",
    primaryClass = "hep-th",
    doi = "10.1007/JHEP01(2022)029",
    journal = "JHEP",
    volume = "01",
    pages = "029",
    year = "2022"
}

@article{Chandrasekaran:2021vyu,
    author = "Chandrasekaran, Venkatesa and Flanagan, Eanna E. and Shehzad, Ibrahim and Speranza, Antony J.",
    title = "{A general framework for gravitational charges and holographic renormalization}",
    eprint = "2111.11974",
    archivePrefix = "arXiv",
    primaryClass = "gr-qc",
    doi = "10.1142/S0217751X22501056",
    journal = "Int. J. Mod. Phys. A",
    volume = "37",
    number = "17",
    pages = "2250105",
    year = "2022"
}

@article{Freidel:2021cjp,
    author = "Freidel, Laurent and Oliveri, Roberto and Pranzetti, Daniele and Speziale, Simone",
    title = "{Extended corner symmetry, charge bracket and Einstein{\textquoteright}s equations}",
    eprint = "2104.12881",
    archivePrefix = "arXiv",
    primaryClass = "hep-th",
    doi = "10.1007/JHEP09(2021)083",
    journal = "JHEP",
    volume = "09",
    pages = "083",
    year = "2021"
}

@article{Speziale:2025zjp,
    author = "Speziale, Simone and Steer, Dani{\`e}le A.",
    title = "{An Introduction to Gravitational Wave Theory}",
    eprint = "2508.21817",
    archivePrefix = "arXiv",
    primaryClass = "gr-qc",
    month = "8",
    year = "2025"
}

@article{Odak:2021axr,
    author = "Odak, Gloria and Speziale, Simone",
    title = "{Brown-York charges with mixed boundary conditions}",
    eprint = "2109.02883",
    archivePrefix = "arXiv",
    primaryClass = "hep-th",
    doi = "10.1007/JHEP11(2021)224",
    journal = "JHEP",
    volume = "11",
    pages = "224",
    year = "2021"
}

@article{Odak:2022ndm,
    author = "Odak, Gloria and Rignon-Bret, Antoine and Speziale, Simone",
    title = "{Wald-Zoupas prescription with soft anomalies}",
    eprint = "2212.07947",
    archivePrefix = "arXiv",
    primaryClass = "hep-th",
    doi = "10.1103/PhysRevD.107.084028",
    journal = "Phys. Rev. D",
    volume = "107",
    number = "8",
    pages = "084028",
    year = "2023"
}

@article{Odak:2023pga,
    author = "Odak, Gloria and Rignon-Bret, Antoine and Speziale, Simone",
    title = "{General gravitational charges on null hypersurfaces}",
    eprint = "2309.03854",
    archivePrefix = "arXiv",
    primaryClass = "gr-qc",
    doi = "10.1007/JHEP12(2023)038",
    journal = "JHEP",
    volume = "12",
    pages = "038",
    year = "2023"
}

@article{Singleton2020,
author = "David Barry Singleton",
title = "{Robinson-Trautman solutions of Einstein's equations}",
year = "1990",
month = "12",
url = "https://bridges.monash.edu/articles/thesis/Robinson-Trautman_solutions_of_Einstein_s_equations/14876076",
doi = "10.26180/14876076.v1"
}

@article{Tod1989,
doi = {10.1088/0264-9381/6/8/015},
url = {https://doi.org/10.1088/0264-9381/6/8/015},
year = {1989},
month = {aug},
publisher = {},
volume = {6},
number = {8},
pages = {1159},
author = {K P Tod},
title = {Analogue of the past horizon in the Robinson-Trautman metrics},
journal = {Classical and Quantum Gravity}
}

@article{Barnich:2026wpw,
    author = "Barnich, Glenn and Seraj, Ali",
    title = "{Memory of Robinson-Trautman waves}",
    eprint = "2604.16703",
    archivePrefix = "arXiv",
    primaryClass = "gr-qc",
    month = "4",
    year = "2026"
}

@article{Barnich:2012aw,
    author = "Barnich, Glenn and Gomberoff, Andres and Gonzalez, Hernan A.",
    title = "{The Flat limit of three dimensional asymptotically anti-de Sitter spacetimes}",
    eprint = "1204.3288",
    archivePrefix = "arXiv",
    primaryClass = "gr-qc",
    doi = "10.1103/PhysRevD.86.024020",
    journal = "Phys. Rev. D",
    volume = "86",
    pages = "024020",
    year = "2012"
}

@article{Nichols:2017rqr,
    author = "Nichols, David A.",
    title = "{Spin memory effect for compact binaries in the post-Newtonian approximation}",
    eprint = "1702.03300",
    archivePrefix = "arXiv",
    primaryClass = "gr-qc",
    doi = "10.1103/PhysRevD.95.084048",
    journal = "Phys. Rev. D",
    volume = "95",
    number = "8",
    pages = "084048",
    year = "2017"
}

@article{Pasterski:2015tva,
    author = "Pasterski, Sabrina and Strominger, Andrew and Zhiboedov, Alexander",
    title = "{New Gravitational Memories}",
    eprint = "1502.06120",
    archivePrefix = "arXiv",
    primaryClass = "hep-th",
    doi = "10.1007/JHEP12(2016)053",
    journal = "JHEP",
    volume = "12",
    pages = "053",
    year = "2016"
}

@article{Bonga:2023eml,
    author = "Bonga, B{\'e}atrice and Bunster, Claudio and P{\'e}rez, Alfredo",
    title = "{Gravitational radiation with {\ensuremath{\Lambda}}{\ensuremath{>}}0}",
    eprint = "2306.08029",
    archivePrefix = "arXiv",
    primaryClass = "gr-qc",
    doi = "10.1103/PhysRevD.108.064039",
    journal = "Phys. Rev. D",
    volume = "108",
    number = "6",
    pages = "064039",
    year = "2023"
}

@article{Geiller:2022vto,
    author = "Geiller, Marc and Zwikel, C{\'e}line",
    title = "{The partial Bondi gauge: Further enlarging the asymptotic structure of gravity}",
    eprint = "2205.11401",
    archivePrefix = "arXiv",
    primaryClass = "hep-th",
    doi = "10.21468/SciPostPhys.13.5.108",
    journal = "SciPost Phys.",
    volume = "13",
    pages = "108",
    year = "2022"
}

@article{Compere:2018ylh,
    author = "Comp\`ere, Geoffrey and Fiorucci, Adrien and Ruzziconi, Romain",
    title = "{Superboost transitions, refraction memory and super-Lorentz charge algebra}",
    eprint = "1810.00377",
    archivePrefix = "arXiv",
    primaryClass = "hep-th",
    doi = "10.1007/JHEP11(2018)200",
    journal = "JHEP",
    volume = "11",
    pages = "200",
    year = "2018",
    note = "[Erratum: JHEP 04, 172 (2020)]"
}

@article{Compere:2018aar,
    author = "Comp\`ere, Geoffrey and Fiorucci, Adrien",
    title = "{Advanced Lectures on General Relativity}",
    eprint = "1801.07064",
    archivePrefix = "arXiv",
    primaryClass = "hep-th",
    month = "1",
    year = "2018"
}

@article{Husnugil:2025edm,
    author = {H{\"u}sn{\"u}gil, Sercan and Lehner, Luis},
    title = "{Sourced Carrollian fluids dual to black hole horizons}",
    eprint = "2508.20284",
    archivePrefix = "arXiv",
    primaryClass = "gr-qc",
    doi = "10.1103/lc67-5ytb",
    journal = "Phys. Rev. D",
    volume = "112",
    number = "10",
    pages = "104043",
    year = "2025"
}

@article{Fernandez-Alvarez:2025qqx,
    author = "Fern{\'a}ndez-{\'A}lvarez, Francisco and Senovilla, Jos{\'e} M. M.",
    title = "{Gravitational radiation at infinity with a negative cosmological constant and AdS4 holography}",
    eprint = "2507.05826",
    archivePrefix = "arXiv",
    primaryClass = "gr-qc",
    doi = "10.1103/9vzd-pvmt",
    journal = "Phys. Rev. D",
    volume = "113",
    number = "6",
    pages = "064029",
    year = "2026"
}

@article{Arenas-Henriquez:2025rpt,
    author = "Arenas-Henriquez, Gabriel and Ciambelli, Luca and Diaz, Felipe and Jia, Weizhen and Rivera-Betancour, David",
    title = "{Radiation in fluid/gravity and the flat limit}",
    eprint = "2508.01446",
    archivePrefix = "arXiv",
    primaryClass = "hep-th",
    doi = "10.1007/JHEP01(2026)086",
    journal = "JHEP",
    volume = "01",
    pages = "086",
    year = "2026"
}

@article{DeLuca:2024cjl,
    author = "De Luca, Valerio and Khoury, Justin and Wong, Sam S. C.",
    title = "{Gravitational memory and soft theorems: The local perspective}",
    eprint = "2412.01910",
    archivePrefix = "arXiv",
    primaryClass = "gr-qc",
    doi = "10.1103/gbg1-mz49",
    journal = "Phys. Rev. D",
    volume = "112",
    number = "2",
    pages = "L021502",
    year = "2025"
}

@article{DeLuca:2024asq,
    author = "De Luca, Valerio and Khoury, Justin and Wong, Sam S. C.",
    title = "{Gravitational memory and Ward identities in the local detector frame}",
    eprint = "2412.12273",
    archivePrefix = "arXiv",
    primaryClass = "gr-qc",
    doi = "10.1103/PhysRevD.112.024032",
    journal = "Phys. Rev. D",
    volume = "112",
    number = "2",
    pages = "024032",
    year = "2025"
}

@article{Mao:2025yne,
    author = "Mao, Pujian",
    title = "{Supertranslation in the bulk for generic spacetime}",
    eprint = "2512.20331",
    archivePrefix = "arXiv",
    primaryClass = "hep-th",
    month = "12",
    year = "2025"
}

@article{Bart:2019gnf,
    author = "Bart, Henk",
    title = "{Gravitational memory in the bulk}",
    eprint = "1908.07505",
    archivePrefix = "arXiv",
    primaryClass = "gr-qc",
    reportNumber = "MPP-2019-180",
    doi = "10.1007/JHEP05(2020)106",
    journal = "JHEP",
    volume = "05",
    pages = "106",
    year = "2020"
}

@inproceedings{Diaz:2026zzp,
    author = "Diaz, Felipe",
    title = "{Nonperfect Carrollian Fluids Through Holography}",
    booktitle = "{VIII international conference ''Models in Quantum Field Theory''}: {dedicated to professor Alexander Nikolaevich Vasiliev}",
    eprint = "2602.00396",
    archivePrefix = "arXiv",
    primaryClass = "hep-th",
    month = "1",
    year = "2026"
}

@article{Campoleoni:2026abr,
    author = "Campoleoni, Andrea and Delfante, Arnaud and Geiller, Marc and Maindiaux, Nicolas",
    title = "{Asymptotically-FLRW$_3$ spacetimes}",
    eprint = "2606.02722",
    archivePrefix = "arXiv",
    primaryClass = "gr-qc",
    month = "6",
    year = "2026"
}

@article{Rahman:2019bmk,
    author = "Rahman, Adel A. and Wald, Robert M.",
    title = "{Black Hole Memory}",
    eprint = "1912.12806",
    archivePrefix = "arXiv",
    primaryClass = "gr-qc",
    doi = "10.1103/PhysRevD.101.124010",
    journal = "Phys. Rev. D",
    volume = "101",
    number = "12",
    pages = "124010",
    year = "2020"
}

@article{Hartong:2026rbr,
    author = "Hartong, Jelle and Have, Emil and Nenmeli, Vijay and Oling, Gerben",
    title = "{The Energy-Momentum-News Complex near Future Null Infinity}",
    eprint = "2607.07872",
    archivePrefix = "arXiv",
    primaryClass = "hep-th",
    month = "7",
    year = "2026"
}

@article{Hartong:2025jpp,
    author = "Hartong, Jelle and Have, Emil and Nenmeli, Vijay and Oling, Gerben",
    title = "{Boundary Energy-Momentum Tensors for Asymptotically Flat Spacetimes}",
    eprint = "2505.05432",
    archivePrefix = "arXiv",
    primaryClass = "hep-th",
    month = "5",
    year = "2025"
}

@article{Kraus:2024gso,
    author = "Kraus, Per and Myers, Richard M.",
    title = "{Carrollian Partition Functions and the Flat Limit of AdS}",
    eprint = "2407.13668",
    archivePrefix = "arXiv",
    primaryClass = "hep-th",
    month = "7",
    year = "2024"
}

@article{Bagchi:2023fbj,
    author = "Bagchi, Arjun and Dhivakar, Prateksh and Dutta, Sudipta",
    title = "{AdS Witten diagrams to Carrollian correlators}",
    eprint = "2303.07388",
    archivePrefix = "arXiv",
    primaryClass = "hep-th",
    doi = "10.1007/JHEP04(2023)135",
    journal = "JHEP",
    volume = "04",
    pages = "135",
    year = "2023"
}

@article{Poole:2025cmv,
    author = "Poole, Aaron and Skenderis, Kostas and Taylor, Marika",
    title = "{Gravitational charges and radiation in asymptotically locally de Sitter spacetimes}",
    eprint = "2512.14243",
    archivePrefix = "arXiv",
    primaryClass = "hep-th",
    month = "12",
    year = "2025"
}

@article{Arenas-Henriquez:2022www,
    author = "Arenas-Henriquez, Gabriel and Gregory, Ruth and Scoins, Andrew",
    title = "{On acceleration in three dimensions}",
    eprint = "2202.08823",
    archivePrefix = "arXiv",
    primaryClass = "hep-th",
    doi = "10.1007/JHEP05(2022)063",
    journal = "JHEP",
    volume = "05",
    pages = "063",
    year = "2022"
}

@article{Cisterna:2023qhh,
    author = "Cisterna, Adolfo and Diaz, Felipe and Mann, Robert B. and Oliva, Julio",
    title = "{Exploring accelerating hairy black holes in 2+1 dimensions: the asymptotically locally anti-de Sitter class and its holography}",
    eprint = "2309.05559",
    archivePrefix = "arXiv",
    primaryClass = "hep-th",
    doi = "10.1007/JHEP11(2023)073",
    journal = "JHEP",
    volume = "11",
    pages = "073",
    year = "2023"
}

@article{Arenas-Henriquez:2023hur,
    author = "Arenas-Henriquez, Gabriel and Cisterna, Adolfo and Diaz, Felipe and Gregory, Ruth",
    title = "{Accelerating Black Holes in $2+1$ dimensions: Holography revisited}",
    eprint = "2308.00613",
    archivePrefix = "arXiv",
    primaryClass = "hep-th",
    doi = "10.1007/JHEP09(2023)122",
    journal = "JHEP",
    volume = "09",
    pages = "122",
    year = "2023"
}

@book{Griffiths:2009dfa,
    author = "Griffiths, Jerry B. and Podolsky, Jiri",
    title = "{Exact Space-Times in Einstein's General Relativity}",
    doi = "10.1017/CBO9780511635397",
    isbn = "978-1-139-48116-8",
    publisher = "Cambridge University Press",
    address = "Cambridge",
    series = "Cambridge Monographs on Mathematical Physics",
    year = "2009"
}

@article{Barnich:2010eb,
    author = "Barnich, Glenn and Troessaert, Cedric",
    title = "{Aspects of the BMS/CFT correspondence}",
    eprint = "1001.1541",
    archivePrefix = "arXiv",
    primaryClass = "hep-th",
    reportNumber = "ULB-TH-09-28",
    doi = "10.1007/JHEP05(2010)062",
    journal = "JHEP",
    volume = "05",
    pages = "062",
    year = "2010"
}

@article{Skenderis:2017dnh,
    author = "Skenderis, Kostas and Withers, Benjamin",
    title = "{Robinson-Trautman spacetimes and gauge/gravity duality}",
    eprint = "1703.10865",
    archivePrefix = "arXiv",
    primaryClass = "hep-th",
    doi = "10.22323/1.292.0097",
    journal = "PoS",
    volume = "CORFU2016",
    pages = "097",
    year = "2017"
}

@article{Bredberg:2010ky,
    author = "Bredberg, Irene and Keeler, Cynthia and Lysov, Vyacheslav and Strominger, Andrew",
    title = "{Wilsonian Approach to Fluid/Gravity Duality}",
    eprint = "1006.1902",
    archivePrefix = "arXiv",
    primaryClass = "hep-th",
    doi = "10.1007/JHEP03(2011)141",
    journal = "JHEP",
    volume = "03",
    pages = "141",
    year = "2011"
}

@inproceedings{Pasterski:2021raf,
    author = "Pasterski, Sabrina and Pate, Monica and Raclariu, Ana-Maria",
    title = "{Celestial Holography}",
    booktitle = "{Snowmass 2021}",
    eprint = "2111.11392",
    archivePrefix = "arXiv",
    primaryClass = "hep-th",
    month = "11",
    year = "2021"
}

@book{Strominger:2017zoo,
    author = "Strominger, Andrew",
    title = "{Lectures on the Infrared Structure of Gravity and Gauge Theory}",
    eprint = "1703.05448",
    archivePrefix = "arXiv",
    primaryClass = "hep-th",
    isbn = "978-0-691-17973-5",
    month = "3",
    year = "2017"
}

@article{donnay2022goldilocks,
  title={Goldilocks modes and the three scattering bases},
  author={Donnay, Laura and Pasterski, Sabrina and Puhm, Andrea},
  journal={Journal of High Energy Physics},
  volume={2022},
  number={6},
  pages={1--50},
  year={2022},
  publisher={Springer}
}

@article{Compere:2019bua,
    author = "Comp\`ere, Geoffrey and Fiorucci, Adrien and Ruzziconi, Romain",
    title = "{The $\Lambda$-BMS$_4$ group of dS$_4$ and new boundary conditions for AdS$_4$}",
    eprint = "1905.00971",
    archivePrefix = "arXiv",
    primaryClass = "gr-qc",
    doi = "10.1088/1361-6382/ab3d4b",
    journal = "Class. Quant. Grav.",
    volume = "36",
    number = "19",
    pages = "195017",
    year = "2019",
    note = "[Erratum: Class.Quant.Grav. 38, 229501 (2021)]"
}

@article{Lipstein:2025jfj,
    author = "Lipstein, Arthur and Ruzziconi, Romain and Yelleshpur Srikant, Akshay",
    title = "{Towards a Flat Space Carrollian Hologram from AdS$_4$/CFT$_3$}",
    eprint = "2504.10291",
    archivePrefix = "arXiv",
    primaryClass = "hep-th",
    month = "4",
    year = "2025"
}

@article{Fiorucci:2025twa,
    author = "Fiorucci, Adrien and Pekar, Simon and Petropoulos, P. Marios and Vilatte, Matthieu",
    title = "{Carrollian-holographic Derivation of BMS Flux-balance Laws}",
    eprint = "2505.00077",
    archivePrefix = "arXiv",
    primaryClass = "hep-th",
    reportNumber = "CPHT-RR035.042025",
    month = "4",
    year = "2025"
}

@article{Duary:2024kxl,
    author = "Duary, Sarthak and Upadhyay, Shivam",
    title = "{Flat limit of AdS/CFT from AdS geodesics: scattering amplitudes and antipodal matching of Li\'enard-Wiechert fields}",
    eprint = "2411.08540",
    archivePrefix = "arXiv",
    primaryClass = "hep-th",
    month = "11",
    year = "2024"
}

@article{Duval:2014uva,
    author = "Duval, C. and Gibbons, G. W. and Horvathy, P. A.",
    title = "{Conformal Carroll groups and BMS symmetry}",
    eprint = "1402.5894",
    archivePrefix = "arXiv",
    primaryClass = "gr-qc",
    doi = "10.1088/0264-9381/31/9/092001",
    journal = "Class. Quant. Grav.",
    volume = "31",
    pages = "092001",
    year = "2014"
}

@article{Donnay:2022aba,
    author = "Donnay, Laura and Fiorucci, Adrien and Herfray, Yannick and Ruzziconi, Romain",
    title = "{Carrollian Perspective on Celestial Holography}",
    eprint = "2202.04702",
    archivePrefix = "arXiv",
    primaryClass = "hep-th",
    doi = "10.1103/PhysRevLett.129.071602",
    journal = "Phys. Rev. Lett.",
    volume = "129",
    number = "7",
    pages = "071602",
    year = "2022"
}

@article{Donnay:2022wvx,
    author = "Donnay, Laura and Fiorucci, Adrien and Herfray, Yannick and Ruzziconi, Romain",
    title = "{Bridging Carrollian and celestial holography}",
    eprint = "2212.12553",
    archivePrefix = "arXiv",
    primaryClass = "hep-th",
    doi = "10.1103/PhysRevD.107.126027",
    journal = "Phys. Rev. D",
    volume = "107",
    number = "12",
    pages = "126027",
    year = "2023"
}

@article{Robinson:1962zz,
    author = "Robinson, I. and Trautman, A.",
    title = "{Some spherical gravitational waves in general relativity}",
    doi = "10.1098/rspa.1962.0036",
    journal = "Proc. Roy. Soc. Lond. A",
    volume = "265",
    pages = "463--473",
    year = "1962"
}

@article{compere2020lambda,
  title={The $\Lambda$-BMS4 charge algebra},
  author={Compere, Geoffrey and Fiorucci, Adrien and Ruzziconi, Romain},
  journal={Journal of High Energy Physics},
  volume={2020},
  number={10},
  pages={1--45},
  year={2020},
  publisher={Springer}
}

@article{Bakas:2014kfa,
    author = "Bakas, Ioannis and Skenderis, Kostas",
    title = "{Non-equilibrium dynamics and $AdS_4$ Robinson-Trautman}",
    eprint = "1404.4824",
    archivePrefix = "arXiv",
    primaryClass = "hep-th",
    doi = "10.1007/JHEP08(2014)056",
    journal = "JHEP",
    volume = "08",
    pages = "056",
    year = "2014"
}

@article{Flanagan:2023jio,
    author = "Flanagan, Eanna E. and Nichols, David A.",
    title = "{Fully nonlinear transformations of the Weyl-Bondi-Metzner-Sachs asymptotic symmetry group}",
    eprint = "2311.03130",
    archivePrefix = "arXiv",
    primaryClass = "gr-qc",
    doi = "10.1007/JHEP03(2024)120",
    journal = "JHEP",
    volume = "03",
    pages = "120",
    year = "2024"
}

@article{Freidel:2021fxf,
    author = "Freidel, Laurent and Oliveri, Roberto and Pranzetti, Daniele and Speziale, Simone",
    title = "{The Weyl BMS group and Einstein{\textquoteright}s equations}",
    eprint = "2104.05793",
    archivePrefix = "arXiv",
    primaryClass = "hep-th",
    doi = "10.1007/JHEP07(2021)170",
    journal = "JHEP",
    volume = "07",
    pages = "170",
    year = "2021"
}

@article{Flanagan:2015pxa,
    author = "Flanagan, {\'E}anna {\'E}. and Nichols, David A.",
    title = "{Conserved charges of the extended Bondi-Metzner-Sachs algebra}",
    eprint = "1510.03386",
    archivePrefix = "arXiv",
    primaryClass = "hep-th",
    doi = "10.1103/PhysRevD.95.044002",
    journal = "Phys. Rev. D",
    volume = "95",
    number = "4",
    pages = "044002",
    year = "2017",
    note = "[Erratum: Phys.Rev.D 108, 069902 (2023)]"
}

@article{Mao:2024urq,
    author = "Mao, Pujian and Zeng, Baijun",
    title = "{Note on post-Minkowskian expansion and Bondi coordinates}",
    eprint = "2405.11953",
    archivePrefix = "arXiv",
    primaryClass = "gr-qc",
    doi = "10.1016/j.nuclphysb.2025.117111",
    journal = "Nucl. Phys. B",
    volume = "1019",
    pages = "117111",
    year = "2025"
}

@article{Ashtekar:1981bq,
    author = "Ashtekar, A. and Streubel, M.",
    title = "{Symplectic Geometry of Radiative Modes and Conserved Quantities at Null Infinity}",
    doi = "10.1098/rspa.1981.0109",
    journal = "Proc. Roy. Soc. Lond. A",
    volume = "376",
    pages = "585--607",
    year = "1981"
}

@article{Bagchi:2025vri,
    author = "Bagchi, Arjun and Banerjee, Aritra and Dhivakar, Prateksh and Mondal, Saikat and Shukla, Ashish",
    title = "{The Carrollian kaleidoscope}",
    eprint = "2506.16164",
    archivePrefix = "arXiv",
    primaryClass = "hep-th",
    doi = "10.1140/epjc/s10052-026-15437-1",
    journal = "Eur. Phys. J. C",
    volume = "86",
    number = "4",
    pages = "429",
    year = "2026"
}

@article{Bagchi:2026emg,
    author = "Bagchi, Arjun and Lipstein, Arthur and Mondal, Saikat and Zhang, Alex Jiayi",
    title = "{Carrollian ABJM: Fermions and Supersymmetry}",
    eprint = "2604.22582",
    archivePrefix = "arXiv",
    primaryClass = "hep-th",
    month = "4",
    year = "2026"
}

@article{Bulunur:2026yav,
    author = "Bulunur, Ilayda and Ergec, Osman and Kasikci, Oguzhan and Ozkan, Mehmet and Zog, Mustafa Salih",
    title = "{A Twisted Origin for Magnetic Carroll Supersymmetry}",
    eprint = "2603.28269",
    archivePrefix = "arXiv",
    primaryClass = "hep-th",
    month = "3",
    year = "2026"
}

@article{Henneaux:2026dfc,
    author = "Henneaux, Marc",
    title = "{Carroll supergravities}",
    eprint = "2607.08329",
    archivePrefix = "arXiv",
    primaryClass = "hep-th",
    month = "7",
    year = "2026"
}

@article{Ballesteros:2026bqe,
    author = "Ballesteros, Romina and Lescano, Eric and Pati{\~n}o-L{\'o}pez, Sergio",
    title = "{Carrollian limit of NS-NS and Heterotic Supergravity}",
    eprint = "2607.09847",
    archivePrefix = "arXiv",
    primaryClass = "hep-th",
    month = "7",
    year = "2026"
}

@article{SenGupta:1966,
  author = {N.~D.~Sen Gupta},
  title = {On an Analogue of the Galilei Group},
  journal = {Nuovo Cim. A},
  volume = {44},
  pages = {512--517},
  year = {1966},
  doi = {10.1007/BF02740871}
}

@article{Nguyen:2025zhg,
    author = "Nguyen, Kevin",
    title = "{Lectures on Carrollian holography}",
    eprint = "2511.10162",
    archivePrefix = "arXiv",
    primaryClass = "hep-th",
    doi = "10.1140/epjc/s10052-026-15952-1",
    journal = "Eur. Phys. J. C",
    volume = "86",
    number = "6",
    pages = "733",
    year = "2026"
}

@article{Mars:1993mj,
    author = "Mars, Marc and Senovilla, Jose M. M.",
    title = "{Geometry of general hypersurfaces in space-time: Junction conditions}",
    eprint = "gr-qc/0201054",
    archivePrefix = "arXiv",
    doi = "10.1088/0264-9381/10/9/026",
    journal = "Class. Quant. Grav.",
    volume = "10",
    pages = "1865--1897",
    year = "1993"
}

@article{Ciambelli:2024kre,
    author = "Ciambelli, Luca and Pasterski, Sabrina and Tabor, Elisa",
    title = "{Radiation in holography}",
    eprint = "2404.02146",
    archivePrefix = "arXiv",
    primaryClass = "hep-th",
    doi = "10.1007/JHEP09(2024)124",
    journal = "JHEP",
    volume = "09",
    pages = "124",
    year = "2024"
}

@article{Arenas-Henriquez:2024ypo,
    author = "Arenas-Henriquez, Gabriel and Diaz, Felipe and Rivera-Betancour, David",
    title = "{Generalized Fefferman-Graham gauge and boundary Weyl structures}",
    eprint = "2411.12513",
    archivePrefix = "arXiv",
    primaryClass = "hep-th",
    doi = "10.1007/JHEP02(2025)007",
    journal = "JHEP",
    volume = "02",
    pages = "007",
    year = "2025"
}

@article{Hu:2023geb,
    author = "Hu, Yangrui and Pasterski, Sabrina",
    title = "{Detector operators for celestial symmetries}",
    eprint = "2307.16801",
    archivePrefix = "arXiv",
    primaryClass = "hep-th",
    doi = "10.1007/JHEP12(2023)035",
    journal = "JHEP",
    volume = "12",
    pages = "035",
    year = "2023"
}

@article{He:2014laa,
    author = "He, Temple and Lysov, Vyacheslav and Mitra, Prahar and Strominger, Andrew",
    title = "{BMS supertranslations and Weinberg{\textquoteright}s soft graviton theorem}",
    eprint = "1401.7026",
    archivePrefix = "arXiv",
    primaryClass = "hep-th",
    doi = "10.1007/JHEP05(2015)151",
    journal = "JHEP",
    volume = "05",
    pages = "151",
    year = "2015"
}

@article{Donnay:2021wrk,
    author = "Donnay, Laura and Ruzziconi, Romain",
    title = "{BMS flux algebra in celestial holography}",
    eprint = "2108.11969",
    archivePrefix = "arXiv",
    primaryClass = "hep-th",
    doi = "10.1007/JHEP11(2021)040",
    journal = "JHEP",
    volume = "11",
    pages = "040",
    year = "2021"
}

@article{Raclariu:2021zjz,
    author = "Raclariu, Ana-Maria",
    title = "{Lectures on Celestial Holography}",
    eprint = "2107.02075",
    archivePrefix = "arXiv",
    primaryClass = "hep-th",
    month = "7",
    year = "2021"
}

@article{Gonzo:2020xza,
    author = "Gonzo, Riccardo and Pokraka, Andrzej",
    title = "{Light-ray operators, detectors and gravitational event shapes}",
    eprint = "2012.01406",
    archivePrefix = "arXiv",
    primaryClass = "hep-th",
    reportNumber = "SAGEX-20-24-E",
    doi = "10.1007/JHEP05(2021)015",
    journal = "JHEP",
    volume = "05",
    pages = "015",
    year = "2021"
}

@article{Podolsky:2016sff,
    author = "Podolsk{\'y}, J. and {\v{S}}varc, R.",
    title = "{Algebraic classification of Robinson-Trautman spacetimes}",
    eprint = "1608.07118",
    archivePrefix = "arXiv",
    primaryClass = "gr-qc",
    doi = "10.1103/PhysRevD.94.064043",
    journal = "Phys. Rev. D",
    volume = "94",
    number = "6",
    pages = "064043",
    year = "2016"
}

\end{document}